\documentclass[onecolumn,showpacs,nofootinbib,prd,10pt,amssymb]{revtex4-1}
\usepackage{latexsym}
\usepackage{graphicx,epsf, epsfig, amssymb, amsmath}
\usepackage{relsize}
\usepackage{dsfont}
\usepackage{todonotes}
\usepackage{textcomp}
\usepackage{float}

\DeclareMathAlphabet\mathbfcal{OMS}{cmsy}{b}{n}

\newcommand{\aap}{Astron.\ Astrophys.}
\newcommand{\mnras}{Mon.\ Not.\ R.\ Astron.\ Soc.}

\newcommand{\apjl}{Astrophys.\ J.\ Lett.}
\newcommand{\apjs}{Astrophys.\ J.\ Suppl.\ Ser.}
\newcommand{\araa}{Ann.\ Rev.\ Astron.\ Astrophys.}

\newcommand{\Fperp}{\mathbin{^{\mathsmaller{\perp}}\mkern-3mu F}}
\newcommand{\Fpar}{\mathbin{^{\mathsmaller{\parallel}}\mkern-3mu F}}

\newcommand{\Gperp}{\mathbin{^{\mathsmaller{\perp}}\mkern-2mu G}}
\newcommand{\Gpar}{\mathbin{^{\mathsmaller{\parallel}}\mkern-1mu G}}

\newcommand{\Vperp}{\mathbin{^{\mathsmaller{\perp}}\mkern-1mu \mathcal{V}}}
\newcommand{\Vpar}{\mathbin{^{\mathsmaller{\parallel}}\mkern-1mu \mathcal{V}}}

\newcommand{\Wperp}{\mathbin{^{\mathsmaller{\perp}}\mkern-1mu \mathcal{W}}}
\newcommand{\Wpar}{\mathbin{^{\mathsmaller{\parallel}}\mkern-1mu \mathcal{W}}}

\newcommand{\Aperp}{\mathbin{^{\mathsmaller{\perp}}\mkern-6mu \mathcal{A}}}
\newcommand{\Apar}{\mathbin{^{\mathsmaller{\parallel}}\mkern-6mu \mathcal{A}}}
\newcommand{\Adual}{\mathbin{^{\star}\mkern-6mu \mathcal{A}}}

\newcommand{\Fdual}{\mathbin{^{\star}\mkern-3mu F}}
\newcommand{\Gdual}{\mathbin{^{\star}\mkern-1mu G}}
\newcommand{\FFdual}{\mathbin{^{\star}\mkern-3mu \mathcal{F}}}
\newcommand{\Vdual}{\mathbin{^{\star}\mkern-1mu \mathcal{V}}}

\newcommand{\GammaMik}{ \Gamma_{ik}^{\mathsmaller{({\rm M})}}}
\newcommand{\GammaEik}{ \Gamma_{ik}^{\mathsmaller{({\rm E})}}}
\newcommand{\GammaMnn}{ \Gamma_{nn}^{\mathsmaller{({\rm M})}}}

\newcommand{\GammaEiktilde}{ \widetilde{\Gamma}_{ik}^{\mathsmaller{({\rm E})}}}
\newcommand{\GammaEitilde}{ \widetilde{\Gamma}_{i}^{\mathsmaller{({\rm E})}}}
\newcommand{\gammaEtilde}{ \widetilde{\gamma}^{\mathsmaller{({\rm E})}}}

\newcommand{\GammaMki}{ \Gamma_{ki}^{\mathsmaller{({\rm M})}}}
\newcommand{\GammaEki}{ \Gamma_{ki}^{\mathsmaller{({\rm E})}}}

\newcommand{\GammaMi}{ \Gamma_{i}^{\mathsmaller{({\rm M})}}}
\newcommand{\GammaMp}{ \Gamma_{p}^{\mathsmaller{({\rm M})}}}
\newcommand{\GammaEi}{ \Gamma_{i}^{\mathsmaller{({\rm E})}}}
\newcommand{\gammaM}{ \gamma^{\mathsmaller{({\rm M})}}}
\newcommand{\gammaE}{ \gamma^{\mathsmaller{({\rm E})}}}

\begin{document}

\title{Relativistic dynamics of superfluid-superconducting mixtures
in the presence of topological defects and an electromagnetic
field with application to neutron stars}

\author{
M.~E.~Gusakov$^{1,2}$, V.~A.~Dommes$^1$} 
\affiliation{$^1$Ioffe Physical-Technical Institute of the Russian Academy
  of Sciences, Polytekhnicheskaya 26, 194021 Saint-Petersburg, Russia}
\affiliation{$^2$Peter the Great Saint-Petersburg Polytechnic University, Polytekhnicheskaya
  29, 195251 Saint-Petersburg, Russia}

\begin{abstract} 
The relativistic dynamic equations are derived for 
a superfluid-superconducting mixture coupled to the electromagnetic field.
For definiteness, and bearing in mind possible applications of our results to neutron stars, 
it is assumed that the mixture is composed of 
superfluid neutrons, superconducting protons, and normal electrons. 
Proton superconductivity of both I and II types is analysed, 
and possible presence of 
neutron 
and proton 
vortices
(or magnetic domains in the case of type-I proton superconductivity) 
is allowed for.
The derived equations 
neglect all dissipative effects 
except for the mutual friction dissipation and 
are valid for arbitrary temperatures
(i.e. they do not imply that all nucleons are paired),
which is especially important 
for magnetar conditions.
It is demonstrated that 
these general equations can be substantially simplified for typical neutron stars, 
for which a kind of magnetohydrodynamic approximation is justified.
Our results are compared to the nonrelativistic formulations existing in the literature
and a number of discrepancies are found.
In particular, it is shown that, generally, 
the electric displacement ${\pmb D}$ 
does not coincide with the electric field ${\pmb E}$,
contrary to what is stated in the previous works.
The relativistic framework developed here
is easily extendable 
to account for
more sophisticated 
microphysics 
models
and 
it provides the necessary basis for realistic modelling of neutron stars.
\end{abstract}
\date{\today}

%
%
%
%
%
%






\maketitle

\section{Introduction}

Assume that we have a relativistic magnetized finite-temperature plasma 
(possibly in the strong gravitational field)
composed of superfluid neutral particles,
superconducting positively charged particles and 
normal (nonsuperconducting) negatively charged particles.
Depending on the density, the positively charged particles 
may form either type-I or type-II superconductor,
and the plasma may contain topological defects 
-- Feynman-Onsager and/or Abrikosov vortices. 
What are the macroscopic dynamic equations
describing such a 
system?

The question is not so far-fetched as it may seem at first glance.
For example, the neutron-proton-electron ($npe$) 
mixture in the outer
neutron-star cores
meets all the conditions formulated above.
First, it is relativistic and magnetized. 
The typical surface magnetic field is 
$B \sim 10^8 \div 10^{15}$~G \cite{vigano_etal_13, hpy07} and is likely to be larger 
in the deeper layers \cite{gwh16}; 
the surface gravitation acceleration
is also huge, $g_{\rm s} \sim 2 \times 10^{14}$~cm~s$^{-2}$ \cite{hpy07},
electrons are ultra-relativistic, while neutrons can be moderately relativistic.
Second, according to microscopic calculations \cite{gps14, ls01}, 
confirmed (to some extent) by observations 
of cooling and glitching neutron stars \cite{yp04, plps13, hm15},
neutrons and protons in their interiors become superfluid/superconducting 
at temperatures $T\lesssim T_{{\rm c}i}$,
where $T_{{\rm c}i} \sim 10^8 \div 10^{10}$~K is the nucleon critical temperature ($i=n$, $p$).
Third, in a rotating magnetized 
neutron star it can be energetically favourable
to form Feynman-Onsager/Abrikosov vortices \cite{sauls89}
(the latter are formed only if protons are type-II superconductor; 
if, instead, they are of type-I, 
different structures appear, 
see Sec.\ \ref{TypeI} for more details).

Thus it is 
not surprising
that the dynamic properties of magnetized superfluid-superconducting 
neutron-star
plasma have been the subject of numerous studies in the past,
both in nuclear matter
(see, e.g., Refs.\ \cite{ep77, als84, vs81, hk87, ml91, mendell91a, mendell91b, ss95, mendell98, gas11, prix05})
and in quark matter
(e.g., Refs.\ \cite{ib02a, ib02b, as10, hac12}).
In particular, Vardanyan and Sedrakyan \cite{vs81} were the first who
generalized hydrodynamics of a mixture of two superfluids \cite{khalatnikov00,ab76} 
to charged superfluids coupled to the electromagnetic field.
These equations were further extended by Holm and Kupershmidt \cite{hk87}
to $N$ charged superfluids, who derived these equations from the Hamiltonian formalism.
Finally, the most general {\it nonrelativistic} finite-temperature equations,
describing charged superfluids and accounting for the mutual friction forces \cite{hall60,hv56}
between various liquid components, 
were formulated by Mendell and Lindblom \cite{ml91}, 
who used in their work the ideas of Refs.\ \cite{bk61, khalatnikov00, hk87}.
This important work was subsequently used by Mendell \cite{mendell91a,mendell91b}
who applied the equations of Ref.\ \cite{ml91} to neutron stars, 
assuming that all neutrons and protons are paired 
(i.e., $T\ll T_{{\rm c}i}$). 
(A little bit later, Sedrakian and Sedrakian \cite{ss95} 
did a similar job by extending the results of Ref.\ \cite{vs81}
to include dissipation and mutual friction forces in their equations.)
In his work, Mendell formulated a set of simplified magnetohydrodynamic equations,
but, unfortunately, incorrectly identified 
the magnetic field ${\pmb H}$ with the magnetic induction ${\pmb B}$
and the electric displacement ${\pmb D}$ with the electric field ${\pmb E}$.
The first of these inaccuracies (identification of ${\pmb H}$ with ${\pmb B}$) 
was noticed in Ref.\ \cite{cpl00} and corrected by Glampedakis, Andersson, and Samuelsson 
\cite{gas11} (hereafter GAS11);
the second inaccuracy (identification of ${\pmb D}$ with ${\pmb E}$) is discussed here 
(see Appendix \ref{vortapp}).
Except for the corrected inaccuracy, 
the GAS11 version of magnetohydrodynamics 
is equivalent (up to notations) 
to that of Mendell \cite{mendell91a}
and is the most advanced treatment 
of superfluid-superconducting mixtures in neutron stars up to date.
It is derived using the variational framework \cite{prix04,prix05} and assuming $T=0$.

All the works discussed by us so far were performed in the nonrelativistic approximation.
This is 
a rather serious shortcoming
because, as we have already mentioned, 
neutron stars are essentially relativistic objects. 
The extension of magnetohydrodynamics of GAS11 
(as well as more general equations of Ref.\ \cite{ml91})
to the relativistic case is 
not trivial.
For uncharged one-component superfluids this problem 
has been addressed in Refs.\ \cite{kl82,lk82,lsc98,carter00,cl95,kg12,dg16,awv16} 
and has recently been ``solved'' in Ref.\ \cite{gusakov16} (hereafter G16).
We are aware of only one attempt \cite{cl98} 
to consider charged mixtures in full relativity. 
This reference neglected all dissipation effects (including mutual friction)
and studied only the low-temperature case $T\ll T_{{\rm c}i}$;
unfortunately, it did not 
provide a nonrelativistic limit for the derived equations
so that it is hard to compare them with 
the formulations available in the literature.
Note that Ref.\ \cite{cl98} adopted 
the 
variational approach 
similar to
that 
developed
in Ref.\ \cite{cl95} 
in application to uncharged superfluids.
This approach was criticised in G16 (see Appendix F there) 
where it was argued that it does not reproduce 
the well established nonrelativistic 
Hall-Vinen-Bekarevich-Khalatnikov superfluid hydrodynamics \cite{bk61, khalatnikov00}.
We believe the same 
conclusion applies also to the results of Ref.\ \cite{cl98}.

The aim of the present study is to fill the existing gaps
and derive a set of relativistic finite-temperature equations 
describing 
superfluid-superconducting mixtures, 
bearing in mind application of these results 
to magnetized rotating neutron stars.
As in Refs.\ \cite{bk61} and G16, 
our derivation rests on the consistency 
between the conservation laws 
and
the entropy generation equation.
For definiteness, in this paper we consider a liquid composed of superfluid neutrons ($n$), 
superconducting protons ($p$), and normal electrons ($e$).
Extension of our results to more complicated compositions is straightforward 
(see, e.g., Refs.\ \cite{gk08, kg09, dg16, hac12}).
Here we are mostly interested in the non-dissipative equations
(but we allow for mutual friction dissipation, 
see Remark~1 in Sec.\ \ref{TypeII}).
Correspondingly, we assume that neutron and proton thermal excitations
as well as electrons move with one and the same ``normal'' four-velocity $u^{\mu}$.
In what follows all thermodynamic quantities are defined in the frame
comoving with the normal (nonsuperfluid) liquid component, 
in which $u^{\mu}=(1,0,0,0)$.
By default, any 3d-vector appearing in the text 
(e.g., magnetic induction ${\pmb B}$)
is written in that frame.

The paper is organized as follows.
Section \ref{Maxwell} introduces Maxwell's equations in the medium
written both in the standard and 
explicitly Lorentz-covariant
form. 
Section \ref{no} considers uncharged and charged mixtures in the absence
of vortices and other magnetic domain structures.
In Sec.\ \ref{setup} we discuss the strategy for generalization of equations of Sec.\ \ref{no}
in order to allow for the topological defects 
and related bound charges and currents
in the mixture.
In Sec.\ \ref{TypeI} 
this strategy
is applied to derive
the corresponding dynamic equations 
under assumption of
type-I superconductivity of protons.
Section \ref{TypeII} is devoted to considering type-II proton
superconductivity 
and 
accounting for
the possible presence 
of both neutron (Feynman-Onsager) and proton (Abrikosov) vortices. 
Section \ref{symmetry} proves that the energy-momentum tensors
obtained in Secs.\ \ref{TypeI} and \ref{TypeII} are symmetric,
and expresses them through a set of phenomenological coefficients
which can be calculated by specifying a microscopic model for the energy-density
of the mixture.
The general dynamic equations of Sec.\ \ref{TypeII}
are simplified for typical neutron-star conditions in Sec.\ \ref{MHDapprox}.
Finally, we sum up in Sec.\ \ref{summary}.

The paper also contains a number of appendices, 
where we present technical, more model-dependent, 
or less important results.
In particular, 
Appendix \ref{notation} introduces 
some basic notation used throughout the paper.
Appendix \ref{corresp}
provides a correspondence table between our notation and that adopted in G16.
Appendix \ref{transformation} contains
an example of the energy density transformation 
used in Secs.\ \ref{TypeI} and \ref{TypeII}.
Appendix \ref{Abraham} reveals the relation between the energy-momentum tensor 
of Sec.\ \ref{TypeI} and the well known Abraham tensor.
Appendix \ref{vortex1} discusses some general relations
characterizing isolated neutron or proton vortices.
Appendix \ref{why} demonstrates that there exist some bound charges associated 
with each moving vortex.
Appendix \ref{deterapp} presents two simple microscopic models 
allowing one to determine the phenomenological coefficients 
from Sec.\ \ref{symmetry}.
Finally, Appendix \ref{sumapp} contains
the full set of dynamic equations derived in Secs.\ \ref{TypeI} and \ref{TypeII},
and Appendix \ref{nonrel} analyses the nonrelativistic limit 
of simplified equations of Sec.\ \ref{MHDapprox}.

Unless otherwise stated, 
in all sections except for Sec.\ \ref{Maxwell} 
and Appendices \ref{vortex1}, \ref{why}, \ref{deterapp}, and \ref{nonrel}
the speed of light $c$, the Planck constant $\hbar$,
and the Boltzmann constant $k_{\rm B}$ are set to unity,
$c=\hbar=k_{\rm B}=1$.
Throughout the paper we assume that the spacetime metric is flat,
$g_{\mu\nu}={\rm diag}(-1,\, 1,\, 1,\, 1)$. 
Generalization of our results to arbitrary $g_{\mu\nu}$ is straightforward
and can be achieved by replacing ordinary derivatives in all equations
with their covariant counterparts.

\section{Maxwell's equations in the medium}
\label{Maxwell}

\subsection{Standard form of Maxwell's equations}

Maxwell's equations in the medium take the form
\begin{eqnarray}
{\rm div} \, {\pmb D} &=&4 \pi \rho_{\rm free},
\label{divE1}\\
{\rm curl} \, {\pmb E} &=& - \frac{1}{c} \, \frac{\partial {\pmb B}}{\partial t},
\label{rotE1}\\
{\rm div} \, {\pmb B} &=&0,
\label{divB1}\\
{\rm curl} \, {\pmb H} &=& \frac{4 \pi}{c} \, {\pmb J}_{\rm free}
+\frac{1}{c} \, \frac{\partial {\pmb D}}{\partial t},
\label{rotB1}
\end{eqnarray}
where ${\pmb E}$ and ${\pmb B}$ are the electric field and magnetic induction, respectively;
${\pmb D}$ and ${\pmb H}$ are the electric displacement and magnetic field, respectively;
$\rho_{\rm free}$ and ${\pmb J}_{\rm free}$ are macroscopic averages
of the free charge and current densities in the medium
(e.g., Ref.\ \cite{jackson98}).
In the absence of bound charges and currents one has 
${\pmb D}={\pmb E}$ and ${\pmb H}={\pmb B}$.

Equations (\ref{divE1})--(\ref{rotB1}) 
contain the continuity equation for the electric charge,
\begin{equation}
\frac{\partial \rho_{\rm free}}{\partial t}+{\rm div}\, {\pmb J_{\rm free}} =0,
\label{cont}
\end{equation}
and the energy equation,
\begin{equation}
\frac{\partial \varepsilon_{\rm EM}}{\partial t} = 
-{\pmb E}{\pmb J}_{\rm free}+\frac{c}{4\pi}\, {\rm div\left[{\pmb H}\times {\pmb E}\right]},
\label{energy}
\end{equation}
where 
\begin{equation}
d \varepsilon_{\rm EM} =\frac{1}{4\pi} \,{\pmb E} d{\pmb D}+\frac{1}{4\pi} \,{\pmb H} d{\pmb B}
\label{eEM0}
\end{equation}
is the differential of the electromagnetic energy density $\varepsilon_{\rm EM}$.

\subsection{Relativistic representation}

Maxwell's equations (\ref{divE1})--(\ref{rotB1}) can be rewritten 
in a manifestly Lorentz-covariant form \cite{ll60, toptygin15}.
To see this let us introduce the tensors $F^{\alpha\beta}$ and $G^{\alpha\beta}$ such that
\begin{eqnarray}
F^{\alpha \beta} \equiv \partial^\alpha A^\beta - \partial^\beta A^\alpha
&=&\left( 
\begin{array}{cccc}
0    & E_1  &  E_2 & E_3\\
-E_1 &  0   &  B_3 & -B_2 \\
-E_2 & -B_3 &  0   & B_1 \\
-E_3 & B_2  & -B_1 & 0 
\end{array} 
\right),
\label{Fik}\\
G^{\alpha\beta}&=&\left( 
\begin{array}{cccc}
0    & D_1  & D_2  & D_3\\
-D_1 &  0   & H_3  &  -H_2 \\
-D_2 & -H_3 & 0    & H_1 \\
-D_3 & H_2  & -H_1 & 0 
\end{array} 
\right),
\label{Gik}
\end{eqnarray}
where $A^{\alpha}=(\phi,\, {\pmb A})$ 
is the electromagnetic four-potential %
%
\footnote{We remind that in a given coordinate system:
\begin{eqnarray}
{\pmb E} &=& -\frac{1}{c} \, \frac{\partial {\pmb A}}{\partial t}- {\pmb \nabla} \phi,
\nonumber\\
{\pmb B}&=& {\rm curl}\, {\pmb A}.
\nonumber
\end{eqnarray}
}.
%
Using the definitions (\ref{Fik})--(\ref{Gik}),
Maxwell's equations (\ref{divE1})--(\ref{rotB1}) can be represented as
\begin{eqnarray}
\partial_{\alpha} 
\Fdual^{\alpha\beta}&=&0,
\label{11}\\
\partial_\alpha G^{\alpha\beta}&=&-4\pi \, J_{({\rm free})}^\beta,
\label{22}
\end{eqnarray}
where $J_{({\rm free})}^\alpha=(\rho_{\rm free},\, {\pmb J_{\rm free}}/c)$ 
is the four-current density of free charges and $\Fdual^{\mu\nu}$
is the tensor dual to $F^{\mu\nu}$ (see Appendix \ref{notation}).

\subsection{Four-vectors $E^\mu$, $B^\mu$, $D^\mu$, and $H^\mu$}
\label{EBDH}

As is shown in Appendix \ref{notation},
for any antisymmetric tensor 
it is possible to
introduce the 
corresponding ``electric'' and ``magnetic'' four-vectors 
[see Eqs.\ (\ref{AE}) and (\ref{AM})].
In the case of electromagnetic tensors $F^{\mu\nu}$ and $G^{\mu\nu}$ 
we shall use 
the following (standard) notation for these vectors,
\begin{eqnarray}
E^\mu &\equiv& F^{\mu}_{(\rm E)} = u_{\nu} F^{\mu\nu},
\label{Emux}\\
D^{\mu}&\equiv& G^{\mu}_{(\rm E)}=u_{\nu} G^{\mu\nu},
\label{Dmux}\\
B^{\mu} &\equiv& F^{\mu}_{(\rm M)}
=u_{\nu} 
\Fdual^{\mu\nu}= 
\frac{1}{2} \, \epsilon^{\mu \nu \lambda \eta} \, u_{\nu} \, F_{\lambda \eta},
\label{Bmux}\\
H^{\mu} &\equiv& G^{\mu}_{(\rm M)}
=u_{\nu} 
\Gdual^{\mu\nu} = 
\frac{1}{2} \, \epsilon^{\mu \nu \lambda \eta} \, u_{\nu} \, G_{\lambda \eta}
\label{Hmux}
\end{eqnarray}
instead of,
respectively, the universal notations 
$F^{\mu}_{(\rm E)}$, $G^{\mu}_{(\rm E)}$, $F^{\mu}_{(\rm M)}$, and $G^{\mu}_{(\rm M)}$ 
suggested in Appendix~\ref{notation}.
In the comoving frame, in which the four-velocity of normal liquid component
is $u^\mu=(1,\, 0,\, 0,\, 0)$
these vectors reduce to $E^{\mu}=(0,\, {\pmb E})$, 
$B^{\mu}=(0,\, {\pmb B})$,
$D^{\mu}=(0,\, {\pmb D})$, and
$H^{\mu}=(0,\, {\pmb H})$.

\section{No vortices, bound charges, and bound currents}
\label{no}

In order to establish notations and get some insight into the problem 
we start with the simplest possible situation
and discuss relativistic 
equations 
for the superfluid-superconducting $npe$-mixture
without vortices, bound charges, and bound currents. 
The latter assumption means that we set ${\pmb D}={\pmb E}$ and ${\pmb H}={\pmb B}$ 
in all equations in this section.

\subsection{General structure}

The relativistic equations describing $npe$-mixture
consist of the energy-momentum conservation,
\begin{equation}
\partial_{\mu}T^{\mu\nu}=0
\label{dTmunu}
\end{equation}
and continuity equations for particle species $j$ 
(here and hereafter index  
$j=n$, $p$, and $e$) %
%
\footnote{We neglect, for clarity, possible sources in these equations due to beta-processes,
thus assuming that the latter are effectively frozen. 
They can be easily accounted for if necessary.},
%
\begin{equation}
\partial_{\mu}j^{\mu}_{(j)}=0.
\label{dj}
\end{equation}
In Eqs.\ (\ref{dTmunu}) and (\ref{dj}) 
$T^{\mu\nu}$ is the total energy-momentum tensor, which is
a sum of fluid and electromagnetic contributions, 
\begin{equation}
T^{\mu\nu}=T^{\mu\nu}_{({\rm fluid})}+T^{\mu\nu}_{(\rm EM)},
\label{Tmunu1}
\end{equation}
and $j^{\mu}_{(j)}$ is the current density for particle species $j$.
These equations should be supplemented by the
second law of thermodynamics, Maxwell's equations (see Sec.\ \ref{Maxwell}), 
as well as by a number of additional equations and constraints describing 
superfluid degrees of freedom (see below).

\subsection{Uncharged mixtures}
\label{uncharged}

Assume for a moment that all the mixture components ($n$, $p$, and  $e$) are uncharged.
The corresponding nondissipative hydrodynamics has been extensively studied, 
e.g., in  Refs.\ \cite{ga06, gusakov07, kg11, gkcg13}. 
It consists of 
the second law of thermodynamics
\begin{equation}
d \varepsilon_{\rm fluid} = T \, d S + \mu_i \, d n_i + \mu_{e} \, d n_{e} 
+ { Y_{ik} \over 2} \, d \left( w^{\alpha}_{(i)} w_{(k) \alpha} \right)
\label{2ndlaw}
\end{equation}
and
Eqs.\ (\ref{dTmunu}), (\ref{dj}),
in which the energy-momentum tensor, 
$T^{\mu\nu}=T^{\mu\nu}_{({\rm fluid})}$, is given by
\begin{equation}
T^{\mu \nu}_{\rm (fluid)} = (P_{\rm fluid}+\varepsilon_{\rm fluid}) \, u^{\mu} u^{\nu} + P_{\rm fluid} \, g^{\mu \nu} 
+ Y_{ik} \left( w^{\mu}_{(i)} w^{\nu}_{(k)} + \mu_i \, w^{\mu}_{(k)} u^{\nu} 
+ \mu_k \, w^{\nu}_{(i)} u^{\mu} \right), 
\label{Tmunufluid}
\end{equation}
and the particle four-currents are
\begin{eqnarray}
j^{\mu}_{(i)} &=& n_i u^{\mu} + Y_{ik} w^{\mu}_{(k)},
\label{ji}\\
j^{\mu}_{({e})} &=& n_{e} u^{\mu}.
\label{je}
\end{eqnarray}
Here and below, the subscripts $i$ and $k$ 
refer to nucleons: $i$, $k=n$, $p$. 
Unless otherwise stated, 
a summation is assumed over repeated spacetime indices 
$\mu$, $\nu$, $\ldots$ (Greek letters) 
and nucleon species indices $i$ and $k$ (Latin letters).

In Eqs.\ (\ref{2ndlaw})--(\ref{je})
$\varepsilon_{\rm fluid}$ and $S$ are the fluid energy density and entropy density, respectively;
$T$ is the temperature;
$\mu_j$ and $n_j$ are the relativistic chemical potential 
and number density for particles $j=n$, $p$, and $e$, respectively;
$P_{\rm fluid}$ is the pressure given by
the standard formula
\begin{equation}
P_{\rm fluid} \equiv -\frac{\partial \left(\varepsilon_{\rm fluid} V \right)}{\partial V} 
= -\varepsilon_{\rm fluid} +\mu_e n_e+\mu_i n_i + TS,
\label{pres}
\end{equation}
where $V$ is the system volume and the partial derivative is taken
at fixed total number of particles $n_j V$ ($j=n$, $p$, $e$)
total entropy $S V$, and fixed scalars $w_{(i)\mu}w^{\mu}_{(k)}$ \cite{gusakov16, ab76, khalatnikov00}.

Further, $Y_{ik}$ in Eqs.\ (\ref{2ndlaw})--(\ref{ji})
is the relativistic entrainment matrix \cite{ga06, gkh09a,gkh09b,ghk14}, 
analogue of the 
superfluid or mass-density matrix $\rho_{ik}$ 
of the non-relativistic theory \cite{ab76, bjk96, gh05, ch06, gusakov10}.
In the non-relativistic limit both matrices 
are related by the formula \cite{ga06}: 
$Y_{ik}=\rho_{ik}/(m_i m_k c^2)$, 
where $m_i$ is the bare nucleon mass ($i=n$ or $p$).
Finally, 
the normal four-velocity $u^{\mu}$ is normalized by the condition
\begin{equation}
u_{\mu}u^{\mu}=-1;
\label{norm}
\end{equation}
and the four-vectors $w^{\mu}_{(i)}$ 
in Eqs.\ (\ref{2ndlaw})--(\ref{ji}) 
describe the superfluid degrees of freedom
and are subject to condition
\begin{equation}
u_{\mu}w^{\mu}_{(i)}=0,
\label{uw}
\end{equation}
which ensures that all the thermodynamic quantities are indeed defined (measured)
in the comoving frame in which $u^{\mu}=(1,\, 0,\, 0,\, 0)$
[see G16 for a detailed discussion].
In particular, using Eq.\ (\ref{uw}) one finds from Eqs.\ (\ref{Tmunufluid}) and (\ref{ji})
\begin{eqnarray}
u_{\mu}u_{\nu}T^{\mu\nu}_{\rm fluid} &=& \varepsilon_{\rm fluid},
\label{uuT}\\
u_{\mu}j^{\mu}_{(i)} &=&-n_i.
\label{uj}
\end{eqnarray}
To close the system of hydrodynamic equations we need 
two additional 
conditions
relating the
four-vectors $w^{\mu}_{(i)}$ 
with the wave function phases $\Phi_i$ of the nucleon Cooper-pair condensates.
These conditions are ($i=n$, $p$)
\begin{equation}
w^{\mu}_{(i)} \equiv \partial^\mu \phi_i  
- \mu_i u^{\mu},
\label{def}
\end{equation}
where the scalar $\phi_i=\Phi_i/2$.
Equations (\ref{def}) can be reformulated exclusively in terms of $w^{\mu}_{(i)}$ as
\begin{equation}
\partial_\mu\left[ w_{(i)\nu}+\mu_i u_{\nu}\right]-\partial_\nu\left[ w_{(i)\mu}+\mu_i u_{\mu}\right]=0.
\label{sfleq}
\end{equation}
It is simply a statement that 
$\partial_\mu\partial_\nu \phi_i -\partial_\nu\partial_\mu \phi_i=0$
(or, equivelently,
$\partial_\mu\partial_\nu \Phi_i -\partial_\nu\partial_\mu \Phi_i=0$).

The system of hydrodynamic equations is now closed and contains, in particular,
the entropy generation equation, 
which can be obtained
by composing a vanishing combination, $u_{\nu}\, \partial_\mu T^{\mu\nu}=0$,
and following the same derivation as that discussed in G16. 
Ignoring for the moment the ``superfluid'' equations (\ref{def}) [or (\ref{sfleq})],
one obtains 
\begin{equation}
T \, \partial_\mu(S u^{\mu}) = 
u^{\nu} \, Y_{ik} w^{\mu}_{(k)}\left\{
\partial_\mu\left[ w_{(i)\nu}+\mu_i u_{\nu}\right]-\partial_\nu\left[ w_{(i)\mu}+\mu_i u_{\mu}\right]
\right\}.
\label{entropy}
\end{equation}
The right-hand side of this equation vanishes in view of Eq.\ (\ref{sfleq}),
so that the system entropy does not increase%
%
\footnote{We remind the reader that 
in this work we are mainly interested in the nondissipative dynamics.}
%
and is carried
with the same velocity $u^{\mu}$ as the normal (nonsuperfluid) liquid component.

\subsection{Charged mixtures}
\label{charged}

How should equations of the previous section be modified for charged mixtures?
Concerning the continuity equations (\ref{dj}),
the corresponding particle current densities are still given by 
Eqs.\ (\ref{ji}) and (\ref{je}),
and should be considered as definitions 
of the four-vectors $u^{\mu}$ and $w^{\mu}_{(i)}$ 
[$u^{\mu}$ is still being normalized by Eq.\ (\ref{norm})].
The condition (\ref{uw}) also remains unchanged since 
it directly follows from
the comoving frame 
definition 
(see Section IIA of G16 for a thorough discussion of this issue).
Next, the second law of thermodynamics (\ref{2ndlaw}) and the pressure definition (\ref{pres})
retain their form, 
because they are written for the fluid energy density and fluid pressure,
and hence should not include field contributions 
%
\footnote{
\label{reasons}
We remind the reader that the situation considered in this section 
(superfluid-superconducting mixture in the absence of vortices and not in the intermediate state) 
allows us to separate fluid and field degrees of freedom.
}.
%
In contrast, the energy-momentum tensor $T^{\mu\nu}$ in Eq.\ (\ref{dTmunu}) should be modified 
in order to account for the electromagnetic field contribution.
It is now given by Eq.\ (\ref{Tmunu1})
with
\begin{equation}
T^{\mu\nu}_{({\rm EM})}= \frac{1}{4\pi}\left(
F^\mu_{\,\,\,\,\gamma} F^{\nu\gamma} 
- \frac{1}{4} \, g^{\mu\nu} \, F_{\gamma\delta}F^{\gamma\delta}
\right).
\label{TmunuEM}
\end{equation}
This standard \cite{weinberg71} electromagnetic 
tensor is
obtained under assumption 
$\pmb{D}=\pmb{E}$ and $\pmb{H}=\pmb{B}$ 
[and hence $G^{\alpha\beta}=F^{\alpha\beta}$, see Eqs.\ (\ref{Fik}) and (\ref{Gik})].
It does not include any ``mixed'' terms depending on both
fluid and field degrees of freedom 
because of the 
same reason 
as that
discussed in the footnote \ref{reasons}.
A more general situation, in which such a decoupling 
is ambiguous (not well-defined),
is considered in Secs.\ \ref{TypeI} and \ref{TypeII}.

It remains to find out how the 
presence of charges
affects the 
superfluid equations (\ref{def}) and/or (\ref{sfleq}).
For that, it is instructive to 
repeat the derivation of the entropy generation equation,
now taking into account the electromagnetic contribution (\ref{TmunuEM}).
Again, composing a vanishing combination
$u_{\nu}\, \partial_\mu T^{\mu\nu}=u_{\nu}\, \partial_\mu T^{\mu\nu}_{({\rm fluid})} 
+u_{\nu}\, \partial_\mu T^{\mu\nu}_{({\rm EM})}=0$
and noting that 
$\partial_\mu T^{\mu\nu}_{({\rm EM})}=-F^{\nu\mu}J_{({\rm free})\mu}$
on account of Maxwell's equations (\ref{11}) and (\ref{22})
(see, e.g., $\S$~8, Chapter 2 of Ref.\ \cite{weinberg71}),
one gets
\begin{equation}
T \, \partial_\mu(S u^{\mu}) = 
u^{\nu} \, Y_{ik} w^{\mu}_{(k)}\left\{
\partial_\mu\left[ w_{(i)\nu}+\mu_i u_{\nu}\right]-\partial_\nu\left[ w_{(i)\mu}+\mu_i u_{\mu}\right]
\right\} 
- u^{\nu} \, F_{\nu\mu}J^\mu_{({\rm free})}, 
\label{entropy2}
\end{equation}
where
the four-current density of free charges is given by the formula
[we use Eqs.\ (\ref{ji}) and (\ref{je})]
\begin{equation}
J^\mu_{({\rm free})}=e_j j^{\mu}_{(j)}=J^{\mu}_{({\rm norm})}+e_i \, Y_{ik}w^{\mu}_{(k)},
\label{jfree2}
\end{equation}
in which $e_j$ is the charge of particle $j$ and 
\begin{equation}
J^\mu_{({\rm norm})}= e_j n_j u^\mu = e_p (n_p-n_e) u^\mu
\label{Jnorm}
\end{equation}
is the normal (non-superconducting) part  of the four-current density.
Correspondingly, 
noticing 
that
$F_{\nu\mu}=\partial_\nu A_\mu-\partial_\mu A_\nu$ 
[see Eq.\ (\ref{Fik})]
and $E^{\mu}=u_\nu F^{\mu\nu}$ [Eq.\ (\ref{Emux})],
Eq.\ (\ref{entropy2}) can be rewritten as
\begin{equation}
T \, \partial_\mu(S u^{\mu}) = 
u^{\nu} \, Y_{ik} w^{\mu}_{(k)}\left\{
\partial_\mu\left[ w_{(i)\nu}+\mu_i u_{\nu}+e_i A_{\nu}\right]
-\partial_\nu\left[ w_{(i)\mu}+\mu_i u_{\mu}+e_i A_{\mu}\right]
\right\} 
+E_{\mu}J^\mu_{({\rm norm})}.
\label{entropy3}
\end{equation}
The last term in the r.h.s.\ of this equation equals zero,
\begin{equation}
E_{\mu}J^\mu_{({\rm norm})}=0,
\label{EJ}
\end{equation}
in view of the 
definitions (\ref{Emux}), (\ref{Jnorm}), 
and the equality 
\begin{equation}
u_{\mu}u_{\nu}F^{\mu\nu}=0,
\label{uuF}
\end{equation}
following from the antisymmetry property of the tensor $F^{\mu\nu}$.
Equation (\ref{entropy3}) then becomes
\begin{equation}
T \, \partial_\mu(S u^{\mu}) = 
u^{\nu} \, Y_{ik} w^{\mu}_{(k)}\left\{
\partial_\mu\left[ w_{(i)\nu}+\mu_i u_{\nu}+e_i A_{\nu}\right]
-\partial_\nu\left[ w_{(i)\mu}+\mu_i u_{\mu}+e_i A_{\mu}\right]
\right\}.
\label{entropy4}
\end{equation}
The r.h.s.\ of this equation must vanish identically because, by assumption,
there should be no entropy generation in the system 
(we disregard all the dissipative corrections).
Using this requirement,
it is tempting to conclude that the new form of the 
superfluid equation in the presence of the electromagnetic field is
\begin{equation}
\partial_\mu\left[ w_{(i)\nu}+\mu_i u_{\nu}+e_i A_{\nu}\right]
-\partial_\nu\left[ w_{(i)\mu}+\mu_i u_{\mu}+e_i A_{\mu}\right]=0
\label{sfleq2}
\end{equation}
or, equivalently,
\begin{equation}
w^{\mu}_{(i)} = \partial^\mu \phi_i  
- \mu_i u^{\mu}
- e_i A^{\mu} ,
\label{def2}
\end{equation}
where, again, the scalar $\phi_i=\Phi_i/2$.
This is indeed the correct equation that could be obtained immediately
from the requirement of gauge invariance of the resulting superfluid hydrodynamics 
(see, e.g., Ref.\ \cite{ga06}). 
As follows from the microscopic theory \cite{ll80}, 
the wave function phase $\Phi_i$ and the four-potential $A^{\mu}$ transform as 
\begin{eqnarray}
A^\mu \rightarrow A^\mu+ \partial^\mu \chi,
\label{A}\\
\Phi_i \rightarrow \Phi_i+2e_i \chi
\label{Phi}
\end{eqnarray}
under gauge transformations ($\chi$ is an arbitrary scalar function). 
The four-vectors $w^{\mu}_{(i)}$ 
and hence Eqs.\ (\ref{sfleq2}), (\ref{def2}), 
and other equations in this section
are thus manifestly gauge-invariant %
%
\footnote{The four-vectors $w^{\mu}_{(i)}$ ($i=n$, $p$)
are observables (i.e., must be gauge-invariant)
since
they define the particle current density $j^{\mu}_{(i)}$ 
in the comoving frame [see Eq.\ (\ref{ji})].
}.
%
The system of relativistic equations formulated here 
reduces to the vortex-free equations of Mendell \cite{mendell91a}
and Sedrakian et al. \cite{ss95}  in the non-relativistic limit 
(see also Ref.\ \cite{putterman74}).

\vspace{0.2 cm}
\noindent
%
{\bf Remark 1.} ---
As noted above, 
a simple problem
considered by us here 
allows to decouple the fluid and field degrees of freedom.
In this approach $\varepsilon_{\rm fluid}$ and $P_{\rm fluid}$ are, respectively, 
the {\it fluid} energy density and pressure, 
while field contributions are treated separately.
Such a decoupling 
is hampered
in more general situations (see Secs.\ \ref{TypeI} and \ref{TypeII}).
To facilitate comparison with the results of Secs.\ \ref{TypeI} and \ref{TypeII}
it is worth to reformulate 
the equations discussed here in terms of the total energy density $\varepsilon$,
\begin{equation}
\varepsilon=\varepsilon_{\rm fluid}+\varepsilon_{\rm EM}
\label{eps}
\end{equation}
and the ``pressure'' $P$, defined as [cf. Eq.\ (\ref{pres})]
\begin{equation}
P \equiv -\frac{\partial \left(\varepsilon V \right)}{\partial V} = -\varepsilon +\mu_e n_e+\mu_i n_i + TS,
\label{pres1}
\end{equation}
where the partial derivative is taken at constant $n_j V$ ($j=n$, $p$, $e$),
$S V$, $w_{(i)\mu} w^{\mu}_{(k)}$, $B^{\mu}$, and $D^{\mu}$ 
($=E^{\mu}$ in this section).
In Eq.\ (\ref{eps}) 
$\varepsilon_{\rm EM}$ is the energy density of the electromagnetic field 
measured in the comoving frame,
\begin{equation}
\varepsilon_{\rm EM}=
\frac{{\pmb E}^2}{8\pi}+\frac{{\pmb B}^2}{8\pi} = 
\frac{E_\alpha E^{\alpha}}{8\pi}+\frac{B_\alpha B^{\alpha}}{8\pi},
\label{epsfield}
\end{equation}
where the four-vectors $E^\alpha$ and $B^{\alpha}$ are 
given
by Eqs.\ (\ref{Emux}) and (\ref{Bmux}); in the comoving frame 
they equal, respectively, 
$(0,\, {\pmb E})$ and $(0,\, {\pmb B})$. 
Using Eq.\ (\ref{epsfield}),  it follows
from Eq.\ (\ref{pres1}) that 
\begin{equation}
P=P_{\rm fluid}-\frac{1}{8\pi}
\left(
E_\alpha E^{\alpha}+B_\alpha B^{\alpha}
\right).
\label{Pres2}
\end{equation}

Before 
reformulating the dynamic equations
it is instructive to note that 
the energy-momentum tensor $T^{\mu\nu}_{({\rm EM})}$ 
of the electromagnetic field
can generally be rewritten as
\begin{equation}
T^{\mu\nu}_{({\rm EM})}= -\frac{1}{8\pi} \, \left( E_{\alpha}E^{\alpha}+B_{\alpha}B^{\alpha}\right)
g^{\mu\nu}
+ T^{\mu\nu}_{({\rm E})} + T^{\mu\nu}_{({\rm M})},
\label{TmunuEM2}
\end{equation}
where the ``electric'' part of the tensor equals
\begin{equation}
T^{\mu\nu}_{({\rm E})}=
-\frac{1}{4\pi} \, 
\left(
E^{\mu} E^{\nu} - \perp^{\mu\nu} E_{\alpha} E^{\alpha} 
\right)
\label{TmunuE}
\end{equation}
and the ``magnetic'' part is 
\begin{equation}
T^{\mu\nu}_{({\rm M})}=\frac{1}{4\pi}
\left(
\perp_{\delta \alpha} F^{\mu\delta} F^{\nu \alpha}
-u^{\mu} u^{\nu} u^{\gamma} u_{\beta} F^{\alpha \beta} F_{\alpha \gamma}
\right).
\label{TmunuM}
\end{equation}
Here $\perp^{\mu\nu} \equiv g^{\mu\nu}+u^{\mu}u^{\nu}$ is the projection operator.
Using Eqs.\ (\ref{eps})--(\ref{TmunuM})
the second law of thermodynamics 
takes the form
[cf.\ Eq.\ (\ref{2ndlaw})]
\begin{equation}
d \varepsilon = T \, d S + \mu_i \, d n_i + \mu_{e} \, d n_{e} 
+ { Y_{ik} \over 2} \, d \left( w^{\alpha}_{(i)} w_{(k) \alpha} \right)
+\frac{1}{4\pi} E_{\alpha} d E^{\alpha} + \frac{1}{4\pi} B_{\alpha} d B^{\alpha},
\label{2ndlaw2}
\end{equation}
while the tensor $T^{\mu\nu}$ 
becomes
[cf.\ Eq.\ (\ref{Tmunu1})]
\begin{equation}
T^{\mu \nu} = (P+\varepsilon) \, u^{\mu} u^{\nu} + P g^{\mu \nu} 
+ Y_{ik} \left( w^{\mu}_{(i)} w^{\nu}_{(k)} + \mu_i \, w^{\mu}_{(k)} u^{\nu} 
+ \mu_k \, w^{\nu}_{(i)} u^{\mu} \right) + T^{\mu\nu}_{({\rm E})} + T^{\mu\nu}_{({\rm M})}.
\label{Tmunu2}
\end{equation}
Because 
\begin{eqnarray}
u_{\mu}u_{\nu}T^{\mu\nu}_{({\rm E})}&=&0,
\label{E}\\
u_{\mu}u_{\nu}T^{\mu\nu}_{({\rm M})}&=&0,
\label{M}
\end{eqnarray}
it satisfies the condition 
\begin{equation}
u_{\mu}u_{\nu}T^{\mu\nu}= \varepsilon.
\label{uuT2}
\end{equation}
All other hydrodynamic equations remain unchanged.

\section{Setting up the problem}
\label{setup}

Simple examples considered in the previous section
suggest a possible general approach to the problem of formulation of the macroscopic 
(smooth-averaged) dynamic equations
in various interesting situations 
(e.g., in the system with vortices 
or in the system with small-scale
domain structure of the magnetic field).
The approach is based on using the entropy generation equation 
to constrain the dynamics of superfluid-superconducting mixtures;
it
has been applied recently 
in G16 
(see also Ref.\ \cite{bk61}) and
we refer the interested reader to those references for more details.
All the quantities in this and subsequent sections are 
assumed to be averaged over the volume containing 
large amount of inhomogeneities (vortices or magnetic domains).

Assume that the second law of thermodynamics takes the form
\begin{equation}
d \varepsilon = T \, d S + \mu_i \, d n_i + \mu_{e} \, d n_{e} 
+ { Y_{ik} \over 2} \, d \left( w^{\alpha}_{(i)} w_{(k) \alpha} \right)
+d \varepsilon_{\rm add},
\label{2ndlaw3}
\end{equation}
where $\varepsilon$ is the {\it total} energy density of the system.
All the terms in the r.h.s.\ of this equation except for the last one 
are the standard terms of superfluid hydrodynamics 
[see Eq.\ (\ref{2ndlaw}) and Refs.\ \cite{ga06, gusakov07, gkcg13, gusakov16}];
an additional term $d \varepsilon_{\rm add}$ 
contains
vortex or electromagnetic contribution to $d \varepsilon$, or both.

Accounting for 
this term in Eq.\ (\ref{2ndlaw3})
should not affect most of the dynamic equations
due to the very same reasons as those discussed in 
the beginning of Sec.\ \ref{charged}
(see also section IIIB in G16, 
where a similar problem is discussed in detail).
In particular, 
the expressions (\ref{ji}) and (\ref{je})
for the free four-current densities 
$j^{\mu}_{(n)}$, $j^{\mu}_{(p)}$, and $j^{\mu}_{(e)}$
[which satisfy the continuity equation (\ref{dj})]
should be considered as the {\it definitions} of 
the four-vectors $w^{\mu}_{(n)}$, $w^{\mu}_{(p)}$, 
and the four-velocity 
$u^{\mu}$, 
normalized by the condition (\ref{norm}).
Thus, they remain unchanged.
Next, the requirement that
all the thermodynamic quantities are measured 
in the (comoving) frame, in which $u^{\mu}=(1,\,0,\,0,\,0)$ %
%
\footnote{Mathematically, this requirement is expressed by the condition (\ref{uj}).},
%
unambiguously leads to the same constraints (\ref{uw}).
Finally, 
the free-charge four-current density $J^{\mu}_{({\rm free})}$ 
and the pressure $P$
will still be defined by Eqs.\ (\ref{jfree2}) and (\ref{pres1}), respectively; 
Maxwell's equations (\ref{divE1})--(\ref{rotB1}) or (\ref{11})--(\ref{22})
will also, of course, retain their form.
 
The only equations that should be modified are the expression for the 
total energy-momentum tensor $T^{\mu\nu}$,
\begin{equation}
T^{\mu \nu} = (P+\varepsilon) \, u^{\mu} u^{\nu} + P g^{\mu \nu} 
+ Y_{ik} \left( w^{\mu}_{(i)} w^{\nu}_{(k)} + \mu_i \, w^{\mu}_{(k)} u^{\nu} 
+ \mu_k \, w^{\nu}_{(i)} u^{\mu} \right) + \Delta T^{\mu\nu}.
\label{Tmunu3}
\end{equation}
[which still satisfies Eq.\ (\ref{dTmunu})], 
and the ``superfluid'' equations for neutrons and protons
[Eqs.\ (\ref{sfleq2}) or (\ref{def2}) 
in the simple example of Sec.\ \ref{charged}].
The 
correction
$\Delta T^{\mu\nu}$ in Eq.\ (\ref{Tmunu3})
must be {\it symmetric}; 
it
includes
vortex and/or electromagnetic contributions to $T^{\mu\nu}$
and is absent in the standard superfluid hydrodynamics 
[see Eq.\ (\ref{Tmunufluid})].
Because in the comoving frame 
the component $T^{00}$ of the tensor $T^{\mu\nu}$
equals, by definition, $\varepsilon$,%
%
\footnote{In an arbitrary frame this requirement translates into
$u_{\mu}u_{\nu} T^{\mu\nu}=\varepsilon$, see Eq.\ (\ref{uuT2}).}
%
one should have there $\Delta T^{00}=0$, or, in an arbitrary frame,
\begin{equation}
u_{\mu}u_{\nu} \, \Delta T^{\mu\nu}=0.
\label{dT}
\end{equation}

To determine 
the correction $\Delta T^{\mu\nu}$ and the form
of superfluid equations, 
we,  
as was already mentioned,
utilize the entropy generation equation.
It can be derived 
using the equations discussed above in this section.
The result is 
[cf.\ Eq.\ (\ref{entropy4}) and also equation (65) in G16]
\begin{equation}
T \,\partial_\mu(S u^{\mu}) = 
u_{\nu} \, Y_{ik} w_{(k)\mu} \, \widetilde{\mathcal{V}}^{\mu\nu}_{(i)}
-u^{\mu} \partial_{\mu}\varepsilon_{\rm add}
+u_{\nu}\partial_{\mu}\Delta T^{\mu\nu},
\label{entropy5}
\end{equation}
where
\begin{equation}
\widetilde{\mathcal{V}}^{\mu\nu}_{(i)} \equiv \partial^\mu\left[ w^\nu_{(i)}+\mu_i u^{\nu}\right]
-\partial^\nu\left[ w^\mu_{(i)}+\mu_i u^{\mu}\right].
\label{tildeV}
\end{equation}
As one sees, Eq.\ (\ref{entropy5}) depends on 
$\varepsilon_{\rm add}$, which is assumed to be specified,
and on $\Delta T^{\mu\nu}$, which is unknown.
Because entropy is conserved in the absence of dissipation, 
the r.h.s.\ of this equation should vanish identically.
As shown in Secs.\ \ref{TypeI} and \ref{TypeII}, 
this requirement is sufficient 
to fully reconstruct dynamics 
of superfluid-superconducting $npe$-mixture.

\section{Relativistic dynamic equations for $npe$-mixture: Type-I proton superconductivity}
\label{TypeI}

In this section we consider 
a nonrotating superfluid-superconducting $npe$-mixture in the absence 
of Feynman-Onsager and Abrikosov (single flux quantum) vortices.
However, in contrast to Sec.\ \ref{charged} we formally assume that 
the magnetic field ${\pmb H}$ does not necessarily coincide with 
the magnetic induction ${\pmb B}$, 
i.e., there are some bound currents in the system.
One can imagine that these currents can be generated
either 
due to 
(very weak, in reality) 
magnetic 
response
of particles in the mixture (e.g., electrons) 
to an applied external magnetic field (case 1), 
or due to appearance of various 
inhomogeneous structures
of the (microscopic) magnetic field in the mixture
similar to 
those
appearing in the intermediate state 
of ordinary type-I superconductors 
(see, e.g., Ref.\ \cite{huebener13} 
and Sec.\ \ref{TypeIinter} below; case 2).
The dynamic equations 
in this latter case are a bit more 
complicated
since 
the proton phase winding around such structures can be nonzero.
Thus, 
for pedagogical reasons
we start with the simplest (but unrealistic) 
situation of a homogeneous 
$npe$-mixture with well-behaved phases $\Phi_i$
and
${\pmb B}\neq {\pmb H}$ (case 1).
In what follows we, for generality, 
assume 
also
that 
the electric displacement ${\pmb D}$
is not equal to the electric field ${\pmb E}$ 
(although we set ${\pmb D}={\pmb E}$ in the final equations, see Sec.\ \ref{TypeIsymmetry}).

\subsection{Case 1: Homogeneous $npe$-mixture with ${\pmb B}\neq {\pmb H}$ }
\label{TypeIfree}

The starting point of our consideration is the expression 
for the electro-magnetic contribution $d \varepsilon_{\rm add}$ 
to the second law of thermodynamics,
\begin{equation}
d \varepsilon_{\rm add} = \frac{1}{4\pi} \, E_{\mu} d D^{\mu} + \frac{1}{4\pi} \, H_{\mu} d B^{\mu}.
\label{deadd1}
\end{equation}
This formula is specialized to the comoving frame, 
which is, generally, non-inertial, 
because $u^{\mu}$ changes in time and space.
In the absence of bound charges and currents one has 
$D^{\mu}=E^{\mu}$ and $H^{\mu}=B^{\mu}$, 
so that 
Eq.\ (\ref{deadd1}) 
reduces to the last two
electromagnetic terms in the r.h.s.\ of Eq.\ (\ref{2ndlaw2}).
In the special case when the comoving frame is inertial,
Eq.\ (\ref{deadd1}) is transformed to the standard form (\ref{eEM0})
[see the definitions (\ref{Emux})--(\ref{Hmux})].
Using Eqs.\ (\ref{Emux})--(\ref{Hmux}), 
the last term in Eq.\ (\ref{deadd1}) can be rewritten as
\begin{eqnarray}
\frac{1}{4 \pi} \, H_{\mu} \, d B^{\mu} &=& 
\frac{1}{4 \pi} \, H_{\mu} \, \frac{1}{2} \, \epsilon^{\mu\nu\alpha\beta} \, u_{\nu} \, d F_{\alpha\beta}
+ \frac{1}{4 \pi} \,H_{\mu} \, \frac{1}{2} \, \epsilon^{\mu\nu\alpha\beta}\, F_{\alpha\beta} \, d u_{\nu}
\nonumber\\
&=& 
\frac{1}{8 \pi} \Gperp^{\alpha\beta} d F_{\alpha\beta}
+ \frac{1}{8 \pi} \, \left( H_{\mu} \, \epsilon^{\mu\nu\alpha\beta}\, F_{\alpha\beta} 
+ \underline{u^{\nu} G_{\alpha \beta}F^{\alpha\beta}} \right) \, d u_{\nu}
\nonumber\\
&=& \frac{1}{8\pi} \, \left( 
\Gperp^{\alpha\beta} d F_{\alpha\beta}
 + 2 F_{\alpha\beta} G^{\alpha\gamma} u^{\beta} \, d u_{\gamma} \right),
\label{magnx}
\end{eqnarray}
where 
the added underlined term vanishes on account of normalization condition (\ref{norm}) 
(hence $u^{\nu}d u_{\nu}=0$)
and we used the notation from Appendix \ref{notation}.
Similarly,
\begin{eqnarray}
\frac{1}{4\pi} \, E_{\mu} d D^{\mu} 
&=& \frac{1}{4 \pi} \, 
\left[
d(D_{\mu}E^{\mu}) - D_{\mu}\, d E^{\mu}
\right]
\nonumber\\
&=& 
\frac{1}{4 \pi} \, 
\left[
d(D_{\mu}E^{\mu}) - D_{\mu} u_{\nu} \, d F^{\mu\nu}
-  D_{\mu} F^{\mu\nu} \, d u_{\nu}
\right]
\nonumber\\
&=& 
\frac{1}{4 \pi} \, 
\left[
d(D_{\mu}E^{\mu}) 
-\frac{1}{2}\left( D_{\mu} u_{\nu}-D_{\nu}u_{\mu}\right) \, d F^{\mu\nu}
- D_{\mu} F^{\mu\nu} \, d u_{\nu}
\right]
\nonumber\\
&=& 
\frac{1}{4 \pi} \, 
\left[
d(D_{\alpha}E^{\alpha}) 
+\frac{1}{2} \left(
\Gpar^{\alpha\beta}
d F_{\alpha\beta}
- 2 \, D_{\alpha} F^{\alpha\gamma} \, d u_{\gamma} \right)
\right]
\nonumber\\
&=& 
\frac{1}{4 \pi} \, 
\left[
d(D_{\alpha}E^{\alpha}) 
+\frac{1}{2} \left(
\Gpar^{\alpha\beta}
d F_{\alpha\beta}
- 2 \, G_{\alpha\beta} F^{\alpha\gamma} \, u^{\beta} \, d u_{\gamma} \right)
\right].
\label{electrx}
\end{eqnarray}
Equations (\ref{magnx}) and (\ref{electrx})
can be further transformed as described in Appendix \ref{transformation}.
For that we identify
\begin{eqnarray}
O^{\alpha\beta}  &=& \frac{1}{4\pi} \, \Gperp^{\alpha\beta},
\nonumber\\
\mathcal{F}^{\alpha\beta} &=& F^{\alpha\beta},
\nonumber\\
\mathcal{B}^{\alpha\beta} &=& F^{\alpha\beta},
\nonumber\\
\mathcal{A}^{\alpha\beta} &=& \frac{1}{4\pi} \, G^{\alpha\beta}
\nonumber
\end{eqnarray}
in case of Eq.\ (\ref{magnx})
and 
\begin{eqnarray}
O^{\alpha\beta}  &=& \frac{1}{4\pi} \, \Gpar^{\alpha\beta},
\nonumber\\
\mathcal{F}^{\alpha\beta} &=& F^{\alpha\beta},
\nonumber\\
\mathcal{B}^{\alpha\beta} &=& -\frac{1}{4\pi} \, G^{\alpha\beta},
\nonumber\\
\mathcal{A}^{\alpha\beta} &=& F^{\alpha\beta}
\nonumber
\end{eqnarray}
in case of Eq.\ (\ref{electrx}).
As a result, the second term in the r.h.s.\ of Eq.\ (\ref{entropy5})
can be presented as [see Eq.\ (\ref{de2})]
\begin{eqnarray}
-u^{\mu} \partial_{\mu}\varepsilon_{\rm add} &=&
u^{\nu} F_{\mu\nu} \, \partial_{\alpha}\left( \frac{1}{4\pi} \Gpar^{\mu\alpha}+
\frac{1}{4\pi} \Gperp^{\mu\alpha}\right)
\nonumber\\
&&- \partial_\mu \left[ u^\nu \left( \mathcal{T}^\mu_{({\rm E}) \, \nu}+
\mathcal{T}^\mu_{({\rm M}) \, \nu}\right)\right]
\nonumber\\
&&+ \partial_{\mu} u^{\nu}\left( 
\mathcal{T}^\mu_{({\rm E}) \, \nu}+
\mathcal{T}^\mu_{({\rm M}) \, \nu}
\right),
\label{ude2}
\end{eqnarray}
where the ``electric'' and ``magnetic'' tensors are given, respectively, by
\begin{eqnarray}
\mathcal{T}^\mu_{({\rm E}) \, \nu} &=& 
\frac{1}{4\pi} \left( \Gpar^{\mu\alpha} F_{\nu \alpha}
+ u^{\mu}u^{\gamma}
\perp_{\nu\beta} F^{\alpha\beta} G_{\alpha\gamma} 
+ g^{\mu}_{\,\,\, \nu} \, D_{\alpha} E^{\alpha}
\right),
\label{Te1}\\
\mathcal{T}^\mu_{({\rm M}) \, \nu} &=& 
\frac{1}{4\pi} \left( \Gperp^{\mu\alpha} F_{\nu \alpha}
- u^{\mu}u^{\gamma}
\perp_{\nu\beta} G^{\alpha\beta} F_{\alpha\gamma} 
\right).
\label{Tm1}
\end{eqnarray}
It can be verified that if $G^{\mu\nu}=F^{\mu\nu}$ then these tensors
reduce to the tensors $T^{\mu\nu}_{({\rm E})}$ and $T^{\mu\nu}_{({\rm M})}$
from Sec.\ \ref{charged} 
[see Eqs.\ (\ref{TmunuE}) and (\ref{TmunuM}) there],
$\mathcal{T}^{\mu\nu}_{({\rm E})}=T^{\mu\nu}_{({\rm E})}$ 
and 
$\mathcal{T}^{\mu\nu}_{({\rm M})}=T^{\mu\nu}_{({\rm M})}$. 
For actual calculations it is convenient to represent the tensors 
(\ref{Te1}) and (\ref{Tm1}) in the form
\begin{eqnarray}
\mathcal{T}^{\mu\nu}_{({\rm E})} &=& \frac{1}{4\pi} \, 
\left(
\perp^{\mu\nu} D^{\alpha}E_{\alpha}-D^{\mu}E^{\nu}
\right),
\label{Te2}\\
\mathcal{T}^{\mu\nu}_{({\rm M})} &=& 
\frac{1}{4\pi} 
\left(
\Gperp^{\mu\alpha}\Fperp^{\nu}_{\,\,\,\alpha} + u^{\nu} \Gperp^{\mu\alpha}E_{\alpha}
+u^{\mu}\Gperp^{\nu\alpha}E_{\alpha}
\right).
\label{Tm2}
\end{eqnarray}

The first term in the r.h.s.\ of Eq.\ (\ref{ude2})
can be further simplified by making use of Eqs.\ (\ref{Adec}), (\ref{22}), 
(\ref{jfree2}), (\ref{Jnorm}), and (\ref{EJ}),
\begin{eqnarray}
&&u^{\nu} F_{\mu\nu} \, \partial_{\alpha}\left( \frac{1}{4\pi} \Gpar^{\mu\alpha}+
\frac{1}{4\pi} \Gperp^{\mu\alpha}\right)
= u^{\nu} F_{\mu\nu} \, \partial_{\alpha}\left( \frac{1}{4\pi} G^{\mu\alpha} \right)
\nonumber\\
&& =  u^{\nu} F_{\mu\nu} \, J^\mu_{({\rm free})} 
=  u^{\nu} F_{\mu\nu} \, e_i Y_{ik} w^{\mu}_{(k)}.
\label{1st}
\end{eqnarray}
Substituting now Eq.\ (\ref{ude2}) into Eq.\ (\ref{entropy5}),
one gets
\begin{eqnarray}
T \,\partial_\mu(S u^{\mu}) &=& 
u_{\nu} \, Y_{ik} w_{(k)\mu} 
\left( \widetilde{\mathcal{V}}^{\mu\nu}_{(i)} + e_i F^{\mu\nu}\right)
\nonumber\\
&&- \partial_\mu \left[ u^\nu \left( \mathcal{T}^\mu_{({\rm E}) \, \nu}+
\mathcal{T}^\mu_{({\rm M}) \, \nu} - \Delta T^{\mu}_{\,\,\,\nu} \right)\right]
\nonumber\\
&&+ \partial_{\mu} u^{\nu}\left( 
\mathcal{T}^\mu_{({\rm E}) \, \nu}+
\mathcal{T}^\mu_{({\rm M}) \, \nu}
- \Delta T^{\mu}_{\,\,\,\nu}
\right),
\label{entropy6}
\end{eqnarray}
from which one can conclude that %
%
\footnote{
Equation (\ref{sfleq3}) coincides with the superfluid Eq.\ (\ref{sfleq2}) from 
Sec.\ \ref{charged}, see the definition (\ref{tildeV}). }
%
\begin{equation}
\widetilde{\mathcal{V}}^{\mu\nu}_{(i)} + e_i F^{\mu\nu}=0
\label{sfleq3}
\end{equation}
and, correspondingly, 
in order to vanish identically the r.h.s.\ of Eq.\ (\ref{entropy6}), 
\begin{eqnarray}
\Delta T^{\mu\nu} &=& \mathcal{T}^{\mu\nu}_{({\rm E})}+
\mathcal{T}^{\mu\nu}_{({\rm M})}.	
\label{DeltaTmunu}
\end{eqnarray}
Note that $\Delta T^{\mu\nu}$ automatically satisfies the condition (\ref{dT});
the fact that the tensor $\Delta T^{\mu\nu}$ is symmetric will be proven in Sec.\ \ref{TypeIsymmetry}.
The physical meaning of Eq.\ (\ref{sfleq3}) is transparent.
Using the definition (\ref{tildeV}) it can be rewritten in the form
of the superfluid Eq.\ (\ref{sfleq2}) from Sec.\ \ref{charged},
or as a gauge-invariant expression (\ref{def2}) for the four-vector $w^{\mu}_{(i)}$, 
$w^{\mu}_{(i)} = \partial^\mu \phi_i  
- \mu_i u^{\mu}
- e_i A^{\mu}$,
where the scalar $\phi_i=\Phi_i/2$ and $\Phi_i$ 
is the smooth-averaged wave function phase 
of the Cooper-pair condensate
%
\footnote{
\label{gaugefootnote}
The fact that 
$w^{\mu}_{(i)}$ 
(and hence all other dynamic equations)
appears to be gauge-invariant, 
is not trivial and is directly related 
to the adopted expression (\ref{deadd1}) for 
$d \varepsilon_{\rm add}$, 
in particular, to the assumption that $H^{\mu}$ 
in this expression
is indeed the magnetic field four-vector given by Eq.\ (\ref{Hmux}).
}.
%
Equation (\ref{sfleq3}) thus states that 
\begin{equation}
(\partial_\mu\partial_\nu-\partial_\nu\partial_\mu)\Phi_i =0,
\label{phase}
\end{equation}
which is quite natural, 
since we assume in this section 
that there are no vortices and nonsuperconducting domains 
in the system (the phases $\Phi_i$ are well-defined everywhere in the mixture).

\subsection{Case 2: $npe$-mixture in the intermediate state}
\label{TypeIinter}

Now let us
discuss how the equations of the previous section should be modified in order
to apply them to $npe$-mixture in the intermediate state.
But first let us clarify what we mean by the term ``intermediate''.

According to some estimates (e.g., GAS11), 
protons in the inner cores of neutron stars 
can form a type-I superconductor.
Upon neutron star cooling the superconducting region expands,
but 
it is generally believed that
this process is not accompanied by the magnetic flux expulsion (the Meissner effect)
because of the huge electric conductivity of the outer core and crust 
(see Refs.\ \cite{bpp69a, bpp69b} and a comment 8 in Ref.\ \cite{ib02b}).
As a result, 
it becomes energetically favourable for 
$npe$-mixture 
to
find
itself in the ``intermediate'' state,
consisting of
alternating domains
of superconducting (field-free) regions 
and nonsuperconducting regions hosting the magnetic field.
The topology of these 
domains 
can be very diverse and depends, in particular, 
on their nucleation history \cite{huebener13, pfcp14, pgpp05, sedrakyan05}.
This complicates substantially the problem of calculation of the total energy density $\varepsilon$ 
for such matter.
However, 
we neglect below the relatively small surface and boundary contributions 
to $\varepsilon$ \cite{degennes99, ll60}.
In this approximation the actual domain structure 
is not important for the energy calculation.
We further assume that 
the produced 
magnetic structures have a {\it closed topology},
i.e., normal domains are completely surrounded by the superconducting phase \cite{huebener13}.
This assumption seems reasonable since
the magnetic field of a typical neutron star, $B\sim10^{12}$~G, 
is much smaller than the 
critical thermodynamic field, 
$H_{\rm c} \sim 10^{14}-10^{15}$~G \cite{sedrakyan05},
while it is well known \cite{ll60, degennes99, huebener13, pgpp05, sedrakyan05} 
that 
it is advantageous for a relatively weak field to penetrate
the superconductor in the form 
of flux tubes, each containing many flux quanta.
For definiteness, 
this very form of normal domains (flux tubes) 
will be assumed by us in what follows.
Note, however, that 
the actual form of normal domains is not really important 
for the subsequent consideration  
(what is important is the closed topology assumption). 

The distance between the neighbouring flux tubes
can be estimated as \cite{degennes99, sedrakyan05} $b \sim \sqrt{R \delta}$, 
where $R$ is the typical size of the intermediate-state region and $\delta$
is the typical width of the normal--superconducting boundary \cite{degennes99}.
Taking $\delta \sim \xi_p \sim 10^{-11}$~cm 
($\xi_p$ is the proton coherence length)
and $R \sim 5$~km,
one obtains
$b \sim 2 \times 10^{-3}$~cm.
Then the flux tube radius is 
$a \approx b \, (B/H_{\rm c})^{1/2} \sim 6\times 10^{-5}$~cm and the number of flux quanta
in a single flux tube $N_{\phi} \approx \pi a^2 H_{\rm c}/\hat{\phi}_{p0}\approx 6\times 10^{13}$,
where $\hat{\phi}_{p0}$ is given by Eq.\ (\ref{phi0p2}), and we choose 
$B=10^{12}$~G and $H_{\rm c} = 10^{15}$~G.

From these estimates one can conclude that the flux tubes are rather large 
objects that should interact efficiently with the surrounding normal matter
(electrons and nucleon Bogoliubov excitations), and hence should move 
(at least, in the nondissipative limit) with the normal liquid component.
In the 
terminology of the Hall-Vinen-Bekarevich-Khalatnikov (HVBK) hydrodynamics
one can say that
the system is in the 
``strong-drag'' regime
(see G16).
Using the strong-drag assumption one can try to derive the dynamic equations for 
$npe$-mixture in the intermediate state.
First, 
note that 
all consideration of Sec.\ \ref{TypeIfree} up to and including Eq.\ (\ref{entropy6}) is applicable
to the intermediate state as well, since it only uses, as a starting point, 
the expression (\ref{deadd1})
for the energy density, which remains correct.
From Eq.\ (\ref{entropy6}) one then deduces the same Eq.\ (\ref{sfleq3}) 
for neutrons (by assumption, there are no Feynman-Onsager vortices in the system!)
and Eq.\ (\ref{DeltaTmunu}) for the electromagnetic correction to the energy-momentum tensor.
However, for {\it protons} Eq.\ (\ref{sfleq3}) cannot be applied and must be modified.
The reason is, as suggested by 
the London argument
(e.g., Ref.\ \cite{ll80}),
there is a non-zero proton phase winding 
$\oint \partial_\mu \Phi_p \, d x^\mu$ around
each flux tube, i.e. the phase $\Phi_p$, 
averaged over the volume containing many flux tubes, does not satisfy the 
``potentiality condition'' (\ref{phase}).
This situation is reminiscent of that 
observed in the HVBK-hydrodynamics (see, e.g., Ref.\ \cite{khalatnikov00} and G16). 
In particular, in G16 it is shown that the strong-drag regime we are interested in,
is realized if one replaces Eq.\ (\ref{sfleq3}) for protons 
with the less restrictive condition,%
%
\footnote{Equation (\ref{sfleq3drag}) is a special case of the more general Eq.\ (\ref{sflrot1}) from the next section, which, although describes a different system ($npe$-mixture with type-II proton superconductivity),
has some mathematical resemblance to what is studied here. 
}
%
\begin{equation}
u_\nu \left(\widetilde{\mathcal{V}}^{\mu\nu}_{(p)} + e_p F^{\mu\nu} \right)=0.
\label{sfleq3drag}
\end{equation}
It is easily verified that with this equation the r.h.s.\ of Eq.\ (\ref{entropy6})
is still zero, as it should be.
Summarizing, we find that to model the $npe$-mixture in the intermediate state
one should use superfluid Eqs.\  (\ref{sfleq3}) for neutrons 
and (\ref{sfleq3drag}) for protons;
the correction $\Delta T^{\mu\nu}$ to the energy-momentum tensor $T^{\mu\nu}$
is given by Eq.\ (\ref{DeltaTmunu}).
The last thing to do in order to close the system of dynamic equations
discussed here
is to specify the relation between the tensors 
$G^{\mu\nu}$ and $F^{\mu\nu}$, 
and to prove that 
the resulting tensor $\Delta T^{\mu\nu}$ is indeed symmetric.
This is done in Sec.\ \ref{TypeIsymmetry}. 
A complete system of equations is summarized in Appendix \ref{sumapp}.

\vspace{0.2 cm}
\noindent
%
{\bf Remark 1.} ---
It may be noted that exactly the same derivation of the tensor 
(\ref{DeltaTmunu}) can be made also for {\it normal} (nonsuperfluid) matter
if we put
$Y_{ik}=0$ in all relevant equations. 
This indicates that the tensor $\Delta T^{\mu\nu}$ must be 
well known
in the electrodynamics of continuous media.
As shown in Appendix \ref{Abraham} 
this is indeed the case and it is 
directly 
related to the so called Abraham 
electromagnetic tensor in the medium (see, e.g., Refs.\ \cite{ginzburg73, gu76, toptygin15}).

\section{Relativistic dynamic equations for $npe$-mixture with neutron and proton vortices: Type-II proton superconductivity}
\label{TypeII}

In this section we consider a region of densities where protons form type-II superconductor
and
allow for the possible presence of neutron and proton vortices in the system.
Since our consideration is very similar to that in G16 
we will be brief here and 
refer the interested reader to this reference for more details %
%
\footnote{Note that G16 uses somewhat different notation. 
The correspondence table between our notation and that of G16 is provided in Appendix \ref{corresp}.}.
%
In the 
system with
vortices the condition 
$(\partial_\mu\partial_\nu-\partial_\nu\partial_\mu)\Phi_i =0$
is not
satisfied at the vortex lines. 
Hence, as in 
Sec.\ \ref{TypeIinter}, 
the macroscopic (smooth-averaged) 
superfluid Eq.\ (\ref{sfleq2}) [or (\ref{sfleq3})]
should be replaced by a weaker constraint [see Eq.\ (\ref{sflrot1}) below].
In what follows it will be convenient 
to use the {\it vorticity} tensor $\mathcal{V}^{\mu\nu}_{(i)}$,
\begin{eqnarray}
\mathcal{V}^{\mu\nu}_{(i)} &\equiv & 
\widetilde{\mathcal{V}}^{\mu\nu}_{(i)} + e_i F^{\mu\nu}
\nonumber\\
&=& \partial^\mu\left[ w^\nu_{(i)}+\mu_i u^{\nu}+e_i A^{\nu}\right]
-\partial^\nu\left[ w^\mu_{(i)}+\mu_i u^{\mu}+e_i A^{\mu}\right],
\label{Vmunu}
\end{eqnarray}
with the obvious property [cf.\ Eq.\ (\ref{11})]
\begin{equation}
\partial_\mu \Vdual^{\mu\nu}_{(i)}=0.
\label{vortstar}
\end{equation}
The tensor $\mathcal{V}^{\mu\nu}_{(i)}$ 
is equivalent to $m_i \,{\rm curl}\, {\pmb V}_{{\rm s}i}+e_i \, {\pmb B}$ 
of the nonrelativistic HVBK-hydrodynamics 
(${\pmb V}_{{\rm s}i}$ is the superfluid velocity) %
%
\footnote{
\label{nonrel2}	
To be more precise, 
the vector $\mathcal{V}^{\mu}_{({\rm M}i)}$ 
[see Eq.\ (\ref{Vmagn}) below],
constructed with the help of this tensor, is equivalent to 
$m_i \,{\rm curl}\, {\pmb V}_{{\rm s}i}+e_i \, {\pmb B}$.}.
The 
geometrical
meaning of 
this tensor
is quite transparent.
Assume we have a surface spanned by some closed contour.
Then $\mathcal{V}^{\mu\nu}_{(i)}$
is related to the number $N_i$ of 
neutron ($i=n$) or proton ($i=p$) vortices 
piercing the surface by the formula
(see G16 for more details 
%
\footnote{Note that the factor 1/2 was inadvertently omitted 
in the corresponding equation (42) in that reference.})
%
\begin{equation}
\frac{1}{2} \, \int d f^{\mu \nu} \, 
\mathcal{V}_{(i)\mu \nu}
= \pi N_i,
\label{int33}
\end{equation}
where an integral is taken over the surface area.
In the absence of vortices one has $\mathcal{V}^{\mu\nu}_{(i)}=0$ 
[see Eqs.\ (\ref{sfleq2}) and (\ref{Vmunu})].
With the tensor $\mathcal{V}^{\mu\nu}_{(i)}$ one can construct, 
using Eqs.\ (\ref{AE}) and (\ref{AM}), 
the ``electric'' and ``magnetic'' four-vectors $\mathcal{V}^{\mu}_{({\rm E}i)}$
and $\mathcal{V}^{\mu}_{({\rm M}i)}$, respectively,
\begin{eqnarray}
\mathcal{V}^{\mu}_{({\rm E}i)} &\equiv&  
u_\nu \mathcal{V}^{\mu\nu}_{(i)},
\label{Velectr}\\
\mathcal{V}^{\mu}_{({\rm M}i)} &\equiv&  
\frac{1}{2} \, \epsilon^{\mu \nu \alpha \beta} \, u_{\nu} \, \mathcal{V}_{(i) \alpha \beta}.
\label{Vmagn}
\end{eqnarray}

In addition to modifying the superfluid equation, 
vortices
affect also
the second law of thermodynamics (\ref{2ndlaw3}),
because 
a certain amount of energy is associated with each vortex.
This energy 
should be accounted for 
in Eq.\ (\ref{2ndlaw3})
together with the electromagnetic contribution. 
The expression for $d\varepsilon_{\rm add}$, 
that includes vortex contribution, reads
\begin{equation}
d \varepsilon_{\rm add} = 
\frac{1}{4\pi} \, E_{\mu} d D^{\mu} + \frac{1}{4\pi} \, H_{\mu} d B^{\mu}
+ \mathcal{V}^{\mu}_{({\rm E}i)} d \mathcal{W}_{({\rm E}i)\mu}
+ \mathcal{W}_{({\rm M}i)\mu} d \mathcal{V}^{\mu}_{({\rm M}i)},
\label{deadd2}
\end{equation}
where
the four-vectors 
$\mathcal{W}^\mu_{({\rm E}i)}$ 
and
$\mathcal{W}^\mu_{({\rm M}i)}$ 
are analogous to 
$D^{\mu}$ 
and $H^{\mu}$, respectively.
As shown in Sec.\ \ref{TypeIIsymmetry} (see below) 
they can generally be presented as
\begin{eqnarray}
\mathcal{W}^{\mu}_{({\rm E}i)} &\equiv&  
u_\nu \mathcal{W}^{\mu\nu}_{(i)},
\label{Welectr}\\
\mathcal{W}^{\mu}_{({\rm M}i)} &\equiv&  \frac{1}{2} \, \epsilon^{\mu \nu \alpha \beta} \, u_{\nu} \, \mathcal{W}_{(i) \alpha \beta}.
\label{Wmagn}
\end{eqnarray}
Here $\mathcal{W}^{\mu\nu}_{(i)}$ is some auxiliary antisymmetric tensor, 
which plays the same role with respect to $\mathcal{V}^{\mu\nu}_{(i)}$
as the tensor $G^{\mu\nu}$ with respect to $F^{\mu\nu}$.
It is easy to see that the third and fourth terms in the r.h.s.\ of Eq.\ (\ref{deadd2})
are written in full analogy with respectively, the first and second electromagnetic terms.
This coincidence is not accidental. 
As detailed 
in Appendix \ref{vortapp} 
the fourth term here is basically 
responsible for 
the vortex energy 
(including its magnetic energy),
while the third term
comes into play 
if one takes into account 
the electric field generated by moving vortices.
Note that in G16 only the fourth term has been allowed for,
since that reference analysed uncharged superfluids.
In addition,
in that reference it was from the very beginning assumed that
$\mathcal{W}^{\mu}_{({\rm M}i)}$ is directly proportional to $\mathcal{V}^{\mu}_{({\rm M}i)}$,
$\mathcal{W}^{\mu}_{({\rm M}i)} \propto \mathcal{V}^{\mu}_{({\rm M}i)}$,
which is the only viable option in the absence of other magnetic vectors
in the problem [cf. Eq.\ (\ref{Wmu2}) in Sec.\ \ref{TypeIIsymmetry}].

Our next step will be to transform the energy $d \varepsilon_{\rm add}$
in a way similar to how it was done in Sec.~\ref{TypeI}.
The first two terms in the r.h.s.\ of Eq.\ (\ref{deadd2})
are transformed exactly 
as in Sec.\ \ref{TypeI},
the result is given by Eq.\ (\ref{ude2}).
Let us analyse the fourth term. 
It reads [cf. Eq.\ (\ref{magnx})]
\begin{eqnarray}
\mathcal{W}_{({\rm M}i)\mu} d \mathcal{V}^{\mu}_{({\rm M}i)}
&=& \mathcal{W}_{({\rm M}i)\mu} 
\, \frac{1}{2} \, \epsilon^{\mu\nu\alpha\beta} \, u_{\nu} \, d \mathcal{V}_{(i)\alpha\beta}
+ \mathcal{W}_{({\rm M}i)\mu} \, 
\frac{1}{2} \, \epsilon^{\mu\nu\alpha\beta}\, \mathcal{V}_{(i)\alpha\beta} \, d u_{\nu}
\nonumber\\
&=& 
\frac{1}{2 } \Wperp^{\alpha\beta}_{(i)} d \mathcal{V}_{(i)\alpha\beta}
+ \frac{1}{2} \, \left( \mathcal{W}_{({\rm M}i)\mu} \, \epsilon^{\mu\nu\alpha\beta}\, \mathcal{V}_{(i)\alpha\beta} 
+ \underline{u^{\nu} \mathcal{W}_{(i)\alpha \beta}\mathcal{V}^{\alpha\beta}_{(i)}} \right) \, d u_{\nu}
\nonumber\\
&=& \frac{1}{2} \, \left( 
\Wperp^{\alpha\beta} _{(i)} d \mathcal{V}_{(i)\alpha\beta}
+ 2 \mathcal{V}_{(i) \alpha\beta} \mathcal{W}^{\alpha\gamma}_{(i)} u^{\beta} \, d u_{\gamma} \right),
\label{third}
\end{eqnarray}
where the underlined term vanishes because $u^{\nu} du_\nu=0$;
$\Wperp^{\alpha\beta} _{(i)}$ is defined by Eq.\ (\ref{Aperp}).
To further transform this equation we make use of Appendix \ref{transformation}.
Comparing Eq.\ (\ref{third}) with (\ref{depart})
allows us to identify
\begin{eqnarray}
O^{\alpha\beta}  &=& \Wperp^{\alpha\beta}_{(i)},
\nonumber\\
\mathcal{F}^{\alpha\beta} &=& \mathcal{V}^{\alpha\beta}_{(i)},
\nonumber\\
\mathcal{B}^{\alpha\beta} &=& \mathcal{V}^{\alpha\beta}_{(i)},
\nonumber\\
\mathcal{A}^{\alpha\beta} &=&  \mathcal{W}^{\alpha\beta}_{(i)},
\nonumber
\end{eqnarray}
hence
\begin{eqnarray}
-u^{\mu} \,\mathcal{W}_{({\rm M}i)\alpha} \partial_{\mu} \mathcal{V}^{\alpha}_{({\rm M}i)}  
&=&
u^{\nu} \mathcal{V}_{(i)\mu\nu} \, \, \partial_{\alpha} \Wperp^{\mu \alpha}_{(i)}
\nonumber\\
&-&\partial_{\mu}\left[ u^{\nu} \left(
\Wperp^{\mu \alpha}_{(i)} \mathcal{V}_{(i)\nu \alpha}
- u^{\mu}u^{\gamma}
\perp_{\nu\beta}  \mathcal{W}^{\alpha\beta}_{(i)} \mathcal{V}_{(i)\alpha\gamma} 
\right)\right]
\nonumber\\
&+&\partial_{\mu}u^{\nu} \left( \Wperp^{\mu\alpha}_{(i)} \mathcal{V}_{(i)\nu\alpha}
- u^{\mu}u^{\gamma}
\perp_{\nu\beta} \mathcal{W}^{\alpha\beta}_{(i)} \mathcal{V}_{(i)\alpha\gamma}
\right).
\label{de3}
\end{eqnarray}
Looking at Eqs.\ (\ref{third}) and (\ref{de3}) 
it may be noted that the transformation of the fourth and second terms in Eq.\ (\ref{deadd2})
are identical 
provided that 
one replaces 
$\mathcal{W}^{\mu}_{({\rm M}i)} \rightarrow H^{\mu}/(4 \pi)$ 
and $\mathcal{V}^{\mu}_{({\rm M}i)} \rightarrow B^{\mu}$ 
[compare Eqs.\ (\ref{magnx}) and (\ref{third})].
Similarly, the transformation of the third and first terms 
in Eq.\ (\ref{deadd2}) can be obtained from one another
by replacing
$\mathcal{W}^{\mu}_{({\rm E}i)} \rightarrow D^{\mu}/(4 \pi)$ 
and
$\mathcal{V}^{\mu}_{({\rm E}i)} \rightarrow E^{\mu}$. 
With these replacements, 
one can use Eq.\ (\ref{electrx}) 
to transform the third term in Eq.\ (\ref{deadd2}).
The result is
\begin{eqnarray}
-u^{\mu} \,\mathcal{V}_{({\rm E}i) \alpha} \partial_\mu \mathcal{W}^{\alpha}_{({\rm E}i)} &=&
u^{\nu} \mathcal{V}_{(i)\mu\nu} 
\, \partial_{\alpha} \Wpar^{\mu\alpha}_{(i)}  
\nonumber\\
&&- \partial_\mu \left[ u^\nu \left( 
\Wpar^{\mu\alpha}_{(i)} \mathcal{V}_{(i) \nu \alpha}
+ u^{\mu}u^{\gamma}
\perp_{\nu\beta} \mathcal{V}^{\alpha\beta}_{(i)} \mathcal{W}_{(i) \alpha\gamma} 
+ g^{\mu}_{\,\,\, \nu} \, \mathcal{W}_{({\rm E}i)\alpha} \mathcal{V}^{\alpha}_{({\rm E}i)}
\right)\right]
\nonumber\\
&&+ \partial_{\mu} u^{\nu}\left( 
\Wpar^{\mu\alpha}_{(i)} \mathcal{V}_{(i) \nu \alpha}
+ u^{\mu}u^{\gamma}
\perp_{\nu\beta} \mathcal{V}^{\alpha\beta}_{(i)} \mathcal{W}_{(i) \alpha\gamma} 
+ g^{\mu}_{\,\,\, \nu} \, \mathcal{W}_{({\rm E}i)\alpha} \mathcal{V}^{\alpha}_{({\rm E}i)}
\right).
\label{ue3}
\end{eqnarray}

Collecting together the electromagnetic terms (\ref{ude2}) 
and the vortex terms (\ref{de3}) and (\ref{ue3}),
one obtains
\begin{eqnarray}
-u^{\mu} \partial_{\mu}\varepsilon_{\rm add} &=&
u^{\nu} F_{\mu\nu} \, \partial_{\alpha}\left( \frac{1}{4\pi} \Gpar^{\mu\alpha}+
\frac{1}{4\pi} \Gperp^{\mu\alpha}\right)
+u^{\nu} \mathcal{V}_{(i)\mu\nu} \, \, \partial_{\alpha} 
\left( \Wpar^{\mu\alpha}_{(i)} + \Wperp^{\mu \alpha}_{(i)}
\right)
\nonumber\\
&&- \partial_\mu \left[ u^\nu \left( \mathcal{T}^\mu_{({\rm E}) \, \nu}+
\mathcal{T}^\mu_{({\rm M}) \, \nu} 
+\mathcal{T}^\mu_{({\rm VE}) \, \nu}
+ \mathcal{T}^\mu_{({\rm VM}) \, \nu}\right)\right]
\nonumber\\
&&+ \partial_{\mu} u^{\nu}\left( 
\mathcal{T}^\mu_{({\rm E}) \, \nu}+
\mathcal{T}^\mu_{({\rm M}) \, \nu} 
+\mathcal{T}^\mu_{({\rm VE}) \, \nu}
+ \mathcal{T}^\mu_{({\rm VM}) \, \nu}
\right),
\label{ude3}
\end{eqnarray}
where the tensors $\mathcal{T}^\mu_{({\rm E}) \, \nu}$ and 
$\mathcal{T}^\mu_{({\rm M}) \, \nu}$ 
are given by Eqs.\ (\ref{Te1}) and (\ref{Tm1}), and
\begin{eqnarray}
\mathcal{T}^{\mu}_{({\rm VE})\,\,\,\nu} &=&
\Wpar^{\mu\alpha}_{(i)} \mathcal{V}_{(i) \nu \alpha}
+ u^{\mu}u^{\gamma}
\perp_{\nu\beta} \mathcal{V}^{\alpha\beta}_{(i)} \mathcal{W}_{(i) \alpha\gamma} 
+ g^{\mu}_{\,\,\, \nu} \, \mathcal{W}_{({\rm E}i)\alpha} \mathcal{V}^{\alpha}_{({\rm E}i)},
\label{vortexE0}\\
\mathcal{T}^{\mu}_{({\rm VM})\,\,\,\nu} &=&\Wperp^{\mu\alpha}_{(i)} \mathcal{V}_{(i)\nu\alpha}
- u^{\mu}u^{\gamma}
\perp_{\nu\beta} \mathcal{W}^{\alpha\beta}_{(i)} \mathcal{V}_{(i)\alpha\gamma}
\label{vortex0}
\end{eqnarray}
are, respectively, the ``electric'' and ``magnetic'' vortex contributions
to the energy-momentum tensor
(note a summation over $i=n$, $p$ here).
Similarly to tensors (\ref{Te1}) and (\ref{Tm1}),
these tensors can be represented as [cf. Eq.\ (\ref{Te2}) and (\ref{Tm2})]
\begin{eqnarray}
\mathcal{T}^{\mu\nu}_{({\rm VE})} &=&  
\perp^{\mu\nu} \mathcal{W}_{({\rm E}i)}^{\alpha}\mathcal{V}_{({\rm E}i)\alpha }
-\mathcal{W}_{({\rm E}i)}^{\mu} \mathcal{V}^{\nu}_{({\rm E}i)},
\label{vortexE}\\
\mathcal{T}^{\mu\nu}_{({\rm VM})} &=& 
\Wperp^{\mu\alpha}_{(i)}\Vperp^{\nu}_{(i)\,\alpha} + u^{\nu} \Wperp^{\mu\alpha}_{(i)} \mathcal{V}_{({\rm E}i)\alpha}
+u^{\mu}\Wperp^{\nu\alpha}_{(i)}\mathcal{V}_{({\rm E}i)\alpha}.
\label{vortex}
\end{eqnarray}

Using Eq.\ (\ref{ude3}), as well as Eq.\ (\ref{1st}), 
the definition (\ref{Vmunu}), and the equality 
$\mathcal{W}^{\mu\alpha}_{(i)}=\Wpar^{\mu\alpha}_{(i)}+\Wperp^{\mu\alpha}_{(i)}$
[see Eq.\ (\ref{Adec})],
the entropy generation equation (\ref{entropy5}) 
becomes
\begin{eqnarray}
T \,\partial_\mu(S u^{\mu}) &=& 
u^{\nu} \, \mathcal{V}_{(i)\mu\nu} 
\left[
Y_{ik} w^{\mu}_{(k)} + \partial_{\alpha} 
\mathcal{W}^{\mu\alpha}_{(i)}
\right]
\nonumber\\
&&- \partial_\mu \left[ u^\nu \left( \mathcal{T}^\mu_{({\rm E}) \, \nu}+
\mathcal{T}^\mu_{({\rm M}) \, \nu} 
+\mathcal{T}^\mu_{({\rm VE}) \, \nu}
+\mathcal{T}^\mu_{({\rm VM}) \, \nu}
- \Delta T^{\mu}_{\,\,\,\nu} \right)\right]
\nonumber\\
&&+ \partial_{\mu} u^{\nu}\left( 
\mathcal{T}^\mu_{({\rm E}) \, \nu}+
\mathcal{T}^\mu_{({\rm M}) \, \nu} 
+\mathcal{T}^\mu_{({\rm VE}) \, \nu}
+\mathcal{T}^\mu_{({\rm VM}) \, \nu}
- \Delta T^{\mu}_{\,\,\,\nu}
\right).
\label{entropy7}
\end{eqnarray}
The r.h.s.\ of this equation has the same structure as 
Eq.\ (77) in G16.
Correspondingly, 
its analysis and the resulting equations are very similar.
Using the argumentation of that reference, 
one finds that, in order for the entropy to be conserved,
it is necessary to have
\begin{eqnarray}
u^{\nu} \, \mathcal{V}_{(i)\mu\nu} 
\left[
Y_{ik} w^\mu_{(k)} + \partial_{\alpha} \mathcal{W}^{\mu \alpha}_{(i)}
\right]=0,
\label{sfleq4}\\
\Delta T^{\mu\nu}=
\mathcal{T}^{\mu\nu}_{({\rm E})}+
\mathcal{T}^{\mu\nu}_{({\rm M})} 
+\mathcal{T}^{\mu\nu}_{({\rm VE})}
+\mathcal{T}^{\mu\nu}_{({\rm VM})}.
\label{dT2}
\end{eqnarray}
Note that $\Delta T^{\mu\nu}$ satisfies the required constraint (\ref{dT})
and is symmetric (see Sec.\ \ref{TypeIIsymmetry}).
As demonstrated in G16, the condition (\ref{sfleq4})
is equivalent to the following equation, which replaces the 
superfluid Eq.\ (\ref{sfleq2}) [or (\ref{sfleq3})] 
of the vortex-free system,
\begin{equation}
u^\nu \mathcal{V}_{(i)\mu\nu}= \mu_i n_i \, f_{(i)\mu},
\label{sflrot1}
\end{equation}
where
\begin{eqnarray}
f^{\mu}_{(i)} &=& \alpha_i \perp^{\mu\nu} \mathcal{V}_{(i)\nu\lambda} \, W_{(i)\delta} \perp^{\lambda \delta},
\label{f2}\\
W^{\mu}_{(i)} &\equiv& \frac{1}{n_i}\left[
Y_{ik} w^\mu_{(k)} + \partial_{\alpha} 
\mathcal{W}^{\mu \alpha}_{(i)} 
\right],
\label{Wmu}
\end{eqnarray}
and $\alpha_i$ is a non-dissipative mutual friction coefficient
[note that there are no summation over $i$ in Eqs.\ (\ref{sflrot1})--(\ref{Wmu})].
The l.h.s.\ of Eq.\ (\ref{sflrot1}) is simply the four-vector 
$\mathcal{V}^{\mu}_{({\rm E}i)}$, so that it can be rewritten as
(now in the dimensional form)
\begin{equation}
\mathcal{V}^{\mu}_{({\rm E}i)} = \frac{\mu_i n_i}{c^3} \, f_{(i)}^\mu.
\label{nuE}
\end{equation}

Equations (\ref{dT2}) and (\ref{sflrot1})  
[or (\ref{nuE})] 
are the main results of this section. 
They show how the energy-momentum tensor and superfluid equation
should be modified in the presence of vortices.
These equations depend on the tensors 
$G^{\mu\nu}$ and 
$\mathcal{W}^{\mu\nu}_{(i)}$,
which will be found in Sec.\ \ref{TypeIIsymmetry}.
The symmetry of the tensor $\Delta T^{\mu\nu}$ 
will be demonstrated in the same section.
The whole system of dynamic equations in the presence of vortices 
is summarized in Appendix~\ref{sumapp}.

\vspace{0.2 cm}
\noindent
%
{\bf Remark 1.} ---
In this work we are mainly interested in the nondissipative dynamic equations.
In particular, we assumed that normal (nonsuperfluid) 
components of all particle species move with one and the same 
velocity $u^{\mu}$.
In principle, this condition does not guarantee that 
there are no dissipation in the system.
Indeed, the entropy can be produced, e.g., because of scattering
of nucleon thermal excitations and/or electrons off the vortex cores.
This mechanism is known as the ``mutual friction'' 
\cite{hv56, hall60, bk61, khalatnikov00, donnelly05, mendell91a, ss95}.
Only this dissipative mechanism has been taken into account in GAS11.
To include mutual friction dissipation into consideration,
we start with Eq.\ (\ref{entropy7}) and require positive definiteness 
of its right-hand side.
Then, following the consideration of G16 
[see the text after Eq.\ (77) in that reference],
we find that Eq.\ (\ref{sfleq4}) should be replaced with
the inequality 
\begin{equation}
u^{\nu} \, \mathcal{V}_{(i)\mu\nu} 
\left[
Y_{ik} w^\mu_{(k)} + \partial_{\alpha} 
\mathcal{W}^{\mu\alpha}_{(i)}
\right] \geq 0,
\label{sfleq5}
\end{equation}
from which one obtains 
the same superfluid equation (\ref{sflrot1}),
but with $f^{\mu}_{(i)}$ given by
\begin{equation}
f^{\mu}_{(i)} = \alpha_i \perp^{\mu\nu} \mathcal{V}_{(i)\nu\lambda} \, W_{(i)\delta} \perp^{\lambda \delta}
+ \frac{\beta_i-\gamma_i}{\mathcal{V}_{({\rm M}i)}}  \perp^{\mu\eta} \perp^{\nu\sigma}  \mathcal{V}_{(i)\eta\sigma} \mathcal{V}_{(i)\lambda\nu}  \,
W_{(i)\delta} \perp^{\lambda \delta}
+\gamma_i  \mathcal{V}_{({\rm M}i)} \, W_{(i)\delta} \perp^{\mu \delta},
\label{f3}
\end{equation}
where $\alpha_i$ is the same non-dissipative coefficient as in Eq.\ (\ref{f2});
$\beta_i\geq0$ and $\gamma_i\geq 0$ are the positive 
dissipative mutual friction coefficients and
\begin{equation}
\mathcal{V}_{({\rm M}i)} \equiv \sqrt{\mathcal{V}^{\mu}_{({\rm M}i)}\mathcal{V}_{({\rm M}i)\mu}}.
\label{VM}
\end{equation}
Note that Eq.\ (\ref{f3}) is not the most general form of 
$f^{\mu}_{(i)}$ satisfying the inequality (\ref{sfleq5}).
In principle there could be cross-terms depending on both
$\mathcal{V}^{\mu\nu}_{(n)}$ and $\mathcal{V}^{\mu\nu}_{(p)}$ 
(see, e.g., Ref.\ \cite{mendell91a} for a non-relativistic analogue of such terms).
These terms are ignored in Eq.\ (\ref{f3})
since we do not see any plausible physical interpretation behind them.
Anyway, 
one should bear in mind the possibility that Eq.\ (\ref{f3}) is not complete.
In the nonrelativistic limit a more general expression for $f^{\mu}_{(i)}$
is contained in Appendix of Ref.\ \cite{mendell91a}. 
Generalization of that result to the relativistic case is straightforward.

\vspace{0.2 cm}
\noindent
%
{\bf Remark 2.} ---
Expression (\ref{f3}) for $f^{\mu}_{(i)}$ 
can be rewritten in terms of the magnetic four-vector $\mathcal{V}_{({\rm M}i)}^{\mu}$
as [see a similar formula (53) in G16]
\begin{equation}
f^{\mu}_{(i)} = -\alpha_i \, X^{\mu}_{(i)}
- \beta_i \, \epsilon^{\mu\nu\lambda\eta} \, u_{\nu} \, {\rm e}_{(i)\lambda} \, X_{(i)\eta}
+ \gamma_i \, {\rm e}^{\mu}_{(i)} \, W^{\lambda}_{(i)} \mathcal{V}_{({\rm M}i)\lambda},
\label{finv}
\end{equation}
where ${\rm e}^{\mu}_{(i)} \equiv \mathcal{V}^{\mu}_{({\rm M}i)}/\mathcal{V}_{({\rm M}i)}$ 
and $X^{\mu}_{(i)} \equiv 
\epsilon^{\mu\nu\lambda\eta} \, u_{\nu} \, \mathcal{V}_{({\rm M}i)\lambda} \, W_{(i)\eta}$.

\vspace{0.2 cm}
\noindent
%
{\bf Remark 3.} ---
As it is argued in Refs.\ \cite{khalatnikov00,ss95}, 
the coefficients $\gamma_i$ ($i=n$ or $p$) 
in Eq.\ (\ref{f3}) are most likely very small
and the corresponding terms can be neglected.
Assume that it is indeed the case and that the tensor
$\mathcal{V}_{(i)}^{\mu\nu}$ satisfies Eq.\ (\ref{sflrot1})
with $f_{(i)}^{\mu}$ defined by Eq.\ (\ref{f3}).
Then it can be shown (see Remark 2 in section IIIA of G16)
that a four-vector $v_{({\rm L}i)}^{\mu}$
exists, given by,
\begin{equation}
v_{({\rm L}i)}^{\mu}= u^{\mu} - \mu_i n_i \, \alpha_i \, W_{(i)\nu} \perp^{\mu\nu}
+\frac{\mu_i n_i \, \beta_i}{\mathcal{V}_{({\rm M}i)}} \, \perp^{\mu \alpha} \perp^{\nu \beta} \, \mathcal{V}_{(i)\alpha\beta} \, W_{(i)\nu},
\label{VlX0}
\end{equation}
such that the combination
$v_{({\rm L}i)}^{\nu} \mathcal{V}_{(i)\mu\nu}$ is identically zero,
\begin{equation}
v_{({\rm L}i)}^{\nu} \mathcal{V}_{(i)\mu\nu}=0
\label{VlX1}
\end{equation}
(no summation over $i$ 
here).
Equation (\ref{VlX1}) is analogous to the vorticity conservation equation
of the non-relativistic HVBK-hydrodynamics (see Appendix A of G16)
and the four-vector $v_{({\rm L}i)}^{\mu}$ 
has the meaning of (non-normalized) vortex velocity.

Using Eqs.\ (\ref{nuE}), (\ref{finv}), and (\ref{VlX0}),
it is straightforward to show that the spatial components
$\pmb{v}_{{\rm L}i}/c$,
$\mathbfcal{V}_{{\rm E}i}$, and 
$\mathbfcal{V}_{{\rm M}i}$
of the four-vectors 
$v^{\mu}_{({\rm L}i)}$,
$\mathcal{V}^{\mu}_{({\rm E}i)}$,
and
$\mathcal{V}^{\mu}_{({\rm M}i)}$
are related, in the comoving frame, 
by the condition
\begin{equation}
\mathbfcal{V}_{{\rm E}i} = \frac{1}{c}\,  \mathbfcal{V}_{{\rm M}i} \times \pmb{v}_{{\rm L}i}.
\label{important}
\end{equation}
For future convenience the latter equation is written in the dimensional form.

\vspace{0.2 cm}
\noindent
%
{\bf Remark 4.} ---
It is notable that the vortex energy-momentum tensors
(\ref{vortexE0}) and (\ref{vortex0}) are obtained in the same way 
and have \underline{exactly} the same structure
as, respectively, 
the electromagnetic tensors (\ref{Te1}) and (\ref{Tm1}).
This is a direct consequence of the striking similarity 
of the electromagnetic and vortex
contributions to the energy density $d \varepsilon_{\rm add}$ in Eq.\ (\ref{deadd2}).

\section{Symmetry of the energy-momentum tensor}
\label{symmetry}

The symmetry of the energy-momentum tensors 
obtained in Secs.\ \ref{TypeI} and \ref{TypeII}
is not manifest.
In this section we prove that they are indeed symmetric.
To do this it is necessary to express
the tensors $G^{\mu\nu}$ and 
$\mathcal{W}^{\mu\nu}_{(i)}$
in Eqs.\ (\ref{Te2}), (\ref{Tm2}), and (\ref{vortexE}), (\ref{vortex})
through the tensors $F^{\mu\nu}$ and $\mathcal{V}^{\mu\nu}_{(i)}$.
This can be done by specifying the expression for the energy density 
$d \varepsilon_{\rm add}$ [see Eqs.\ (\ref{deadd1}) and (\ref{deadd2})],
which is different for the situations considered in Secs.\ \ref{TypeI} and \ref{TypeII}.

\subsection{npe-mixture in the intermediate state (type-I proton superconductivity)}
\label{TypeIsymmetry}

We start with the intermediate state model of Sec.\ \ref{TypeIinter}.
Generally, since there are no vortices in the system, 
the energy density $\varepsilon$
can be a function of 
$S$, $n_i$, $n_e$, $w_{(i)\mu} w^\mu_{(k)}$, and 
various invariants composed of the four-vectors $D^\alpha$ and
$B^{\alpha}$ in combination with the four-vectors $u^{\mu}$ and $w^{\mu}_{(i)}$
characterizing the system in the field-free case 
%
\footnote{We remind the reader that $\varepsilon$ 
is a scalar defined in the comoving frame;
it is thus invariant under Lorentz transformations.}.
%
In what follows, we assume that there are no bound charges 
in the system 
(i.e., nonsuperconducting domains move with the normal liquid component), 
so that $E^{\mu}=D^{\mu}$, 
that is 
$\varepsilon$ depends on $D^{\mu}$  through
the term $D_{\mu}D^{\mu}/(8\pi)$.
Concerning magnetic contribution, 
the simplest (and largest) invariant allowed by the symmetry
%
\footnote{Other possible invariants, for example, 
$B_{\mu}w^{\mu}_{(i)}\, B_{\nu}w^{\nu}_{(k)}$, 
$\epsilon_{\alpha \beta \gamma \delta} u^{\beta} w^{\gamma}_{(i)}B^{\delta} \,
\epsilon^{\alpha b c d} u_b w_{(i)c} B_d$ etc.\
are small, because the four-vector $w^{\mu}_{(i)}$
is proportional to the generally small difference between the 
normal and superfluid velocities, see, e.g., G16.
Note also that the invariant $u_{\mu} B^{\mu}$, 
which could be used as a building brick for constructing other invariants, is zero, 
$u_{\mu} B^{\mu}=0$, see Eq.\ (\ref{Bmux}). }
%
is $x\equiv B_\mu B^{\mu}/(8\pi)$ 
[the factor $1/(8\pi)$ is introduced for further convenience].
We thus have for $d \varepsilon$ the same equation (\ref{2ndlaw3})
with
\begin{equation}
d\varepsilon_{\rm add}= 
\frac{1}{4\pi}D_{\mu} d D^{\mu} + 
\frac{1}{4\pi} \, 
\frac{\partial \varepsilon}{\partial x} \, B_{\mu}dB^{\mu}, 
\label{deadd4}
\end{equation}
where the partial derivative is taken at constant
$S$, $n_i$, $n_e$, $w_{(i)\mu} w^\mu_{(k)}$, and $D^{\mu}$.
Comparing this equation with Eq.\ (\ref{deadd1}),
one finds that, indeed,
\begin{equation}
E^{\mu}=D^{\mu} 
\label{Emux2}
\end{equation}
and
\begin{equation}
H^{\mu} = \gamma B^{\mu}
\quad\quad {\rm with } \quad \quad
\gamma\equiv \frac{\partial \varepsilon}{\partial x}.
\label{Hmux2}
\end{equation}
Equations (\ref{Emux2}) and (\ref{Hmux2}) completely
determine the tensors $\Gpar^{\mu\nu}=\Fpar^{\mu\nu}$ 
and 
$\Gperp^{\mu\nu}=\gamma \, \Fperp^{\mu\nu}$,
and hence the tensor $G^{\mu\nu}=\Gpar^{\mu\nu}+\Gperp^{\mu\nu}$ 
[see Eqs.\ (\ref{Apar}), (\ref{Aperp}), and (\ref{Adec})].
Using them, the electro-magnetic tensor (\ref{DeltaTmunu})
can be presented in the manifestly symmetric form,
\begin{eqnarray}
\Delta T^{\mu\nu} &=&-\frac{1}{4\pi} \, 
\left(E^{\mu} E^{\nu} -  \perp^{\mu\nu} \, E_{\alpha}E^{\alpha}\right)
\nonumber\\
&+& 
\frac{\gamma}{4\pi}\, 
\,\left(  \perp_{\delta \alpha} F^{\mu\delta} F^{\nu \alpha}
-u^{\mu} u^{\nu} u^{\gamma} u_{\beta} F^{\alpha \beta} F_{\alpha \gamma} \right).
\label{DeltaTmunu2}
\end{eqnarray}
The phenomenological coefficient $\gamma$ is calculated 
for a simple model in Appendix \ref{interapp}.

\subsection{npe-mixture with neutron and proton vortices (type-II proton superconductivity)}
\label{TypeIIsymmetry}

In this case $\varepsilon$ can depend on additional invariants
composed of the four-vectors 
$D^{\mu}$, $B^{\mu}$, $\mathcal{W}^{\mu}_{({\rm E}n)}$, 
$\mathcal{W}^{\mu}_{({\rm E}p)}$,
$\mathcal{V}^{\mu}_{({\rm M}n)}$, 
and $\mathcal{V}^{\mu}_{({\rm M}p)}$ 
[see Eq.\ (\ref{deadd2})].
One can construct the following invariants from these vectors %
%
\footnote{Of course, the number of possible invariants is much larger.
Here we only write out those invariants whose physical meaning is clear to us 
(see Appendix \ref{vortapp}), 
but one should bear in mind that it is straightforward to consider other possibilities.}:
%
$z=D_{\mu}D^{\mu}/(8\pi)$,
$z_i=D_{\mu}\mathcal{W}^{\mu}_{({\rm E}i)}$,
$z_{ik}=\mathcal{W}_{({\rm E}i)\mu} \mathcal{W}^{\mu}_{({\rm E}k)}/2$,
$x=B_{\mu}B^{\mu}/(8\pi)$,
$x_i=B_{\mu}\mathcal{V}^{\mu}_{({\rm M}i)}$,
and
$x_{ik}=\mathcal{V}_{({\rm M}i)\mu} \mathcal{V}^{\mu}_{({\rm M}k)}/2$ 
($i$, $k=n$ or $p$). 
Correspondingly, 
the differential of the energy density 
$\varepsilon(S, \, n_i, \,n_e, \, w_{(i)\mu} w^\mu_{(k)},
\,z,\, z_i,\, z_{ik}, \,x, \, x_i, \, x_{ik})$ 
is given by Eq.\ (\ref{2ndlaw3}),
in which 
\begin{eqnarray}
d\varepsilon_{\rm add} &=& 
\frac{1}{4\pi} \, 
\frac{\partial \varepsilon}{\partial z} \, D_{\mu}dD^{\mu}
+\frac{\partial \varepsilon}{\partial z_i} \,d\left[ D_{\mu} \mathcal{W}^{\mu}_{({\rm E}i)}\right]
+\frac{\partial \varepsilon}{\partial z_{ik}} 
\mathcal{W}_{({\rm E}k)\mu} d\mathcal{W}^{\mu}_{({\rm E}i)}
\nonumber\\
&+&\frac{1}{4\pi} \, 
\frac{\partial \varepsilon}{\partial x} \, B_{\mu}dB^{\mu}
+\frac{\partial \varepsilon}{\partial x_i} \,d\left[ B_{\mu} \mathcal{V}^{\mu}_{({\rm M}i)}\right]
+\frac{\partial \varepsilon}{\partial x_{ik}} 
\mathcal{V}_{({\rm M}k)\mu} d\mathcal{V}^{\mu}_{({\rm M}i)} 
\nonumber\\
&=& 
\frac{1}{4\pi} \, 
\left[ 
\gammaE \, D_{\mu} 
+ 4 \pi \, \GammaEi \mathcal{W}_{({\rm E}i)\mu} 
\right] d D^\mu
+ \left[ \GammaEik \mathcal{W}_{({\rm E}k)\mu}    
+ \GammaEi  D_\mu
\right]
d \mathcal{W}^{\mu}_{({\rm E}i)}
\nonumber\\
&+&
\frac{1}{4\pi} \, 
\left[ 
\gammaM \, B_{\mu} 
+ 4 \pi \, \GammaMi \mathcal{V}_{({\rm M}i)\mu} 
\right] d B^\mu
+ \left[ \GammaMik \mathcal{V}_{({\rm M}k)\mu}    
+ \GammaMi  B_\mu
\right]
d \mathcal{V}^{\mu}_{({\rm M}i)},
\label{deadd5}
\end{eqnarray}
where
\begin{eqnarray}
\gammaE &\equiv& \frac{\partial \varepsilon}{\partial z}, \quad \quad
\gammaM \equiv \frac{\partial \varepsilon}{\partial x}, \quad \quad
\label{gammaEM}\\
\GammaEi &\equiv&  \frac{\partial \varepsilon}{\partial z_i}, \quad \quad 
\GammaMi \equiv  \frac{\partial \varepsilon}{\partial x_i},
\label{Gammai}\\
\GammaEik = \GammaEki &\equiv& \frac{\partial \varepsilon}{\partial z_{ik}},\quad \quad
\GammaMik=\GammaMki \equiv \frac{\partial \varepsilon}{\partial x_{ik}}.
\label{Gammaik}
\end{eqnarray}
Comparing Eqs.\ (\ref{deadd5}) and (\ref{deadd2}),
one identifies 
\begin{eqnarray}
E^{\mu}&=&\gammaE \, D^{\mu} 
+ 4\pi \, \GammaEi \mathcal{W}^{\mu}_{({\rm E}i)}, 
\label{Emu2}\\
\mathcal{V}^{\mu}_{({\rm E}i)}&=&\GammaEik \mathcal{W}^{\mu}_{({\rm E}k)}    
+ \GammaEi  D^\mu,
\label{WmuE2}\\
H^{\mu}&=&\gammaM \, B^{\mu} 
+ 4\pi \, \GammaMi \mathcal{V}^{\mu}_{({\rm M}i)}, 
\label{Hmu2}\\
\mathcal{W}^{\mu}_{({\rm M}i)}&=&\GammaMik \mathcal{V}^{\mu}_{({\rm M}k)}    
+ \GammaMi  B^\mu.
\label{Wmu2}
\end{eqnarray}
The system of Eqs.\ (\ref{Emu2}) and (\ref{WmuE2}) can be inverted 
and
the four-vectors $D^{\mu}$ and $\mathcal{W}_{({\rm E}i)}^{\mu}$
can be presented as
\begin{eqnarray}
D^{\mu}&=&\gammaEtilde \, E^{\mu} 
+ 4\pi \, \GammaEitilde \mathcal{V}^{\mu}_{({\rm E}i)}, 
\label{Emu3}\\
\mathcal{W}^{\mu}_{({\rm E}i)}&=& \GammaEiktilde \mathcal{V}^{\mu}_{({\rm E}k)}    
+ \GammaEitilde  E^\mu,
\label{WmuE3}
\end{eqnarray}
where the quantities $\gammaEtilde$, $\GammaEitilde$,
and $\GammaEiktilde$
can easily be expressed through 
$\gammaE$, $\GammaEi$, and $\GammaEik$
using  Eqs.\ (\ref{Emu2}) and (\ref{WmuE2}).
From Eqs.\ (\ref{Wmu2}) and (\ref{WmuE3}) 
one now sees that the four-vectors $\mathcal{W}^{\mu}_{({\rm E}i)}$ and
$\mathcal{W}^{\mu}_{({\rm M}i)}$ indeed have the form 
assumed in Eqs.\ (\ref{Welectr}) and (\ref{Wmagn}).

Using Eqs.\ (\ref{Hmu2})--(\ref{WmuE3}),
as well as 
Eqs.\ (\ref{Aperp}), (\ref{Apar}), and (\ref{Adec}), 
one can find the tensors $G^{\mu\nu}$ 
and $\mathcal{W}^{\mu\nu}_{(i)}$ %
%
\footnote{In particular, $\mathcal{W}^{\mu\nu}_{(i)}=\Wperp^{\mu\nu}_{(i)}+\Wpar^{\mu\nu}_{(i)}$, 
	where
	$\Wperp^{\mu\nu}_{(i)}=\epsilon^{\alpha\beta\mu\nu}u_{\beta}
(\GammaMik \mathcal{V}_{({\rm M}k)\alpha}+ \GammaMi  B_\alpha)
= \GammaMik \Vperp^{\mu\nu}_{(k)}+\GammaMi \Fperp^{\mu\nu}$ 
and
$\Wpar^{\mu\nu}_{(i)}= \GammaEiktilde \Vpar^{\mu\nu}_{(k)}    
+ \GammaEitilde  \Fpar^{\mu\nu}$.
},
%
and to present
the tensor $\Delta T^{\mu\nu}$ (\ref{dT2})
in the manifestly symmetric form,
\begin{eqnarray}
\Delta{T}^{\mu\nu} &=&
-\frac{\gammaEtilde}{4\pi} \, 
\left(
E^\mu E^\nu - \perp^{\mu\nu} \, E_\alpha E^{\alpha}
\right)
\nonumber\\
&-& \GammaEitilde \left[
\left(
\mathcal{V}^{\mu}_{({\rm E}i)} E^{\nu} + E^{\mu} \mathcal{V}^{\nu}_{({\rm E}i)}
\right)
-2 \perp^{\mu\nu} \, E_\alpha \mathcal{V}^{\alpha}_{({\rm E}i)}
\right]
\nonumber\\
&-& \frac{\GammaEiktilde}{2} \left[
\left(
\mathcal{V}^{\mu}_{({\rm E}i)} \mathcal{V}^{\nu}_{({\rm E}k)} + \mathcal{V}^{\mu}_{({\rm E}k)} \mathcal{V}^{\nu}_{({\rm E}i)}
\right)
-2 \perp^{\mu\nu} \, \mathcal{V}_{({\rm E}i)\alpha} \mathcal{V}^{\alpha}_{({\rm E}k)}
\right]
\nonumber\\
&+&\frac{\gammaM}{4\pi} \left(  \perp_{\delta \alpha} F^{\mu\delta} F^{\nu \alpha}
-u^{\mu} u^{\nu} u^{\gamma} u_{\beta} \, F^{\alpha \beta} F_{\alpha \gamma} \right)
\nonumber\\
&+& \GammaMi \,\left[  \perp_{\delta \alpha} 
\left(\mathcal{V}^{\mu\delta}_{(i)} F^{\nu \alpha} + F^{\mu\delta} \mathcal{V}^{\nu \alpha}_{(i)} \right)
-2 \, u^{\mu} u^{\nu} u^{\gamma} u_{\beta} \, \mathcal{V}^{\alpha \beta}_{(i)} F_{\alpha \gamma} \right]
\nonumber\\
&+& \frac{\GammaMik}{2} \,\left[  \perp_{\delta \alpha} 
\left(\mathcal{V}^{\mu\delta}_{(i)} \mathcal{V}^{\nu \alpha}_{(k)} + \mathcal{V}^{\mu\delta}_{(k)} \mathcal{V}^{\nu \alpha}_{(i)} \right)
-2 \, u^{\mu} u^{\nu} u^{\gamma} u_{\beta} \,
\mathcal{V}^{\alpha \beta}_{(i)} \mathcal{V}_{(k) \alpha \gamma} \right].
\label{Tmunufrak3}
\end{eqnarray}
In the absence of vortices this tensor reduces to that in Sec.\ \ref{TypeIsymmetry} 
[see Eq.\ (\ref{DeltaTmunu2})].
In another limiting case of only one neutral superfluid particle species (e.g., $i=n$)
it
reproduces the tensor presented in G16 if one sets 
all the coefficients except for $\GammaMnn$ 
to zero
[see equation (79) in that reference].
A simple microscopic model allowing to calculate the 
phenomenological coefficients $\gammaEtilde$, $\gammaM$, $\GammaEitilde$, $\GammaMi$,
$\GammaEiktilde$, and $\GammaMik$ 
in Eq.\ (\ref{Tmunufrak3})
is considered in Appendix \ref{vortapp}.
This model is analogous to the model discussed in detail in GAS11.

\section{``Magnetohydrodynamic'' approximation for $npe$-mixture with neutron and proton vortices (type-II proton superconductivity)}
\label{MHDapprox}

General equations of Secs.\ \ref{TypeII} and \ref{TypeIIsymmetry}
can be substantially simplified if 
the magnetic induction ${\pmb B}$ 
is much larger than the fields 
${\pmb E}$, ${\pmb D}$, and ${\pmb H}$ in the comoving frame 
(hereafter the {\it magnetohydrodynamic} approximation) %
%
\footnote{In what follows we assume
that the relative velocity between the normal and superfluid components 
is much smaller than the speed of light $c$.
As is argued in G16 (see Appendix D there), 
this is not a very restrictive requirement.}.
%
As it is discussed in Appendix \ref{vortapp} (see Remark 1 there), 
as well as in GAS11, 
this is a typical situation in real neutron stars.
Note also that in the comoving frame 
$\mathcal{V}_{({\rm E}i)} \equiv
(\mathcal{V}_{({\rm E}i) \mu} \mathcal{V}^{\mu}_{({\rm E}i)})^{1/2}
\sim (1/c) \mathcal{V}_{({\rm M}i)}$
and can be neglected in comparison to $\mathcal{V}_{({\rm M}i)}$ 
[this follows from the analysis of the superfluid Eq.\ (\ref{nuE}) and
its nonrelativistic counterpart
in Appendix \ref{nonrel}].
In addition, one can neglect the neutron-related four-vector $\mathcal{V}^{\mu}_{({\rm M}n)}$
in comparison to the proton four-vector $\mathcal{V}^{\mu}_{({\rm M}p)}$ 
in Eq.\ (\ref{Hmu2}), 
because the lengths of these vectors are proportional to the vortex density $N_{{\rm V}i}$ 
[see Eq.\ \ref{Nvi1}], 
which is larger for protons by more than ten orders of magnitude.
Using these facts, a number of simplifications are possible:

(1) One can omit $H^{\mu}$ 
(and $\mathcal{V}^{\mu}_{({\rm M}n)}$, as we have already mentioned)
in Eq.\ (\ref{Hmu2}). 
This leads to the condition relating $B^{\mu}$ and $\mathcal{V}^{\mu}_{({\rm M}p)}$
(here and below in this section we, for definiteness,
use the parameters $\gammaM$, $\GammaMi$, $\GammaMik$, etc.\ calculated for 
a simple microscopic model of Appendix \ref{vortapp}),
\begin{equation}
\mathcal{V}^{\mu}_{({\rm M}p)} \approx -\frac{\gammaM}{4 \pi \GammaMp}\, B^{\mu}
= \frac{\pi}{\hat{\phi}_{p0}} \, B^{\mu} = e_p \, B^{\mu}.
\label{BVprelation}
\end{equation}
Physically, this condition means that almost all the magnetic induction 
is produced by the proton vortices.
Note that, from Eq.\ (\ref{Vmunu}) it follows 
\begin{equation}
\mathcal{V}^{\mu}_{({\rm M}p)} = 
\widetilde{\mathcal{V}}^{\mu}_{({\rm M}p)}
 + e_p B^{\mu}.
\label{Vmunu2}
\end{equation}
Comparing this equation with Eq.\ (\ref{BVprelation})
one sees that in the adopted approximation 
the vector $\widetilde{\mathcal{V}}^{\mu}_{({\rm M}p)}$,
which reduces to $(0,\, m_p \, {\rm curl} \, {\pmb V}_{{\rm s}p})$ 
in the nonrelativistic limit
(see footnote \ref{nonrel2}),
should be neglected in comparison to $e_p \, B^{\mu}$.

(2) Because ${\pmb H}$ and ${\pmb D}$ are small by assumption,
one can discard Maxwell's equation (\ref{22}), 
setting to zero
the four-current density $J^\mu_{({\rm free})}$
in all other equations,
\begin{equation}
J^\mu_{({\rm free})}=e_j j^{\mu}_{(j)}=e_p (n_p-n_e) u^\mu +e_i \, Y_{ik}w^{\mu}_{(k)} = 0,
\label{jfree3}
\end{equation}
that is, since $u_{\mu} w^{\mu}_{(i)}=0$ [see Eq.\ (\ref{uw})],
\begin{eqnarray}
n_e &=& n_p,
\label{quasi}\\
e_i \, Y_{ik}w^{\mu}_{(k)} &=& 0.
\label{jfree4}
\end{eqnarray}

(3) One can ignore the first 3 terms in the r.h.s.\
of the expression (\ref{deadd2}) for the energy density 
$d \varepsilon_{\rm add}$, 
because they depend on small four-vectors
$E^{\mu}$, $D^{\mu}$, $H^{\mu}$, $\mathcal{V}^{\mu}_{({\rm E}i)}$,
and $\mathcal{W}^{\mu}_{({\rm E}i)}$%
%
\footnote{
\label{WeWpar}	
The four-vector $\mathcal{W}^{\mu}_{({\rm E}i)}$ is expressed through
$\mathcal{V}^{\mu}_{({\rm E}i)}$ and $E^\mu$ by Eq.\ (\ref{WmuE3}) and hence is small.
Note that the tensor $\Wpar^{\mu\nu}_{(i)}$ 
is also small since it is in turn related to $\mathcal{W}^{\mu}_{({\rm E}i)}$
by Eq.\ (\ref{Welectr}).}.
%
The last term in Eq.\ (\ref{deadd2}) 
is large in comparison to the neglected terms, 
since it is independent of these small vectors, 
as is shown below.
Using Eq.\ 
(\ref{Hmu2}) with $H^{\mu}=0$, 
as well as Eqs.\ (\ref{Wmu2}), 
and (\ref{gamma})--(\ref{Gammaik3}),
one obtains (no summation over $i$ here) %
%
\footnote{We emphasize once again that the relation (\ref{WM4}) is 
only valid in the (usually adopted) approximation of noninteracting vortices, 
see Appendix \ref{vortapp}.}
%
\begin{eqnarray}
\mathcal{W}^{\mu}_{({\rm M}i)} = \frac{\lambda_i}{m_i \mathcal{V}_{({\rm M}i)}} \,
\mathcal{V}^{\mu}_{({\rm M}i)},
\label{WM4}
\end{eqnarray}
so that this last term can be approximately presented as
\begin{equation}
d \varepsilon_{\rm add} \approx 
\sum_{i=n,\, p} \frac{\lambda_i}{m_i \mathcal{V}_{({\rm M}i)}} \, \mathcal{V}_{({\rm M}i)\mu}
 d \mathcal{V}^{\mu}_{({\rm M}i)},
\label{deadd7}
\end{equation}
where $\lambda_p$ and $\lambda_n$ are given, respectively, by Eqs.\ (\ref{lambdap1}) and (\ref{lambdan1}) %
%
\footnote{In principle, the term with $i=n$
in Eq.\ (\ref{deadd7}) 
could be neglected 
in comparison to the $i=p$ term,
since in neutron stars 
$\mathcal{V}_{({\rm M}n)}\ll\mathcal{V}_{({\rm M}p)}$ and $\lambda_n \sim \lambda_p$.
However, we prefer to retain this term here 
in order to describe situations when protons are normal
and
$i=p$ term 
is absent.}.
%
Note that the proton-related term ($i=p$) in Eq.\ (\ref{deadd7})
reduces to $e_p \lambda_p/(m_p B) \, B_{\mu} dB^{\mu}$
in view of Eq.\ (\ref{BVprelation}) [here $B\equiv (B_{\mu}B^{\mu})^{1/2}$].

(4) One can repeat the derivation of Sec.\ \ref{TypeII}
with $d \varepsilon_{\rm add}$ 
from
Eq.\ (\ref{deadd7}).
As a result, one will 
derive
Eqs.\ (\ref{dT2}) and (\ref{sflrot1})
with the following modifications:
(i) The first three tensors in the r.h.s.\ of Eq.\ (\ref{dT2}) 
will not appear in the approximation adopted here,
since they are smaller than the fourth term 
(in principle, this can be independently checked by direct comparison
of the elements of four these tensors).
We thus left with
\begin{equation}
\Delta T^{\mu\nu} = \mathcal{T}^{\mu\nu}_{({\rm VM})}.
\label{dTmunu2}
\end{equation}
(ii) The four-vector $W^{\mu}_{(i)}$,
entering the definition 
of 
$f^{\mu}_{(i)}$ in Eq.\ (\ref{sflrot1}),
will be modified (no summation over $i$ is assumed), 
\begin{eqnarray}
W^{\mu}_{(i)} &=& 
\frac{1}{n_i}\left[
Y_{ik} w^\mu_{(k)} + \partial_{\alpha} 
\mathcal{W}^{\mu \alpha}_{(i)}
\right]
= \frac{1}{n_i}\left\{
Y_{ik} w^\mu_{(k)} + \partial_{\alpha} 
\left[ \Wperp^{\mu \alpha}_{(i)}+\Wpar^{\mu \alpha}_{(i)} \right]
\right\}
\nonumber\\
&\approx&\frac{1}{n_i}\left[
Y_{ik} w^\mu_{(k)} + \partial_{\alpha} 
\Wperp^{\mu \alpha}_{(i)}
\right]=
\frac{1}{n_i}\left\{
Y_{ik} w^\mu_{(k)} + \partial_{\alpha} 
\left[\epsilon^{\gamma\beta\mu \alpha} u_\beta \mathcal{W}_{({\rm M}i)\gamma}\right]
\right\}
\label{Wmu3}
\end{eqnarray}
[see the footnote \ref{WeWpar}, Eqs.\ (\ref{Aperp}), (\ref{Apar}), and (\ref{Adec})
and note that we neglect the small term depending on $\mathcal{W}^{\mu}_{({\rm E}i)}$ here].
All other equations remain exactly the same.

Summarizing, the system of simplified ``magnetohydrodynamic'' equations
for $npe$-mixture
consists of the energy-momentum and particle conservation laws 
(\ref{dTmunu}) and (\ref{dj}) 
with $j^{\mu}_{(j)}$ given by Eqs.\ (\ref{ji}), (\ref{je}) 
and $T^{\mu\nu}$ given by Eq.\ (\ref{Tmunu3}) with $\Delta T^{\mu\nu}$
from Eq.\ (\ref{dTmunu2}).
When calculating $\Delta T^{\mu\nu}$ one should express $\Wperp^{\mu\nu}_{(i)}$
through $\mathcal{W}^{\mu}_{({\rm M}i)}$, 
which is in turn should be found from Eq.\ (\ref{WM4}).
These equations should be supplemented by 
Maxwell's equation (\ref{11}), 
the second law of thermodynamics (\ref{2ndlaw3}) 
with $d\varepsilon_{\rm add}$ defined in Eq.\ (\ref{deadd7}), 
and by the conditions (\ref{norm}), (\ref{uw}), (\ref{quasi}), and (\ref{jfree4}).
Finally, the system is closed by the neutron and proton 
superfluid equations (\ref{sflrot1}) 
[or (\ref{nuE})],
in which $f^{\mu}_{(i)}$ is defined by Eq.\ (\ref{f3}) [or, equivalently, by Eq.\ (\ref{finv})]
and $W^{\mu}_{(i)}$ is given by Eq.\ (\ref{Wmu3}).
The nonrelativistic version of some of these equations 
is presented in Appendix \ref{nonrel}.

\vspace{0.2 cm}
\noindent
%
{\bf Remark 1.} ---
It is interesting that, 
using Eqs.\ (\ref{finv}) and (\ref{BVprelation}),
the proton four-vector $f^{\mu}_{(p)}$
can be 
represented in terms of
$B^{\mu}$,
\begin{equation}
f^{\mu}_{(p)} \approx -\alpha_p \, X^{\mu}_{(p)}
- \beta_p \, \epsilon^{\mu\nu\lambda\eta} \, u_{\nu} \, {\rm e}_{(p)\lambda} \, X_{(p)\eta}
+ e_p \, \gamma_p \, {\rm e}^{\mu}_{(p)} \, W^{\lambda}_{(p)} B_\lambda,
\label{finvp}
\end{equation}
where ${\rm e}^{\mu}_{(p)} \approx B^{\mu}/B$ 
and $X^{\mu}_{(i)} \approx 
e_p \, \epsilon^{\mu\nu\lambda\eta} \, u_{\nu} \, B_\lambda \, W_{(i)\eta}$.

\vspace{0.2 cm}
\noindent
%
{\bf Remark 2.} ---
Note that the proton four-vector $w^{\mu}_{(p)}$ 
can be found from 
the condition (\ref{jfree4}).
The proton superfluid equation can thus be
used to express the electric four-vector $E^{\mu}$.
Using Eq.\ (\ref{Vmunu}) 
in order to present $\mathcal{V}^{\mu}_{({\rm E}i)}$ 
as
$\mathcal{V}^{\mu}_{({\rm E}i)} = 
\widetilde{\mathcal{V}}^{\mu}_{({\rm E}i)} + e_i E^{\mu}$,
and substituting this expression (for $i=p$) 
into Eq.\ (\ref{nuE}), one finds
\begin{equation}
E^{\mu} = -\frac{1}{e_p}\, \widetilde{\mathcal{V}}^{\mu}_{({\rm E}p)}
+\frac{\mu_p n_p}{e_p} \, f^{\mu}_{(p)},
\label{}
\end{equation}
where $f^{\mu}_{(p)}$ is given by Eq.\ (\ref{finvp}).
Together with Maxwell's equation (\ref{11})
this equation allows one, in principle, 
to exclude $E^{\mu}$ and obtain a closed equation for $B^{\mu}$ only
(see Remark 1 in Appendix \ref{nonrel}, 
where such an equation is derived in the nonrelativistic limit).

\section{Summary and conclusions}
\label{summary}

This paper is devoted to 
studying 
the dynamic properties of
superfluid-superconducting mixtures in neutron stars
accounting for the possible presence
of electric and magnetic fields, 
as well as neutron (Feynman-Onsager) 
and proton (Abrikosov) vortices.
Our results and main conclusions are summarized as follows:

1. Using the method and ideas from Refs.\ \cite{bk61} and G16, we derived 
a set of fully {\it relativistic} equations 
(see Appendix \ref{sumapp}) 
describing a charged mixture composed of superfluid neutrons, superconducting protons,
and electrons (the simplest neutron-star composition). 
Generalization of these equations to more exotic compositions 
(including, e.g., muons, hyperons, etc.) 
is straightforward \cite{dg16,kg09,gk08, hac12}.

2. The proposed equations can be used 
at {\it finite} temperatures, i.e., 
they allow for 
the possible presence of neutron and proton (Bogoliubov) thermal excitations.
This is especially important for a sufficiently hot neutron stars, 
such as magnetars, whose internal temperatures can be $\sim 10^8$~K, 
i.e., of the order of the nucleon critical temperatures $T_{{\rm c}i}$ \cite{kkpy14, vigano_etal_13}
(we remind that at $T>T_{{\rm c}i}$ nucleon species $i=n$, $p$ is completely nonsuperfluid). 

3. The derived dynamic equations are ``nondissipative'' in a sense that to obtain
them we assume that normal (nonsuperfluid) liquid components
(electrons, nucleon thermal excitations, and entropy) move with one and the same velocity 
(i.e., diffusion effects 
are ignored).
However, we do take into account the {\it mutual friction dissipation } 
[see Eqs.\ (\ref{sfleq5}) and (\ref{f3})].
Extension of our results to a fully dissipative problem is rather 
easy
and will be reported elsewhere.

4. Estimates show that protons form type-II superconductor in the 
outer neutron-star core, 
but become of type-I in the inner core (e.g., GAS11, \cite{aw08,ssz97,ss15,sedrakyan05}).
The dynamic equations are derived and analysed in both these cases
with the special emphasis on the more elaborated type-II case. 
It seems that the dynamics of type-I superconductor is discussed
for the first time (in the astrophysical context), but 
the analysis presented is rather brief and simplified, 
and should 
be considered as a first step towards
the solution of this complex problem.

5. Our main results include the ``electromagnetic'' energy-momentum tensors 
$\mathcal{T}_{({\rm E}) }^{\mu\nu}$
(\ref{Te1app}) and $\mathcal{T}_{({\rm M}) }^{\mu\nu}$ (\ref{Tm1app}), 
and the nucleon
``vortex'' energy-momentum tensors $\mathcal{T}_{({\rm VE}) }^{\mu\nu}$ (\ref{vortexEapp}) 
and $\mathcal{T}_{({\rm VM}) }^{\mu\nu}$ (\ref{vortexapp}),
as well as the ``superfluid'' equations for the cases of type-I (\ref{Vappn}), (\ref{Vappp}) 
and type-II (\ref{sflrot1app}) proton superconductivities.
Remarkably, the vortex energy-momentum tensors 
have the same structure and are obtained exactly in the same way as the electromagnetic tensors 
(\ref{Te1app}) and (\ref{Tm1app}) 
(see Remark 4 in Sec.\ \ref{TypeII}).

6. As a by-product of our work it is shown that for normal matter 
the sum $\mathcal{T}_{({\rm E}) }^{\mu\nu}+\mathcal{T}_{({\rm M}) }^{\mu\nu}$ 
of the electromagnetic energy-momentum tensors is directly related 
to the so called Abraham tensor $T^{\mu\nu}_{\rm Abraham}$
of the standard electrodynamics of continuous media \cite{ginzburg73, gu76, toptygin15}.
Thus, our results can be considered as one more derivation of this tensor
based on the conservation laws and the requirement that the entropy of a nondissipative 
closed system remains constant.

7. The equations derived in this paper 
[in particular, 
the expressions for the electromagnetic and vortex energy-momentum tensors 
$\mathcal{T}_{({\rm E}) }^{\mu\nu}$, $\mathcal{T}_{({\rm M}) }^{\mu\nu}$,
$\mathcal{T}_{({\rm VE}) }^{\mu\nu}$, and $\mathcal{T}_{({\rm VM}) }^{\mu\nu}$]
depend on the four-vectors $E^{\mu}$, $B^{\mu}$, 
$\mathcal{V}^{\mu}_{({\rm E}i)}$, $\mathcal{V}^{\mu}_{({\rm M}i)}$
and the complementary four-vectors 
$D^{\mu}$, $H^{\mu}$, 
$\mathcal{W}^{\mu}_{({\rm E}i)}$, $\mathcal{W}^{\mu}_{({\rm M}i)}$.
The physical meaning of these four-vectors is described in detail in the text.
For example, the spatial components of
$E^{\mu}$, $B^{\mu}$, $D^{\mu}$, and $H^{\mu}$
reduce, respectively, to the electric field, magnetic induction,
electric displacement, and magnetic field in the comoving frame 
moving with the normal liquid component (see Sec.\ \ref{EBDH});
the other four-vectors are related to vortices.

The four-vectors mentioned above are not all independent.
To express the quantities $D^{\mu}$, $H^{\mu}$, 
$\mathcal{W}^{\mu}_{({\rm E}i)}$, $\mathcal{W}^{\mu}_{({\rm M}i)}$
through 
$E^{\mu}$, $B^{\mu}$, 
$\mathcal{V}^{\mu}_{({\rm E}i)}$, $\mathcal{V}^{\mu}_{({\rm M}i)}$
one should specify, as in the usual electrodynamics of continuous media,
the microphysics model for the mixture.
This is done, for two simple models,
in Appendix \ref{deterapp}
(in particular, one of these models analyses the system of noninteracting vortices).
However, it is important to point out that the general 
equations obtained here
will likely 
remain unchanged 
if one considers
more complex models. 
The only thing that should be 
modified
in the latter case
is 
the relations between the fields
$D^{\mu}$, $H^{\mu}$, 
$\mathcal{W}^{\mu}_{({\rm E}i)}$, $\mathcal{W}^{\mu}_{({\rm M}i)}$
and
$E^{\mu}$, $B^{\mu}$, 
$\mathcal{V}^{\mu}_{({\rm E}i)}$, $\mathcal{V}^{\mu}_{({\rm M}i)}$.

8. It is instructive to compare our results with the
most advanced nonrelativistic magnetohydrodynamics of GAS11,
describing superfluid-superconducting mixtures. 
In comparison to GAS11 we: 
(i) take into account the relativistic and finite-temperature effects;
(ii) provide a general framework allowing one to easily incorporate new physics
into the existing dynamic equations;
and (iii) demonstrate that the electric displacement field ${\pmb D}$
is {\it not} generally equal to the electric field ${\pmb E}$,
contrary to what was assumed
in GAS11 and some other papers 
starting from the work by Mendell \cite{mendell91a}
(see also Ref.\ \cite{ss95}).

9. The rather complex general system of equations derived
in this work can be substantially simplified for typical neutron-star conditions,
for which 
a kind of
``magnetohydrodynamic'' approximation
is justified.
This approximation is analogous to the usual magnetohydrodynamic approximation
for ordinary stars. 
The corresponding equations 
are derived and analysed 
for a simple model of Appendix \ref{vortapp} in Sec.\ \ref{MHDapprox};
their 
nonrelativistic limit 
is
presented in Appendix \ref{nonrel},
where we also derive a ``magnetic field evolution equation'' (\ref{MagnEvol2}).
It is shown that 
the latter equation
coincides with that proposed in Ref.\ \cite{kg01}, 
but differs from the evolution equation derived in Ref.\ \cite{gagl15} 
using magnetohydrodynamics of GAS11.

\begin{acknowledgments}
	
The authors are deeply grateful to Elena Kantor 
and Andrey Chugunov for numerous useful discussions,
to D.~G.~Yakovlev for encouragement, 
and to Kostas Glampedakis for the discussion 
of the magnetic field evolution equation (\ref{MagnEvol2}).
This study was supported by the Russian Science Foundation 
(grant \textnumero14-12-00316).

\end{acknowledgments}

\appendix

\section{Some useful definitions}
\label{notation}

Assume we have an arbitrary antisymmetric tensor $\mathcal{A}^{\mu\nu}$, 
which can be represented in the matrix form as
\begin{equation}
\mathcal{A}^{\mu \nu} 
=\left( 
\begin{array}{cccc}
0    & \mathcal{A}_{01}  &  \mathcal{A}_{02} & \mathcal{A}_{03}\\
-\mathcal{A}_{01} &  0   &  \mathcal{A}_{12} & \mathcal{A}_{13} \\
-\mathcal{A}_{02} & -\mathcal{A}_{12} &  0   & \mathcal{A}_{23} \\
-\mathcal{A}_{03} & -\mathcal{A}_{13}  & -\mathcal{A}_{23} & 0 
\end{array} 
\right).
\label{Amunu}\\
\end{equation}
Here and below all matrix representations of tensors/vectors are given in the comoving frame,
i.e.\ in the frame, in which
the normal four-velocity is  $u^\mu=(1,\, 0,\, 0,\, 0)$.

The tensor $\Adual^{\mu \nu}$,  dual to the tensor $\mathcal{A}^{\mu\nu}$, is
\begin{equation}
\Adual^{\mu \nu} 
\equiv \frac{1}{2}\, \epsilon^{\mu \nu \alpha \beta} \mathcal{A}_{\alpha \beta}
=\left( 
\begin{array}{cccc}
0    & \mathcal{A}_{23}  &  -\mathcal{A}_{13} & \mathcal{A}_{12}\\
-\mathcal{A}_{23} &  0   &  -\mathcal{A}_{03} & \mathcal{A}_{02} \\
\mathcal{A}_{13} & \mathcal{A}_{03} &  0   & -\mathcal{A}_{01} \\
-\mathcal{A}_{12} & -\mathcal{A}_{02}  & \mathcal{A}_{01} & 0 
\end{array} 
\right).
\label{Amunu2}\\
\end{equation}
Using these tensors, one can construct the ``electric'' $\mathcal{A}_{({\rm E})}^\mu$ and ``magnetic''  
$\mathcal{A}_{({\rm M})}^\mu$ four-vectors \cite{lichnerowicz67}
\begin{eqnarray}
\mathcal{A}_{({\rm E})}^\mu &\equiv&
u_{\nu} \mathcal{A}^{\mu\nu}
=(0,\, \mathcal{A}_{01},\,\mathcal{A}_{02},\,\mathcal{A}_{03}),
\label{AE}\\
\mathcal{A}_{({\rm M})}^\mu &\equiv& u_{\nu} 
\Adual^{\mu\nu}
= \frac{1}{2} \, \epsilon^{\mu \nu \alpha \beta} \, u_{\nu} \, \mathcal{A}_{\alpha \beta}
=(0,\, \mathcal{A}_{23},\,-\mathcal{A}_{13},\,\mathcal{A}_{12}),
\label{AM}
\end{eqnarray}
and two additional tensors 
\begin{eqnarray}
\Aperp^{\mu\nu}&=& \epsilon^{\alpha\beta\mu\nu} u_{\beta} \, \mathcal{A}_{({\rm M})\, \alpha}
=\perp^{\mu\alpha}\perp^{\nu\beta}\mathcal{A}_{\alpha\beta}
=\left( 
\begin{array}{cccc}
0    & 0  &  0 & 0\\
0 &  0   &  \mathcal{A}_{12} & \mathcal{A}_{13} \\
0& -\mathcal{A}_{12} &  0   & \mathcal{A}_{23} \\
0& -\mathcal{A}_{13}  & -\mathcal{A}_{23} & 0 
\end{array} 
\right),
\label{Aperp}\\
\Apar^{\mu\nu} &=& 
-u^\nu \mathcal{A}_{({\rm E})}^\mu +u^\mu \mathcal{A}_{({\rm E})}^\nu
=-u^\nu u_\alpha \, \mathcal{A}^{\mu\alpha}+u^\mu u_\alpha \, \mathcal{A}^{\nu\alpha}
=\left( 
\begin{array}{cccc}
0    & \mathcal{A}_{01}  &  \mathcal{A}_{02} & \mathcal{A}_{03}\\
-\mathcal{A}_{01} &  0   &  0 & 0 \\
-\mathcal{A}_{02} & 0 &  0   & 0\\
-\mathcal{A}_{03} & 0  & 0 & 0 
\end{array} 
\right)
\label{Apar}
\end{eqnarray}
with the properties
\begin{eqnarray}
u_{\nu} \Aperp^{\mu\nu}&=&0,
\label{Perp_prop}\\
\perp_{\mu\nu} \Apar^{\mu\nu} &=& 0,
\label{Par_prop}
\end{eqnarray}
where $\perp^{\mu\nu}=g^{\mu\nu}+u^\mu u^\nu$ is the projection operator and 
$\epsilon^{\alpha \beta \mu \nu}$ is the Levi-Civita tensor, $\epsilon^{0123}=1$.
One can see that the tensor $A^{\mu\nu}$ can be decomposed as
\begin{equation}
\mathcal{A}^{\mu\nu}=\Aperp^{\mu\nu}+\Apar^{\mu\nu}.
\label{Adec}
\end{equation}
%

\section{Comparison of notation used in this paper and in G16}
\label{corresp}

Some of the parameters introduced in G16 and in the present paper
differ only by the index $i$,
since here we have two superfluid/superconducting particle species 
[neutrons ($i=n$) and protons ($i=p$)],
whereas G16 deals with 
one particle species.
Such parameters are not provided in the table below.

\begin{center}
\begin{tabular}{|c|c|c|}
	\hline \rule[-2ex]{0pt}{5.5ex} \quad \quad G16 \quad \quad         & \quad \quad This work \quad\quad& Parameter name \\ 
	\hline \rule[-2ex]{0pt}{5.5ex} $F^{\mu\nu}$ & $\mathcal{V}^{\mu\nu}_{(i)}$ &  Vorticity tensor \\ 
	\hline \rule[-2ex]{0pt}{5.5ex} $O^{\mu\nu}$ & $\Vperp^{\mu\nu}_{(i)}$ & ``Magnetic'' part of the vorticity tensor \\ 		
	\hline \rule[-2ex]{0pt}{5.5ex} $H^{\mu}$    & $\mathcal{V}_{({\rm M}i)}^{\mu}$ & ``Magnetic'' vorticity-related vector \\ 		
	\hline \rule[-2ex]{0pt}{5.5ex} $-E^{\mu}$    & $\mathcal{V}_{({\rm E}i)}^{\mu}$ &``Electric'' vorticity-related vector   \\ 		
	\hline \rule[-2ex]{0pt}{5.5ex} $V_{({\rm L})}^{\mu}$ & $v_{({\rm L}i)}^{\mu}$ & Vortex four-velocity (non-normalized)\\ 		
	\hline \rule[-2ex]{0pt}{5.5ex} $H$          & $\mathcal{V}_{({\rm M}i)}$ &   Length of the four-vector 	$\mathcal{V}_{({\rm M}i)}^{\mu}$ (or $H^{\mu}$)	\\		
	\hline \rule[-2ex]{0pt}{5.5ex} ${\pmb H}$ & $\mathbfcal{V}_{{\rm M}i}$ & Spatial part of the four-vector  $\mathcal{V}_{({\rm M}i)}^{\mu}$ (or $H^{\mu}$) \\ 		
	\hline \rule[-2ex]{0pt}{5.5ex} $-{\pmb E}$ & $\mathbfcal{V}_{{\rm E}i}$ &   Spatial part of the four-vector  $\mathcal{V}_{({\rm E}i)}^{\mu}$ (or $-E^{\mu}$) \\ 				
	\hline \rule[-2ex]{0pt}{5.5ex} ${\pmb V}_{\rm L}$ & ${\pmb v}_{{\rm L}i}$ & Spatial part of the vortex four-velocity $v_{({\rm L}i)}^{\mu}$  [or $V_{({\rm L})}^{\mu}$] \\ 		
				
	\hline 
\end{tabular} 
\end{center}

\section{Energy density transformation}
\label{transformation}

Assume we have a term in the expression for the energy density
which takes the form
\begin{equation}
d \varepsilon_{\rm part}=\frac{1}{2} \, \left( 
O^{\alpha \beta} \, d \mathcal{F}_{\alpha \beta} + 2 \mathcal{B}_{\alpha\beta} \mathcal{A}^{\alpha\gamma} u^{\beta} \, d u_{\gamma}
\right), 
\label{depart}
\end{equation}
where $O^{\alpha \beta}$, $\mathcal{A}^{\alpha\beta}$, 
and $\mathcal{B}^{\alpha\beta}$ are some arbitrary 
{\it antisymmetric} tensors; 
$\mathcal{F}^{\alpha\beta}$ is the antisymmetric tensor satisfying the condition 
%
\footnote{For example, it can be the electromagnetic tensor 
$F^{\alpha\beta}$ or the vorticity tensor $\mathcal{V}^{\mu\nu}_{(i)}$, 
see Eqs.\ (\ref{11}) and (\ref{vortstar}).}
%
\begin{equation}
\partial_{\alpha} 
\FFdual^{\alpha\beta}=0;
\label{111}
\end{equation}
and
$u^{\mu}$ is the four-velocity of normal liquid component.
Our aim will be to transform 
the expression 
$-u^{\mu} \partial_\mu \varepsilon_{\rm part}$
to some standard form [see Eq.\ (\ref{de2}) in what follows];
this transformation
is used
several times in the main text of the paper (see also G16).
Using (\ref{depart}), one has
\begin{equation}
-u^{\mu} \, \partial_{\mu} \varepsilon_{\rm part}
= -\frac{1}{2} \,
u^{\mu}\, O^{\alpha \beta} \, \partial_{\mu} \mathcal{F}_{\alpha \beta} 
- u^{\mu} u^{\delta} \mathcal{B}_{\alpha\delta} \mathcal{A}^{\alpha\nu} \, \partial_{\mu} u_{\nu}.
\label{ude1}
\end{equation}
The first term in the r.h.s. 
of Eq.\ (\ref{ude1}) can be transformed as
\begin{eqnarray}
-\frac{1}{2}\, u^{\mu} \, O^{\alpha \beta} \partial_\mu \mathcal{F}_{\alpha \beta}
&=& u^{\nu} \mathcal{F}_{\mu\nu} \, \, \partial_{\alpha}O^{\mu\alpha}
\nonumber\\
&-&\partial_{\mu}\left( u^{\nu}  \,  O^{\mu\alpha} \mathcal{F}_{\nu\alpha} \right)
\nonumber\\
&+&\partial_{\mu}u^{\nu} \left(O^{\mu\alpha} \mathcal{F}_{\nu \alpha}
\right).
\label{1term}
\end{eqnarray}
To obtain this expression we used Eq.\ (\ref{111}),
which is equivalent to
\begin{equation}
\partial_\mu \mathcal{F}_{\alpha \beta} = 
\partial_{\alpha} \mathcal{F}_{\mu \beta}+\partial_{\beta} \mathcal{F}_{\alpha \mu},
\label{equality}
\end{equation}
and 
the fact that both
tensors $\mathcal{F}^{\mu \nu}$ and $O^{\mu\nu}$ are antisymmetric.

The second term in the r.h.s. of Eq.\ (\ref{ude1})
can be rewritten as
\begin{eqnarray}
&&-  u^{\mu} u^{\delta}\, \mathcal{B}_{\alpha\delta} \mathcal{A}^{\alpha \nu} \, \partial_{\mu} u_{\nu} 
= 
- \left[u^{\mu}u^{\delta}\, \mathcal{B}_{\alpha \delta} \mathcal{A}^{\alpha \nu} 
+ \underline{u^{\mu} u^{\nu} u^{\beta} u_{\gamma} \, \mathcal{B}_{\alpha \beta} \mathcal{A}^{\alpha \gamma} } \right]
\partial_{\mu} u_\nu
\nonumber\\
&&= - u^{\mu}u^{\gamma}
\perp_{\nu\beta} \mathcal{A}^{\alpha\beta} \mathcal{B}_{\alpha\gamma} 
\,\,\partial_{\mu} u^\nu
\nonumber\\
&&= - u^{\mu}u^{\gamma}
\perp_{\nu\beta} \mathcal{A}^{\alpha\beta} \mathcal{B}_{\alpha\gamma} 
\,\,\partial_{\mu} u^\nu
+ \underline{\partial_{\mu} \left(u^{\nu}
	u^{\mu} u^{\gamma} \perp_{\nu\beta}  \mathcal{A}^{\alpha\beta} \mathcal{B}_{\alpha\gamma} \right)},
\label{2term}
\end{eqnarray}
where the underlined terms equal zero 
(because $u_{\nu} \partial_{\mu}u^{\nu}=0$ and $u^{\nu} \perp_{\nu\beta}=0$);
they are added here 
in order to symmetrize the corresponding energy-momentum tensor $T^{\mu\nu}$
and to satisfy the condition $u_{\mu} u_{\nu} \Delta T^{\mu\nu}=0$ 
(see the main text).
Combining Eqs.\ (\ref{1term}) and (\ref{2term}), one obtains
\begin{eqnarray}
-u^{\mu} \, \partial_{\mu} \varepsilon_{\rm part} &=&
u^{\nu} \mathcal{F}_{\mu\nu} \, \, \partial_{\alpha} O^{\mu\alpha}
\nonumber\\
&-&\partial_{\mu}\left[ u^{\nu} \left(
O^{\mu\alpha} \mathcal{F}_{\nu\alpha} 
- u^{\mu}u^{\gamma}
\perp_{\nu\beta} \mathcal{A}^{\alpha\beta} \mathcal{B}_{\alpha\gamma} 
\right)\right]
\nonumber\\
&+&\partial_{\mu}u^{\nu} \left(O^{\mu\alpha} \mathcal{F}_{\nu \alpha}
- u^{\mu}u^{\gamma}
\perp_{\nu\beta} \mathcal{A}^{\alpha\beta} \mathcal{B}_{\alpha\gamma} 
\right).
\label{de2}
\end{eqnarray}
%

\section{Energy-momentum tensor (\ref{DeltaTmunu}) and its relation to the Abraham tensor}
\label{Abraham}

As mentioned in Sec.\ \ref{TypeI}, 
the derivation of
the energy-momentum tensor (\ref{DeltaTmunu})
can also be 
applied to ordinary (nonsuperfluid) matter.
In other words, this tensor should have a well known counterpart in the literature.
Here we explore this issue in more detail.

We consider a normal (isotropic and homogeneous in the comoving frame)
dielectric ``fluid'' 
with the energy-momentum tensor
\begin{equation}
T^{\mu \nu} = (P+\varepsilon) \, u^{\mu} u^{\nu} + P g^{\mu \nu} 
+ \Delta T^{\mu\nu}
\label{Tmunu5}
\end{equation}
and the second law of thermodynamics
\begin{eqnarray}
d \varepsilon = T \, d S + \mu \, d n 
+ \frac{1}{4\pi} \, E_{\mu} d D^{\mu} + \frac{1}{4\pi} \, H_{\mu} d B^{\mu}.
\label{2ndlaw4}
\end{eqnarray}
In Eqs.\ (\ref{Tmunu5}) and (\ref{2ndlaw4})
$\Delta T^{\mu\nu}$ is given by Eq.\ (\ref{DeltaTmunu}) %
%
\footnote{Note that for a dielectric fluid the free-charge four-current density
	$J^{\mu}_{\rm (free)}$ in Eq.\ (\ref{22}) equals zero, 
	$J^{\mu}_{\rm (free)}=0$, 
	hence the first line in the r.h.s.\ of Eq.\ (\ref{ude2}) is zero too and 
	the derivation of Sec.\ \ref{TypeIfree} can indeed be used to obtain $\Delta T^{\mu\nu}$ 
	in the form (\ref{DeltaTmunu}).};
%
$n$ is the ``particle'' number density 
[it can be composite particles; in the case of a few particle species $j$ 
the second term in Eq.\ (\ref{2ndlaw4}) should be replaced with $\sum_j \, \mu_j d n_j$];
$\mu$ is the relativistic chemical potential;
and $P$ is the pressure,
\begin{equation}
P = -\varepsilon+\mu n + TS.
\label{Pres4}
\end{equation}
Since the medium is isotropic and homogeneous, 
the displacement vector ${\pmb D}$ 
and magnetic induction ${\pmb B}$ 
can be presented, 
in the comoving frame, as
\begin{eqnarray}
{\pmb D} &=&\widehat{\varepsilon} \,  {\pmb E},
\label{D2}\\
{\pmb B}&=&\widehat{\mu} \, {\pmb H},
\label{B2}
\end{eqnarray}
where $\widehat{\varepsilon}$ and $\widehat{\mu}$
are the corresponding permeabilities (scalars).
We assume, in addition, that the permeabilities are {\it field-independent},
but can generally be  
functions of $n$ and $S$.
Because the time components of the four-vectors $D^{\mu}$, $E^{\mu}$, 
$B^{\mu}$, and $H^{\mu}$ all vanish in the comoving frame, 
it follows from Eqs.\ (\ref{D2}) and (\ref{B2}) that
\begin{eqnarray}
D^\mu &=&\widehat{\varepsilon} \, E^\mu,
\label{D3}\\
B^\mu&=&\widehat{\mu} \, H^\mu.
\label{B3}
\end{eqnarray}
Using Eqs.\ (\ref{D3}) and (\ref{B3}),
Eq.\ (\ref{2ndlaw4}) can be readily integrated
and presented as
\begin{equation}
\varepsilon = \varepsilon_{\rm fluid}(n,\, S) +
\frac{1}{8\pi}
\left( 
E_\alpha D^{\alpha}+H_{\alpha}B^{\alpha}
\right)
= \varepsilon_{\rm fluid}(n,\, S) 
+\frac{1}{8\pi}
\left( 
\widehat{\varepsilon} \, E_{\alpha}E^{\alpha}+\widehat{\mu} \, H_{\alpha}H^{\alpha}
\right),
\label{eEM}
\end{equation}
where $\varepsilon_{\rm fluid}(n,\, S)$ is the fluid energy density, 
the same function of $n$ and $S$
as in the absence of the electromagnetic field.
Combining Eqs.\ (\ref{Pres4}) and (\ref{eEM}), one obtains
\begin{equation}
P=-\varepsilon_{\rm fluid}(n,\, S) 
+\mu n +TS
-\frac{1}{8\pi}
\left(
E_\alpha D^{\alpha}+H_\alpha B^{\alpha}
\right).
\label{Pres3}
\end{equation}
The chemical potential 
$\mu$
and temperature 
$T$
in this equation
still depend on the fields $D^{\alpha}$ and $B^{\alpha}$. 
As follows from Eqs.\ (\ref{2ndlaw4}) and (\ref{eEM}),
\begin{eqnarray}
\mu(n,\,S,\, D_\alpha D^{\alpha},\, B_{\alpha}B^{\alpha} ) 
&=& \frac{\partial \varepsilon(n,\, S, \, D_\alpha D^{\alpha},\, B_{\alpha}B^{\alpha})}{\partial n}
\nonumber\\
&=& \frac{\partial \varepsilon_{\rm fluid}(n,\, S)}{\partial n} -
\frac{1}{8\pi} 
\left( 
\frac{\partial \widehat{\varepsilon}(n,\, S)}{\partial n} \, E_{\alpha} E^{\alpha}+
\frac{\partial \widehat{\mu}(n,\, S)}{\partial n} \, H_{\alpha} H^{\alpha}
\right) 
\nonumber\\
&\equiv& \mu_{\rm fluid}(n,\, S)+\delta \mu,
\label{mu}
\end{eqnarray}
where
$\mu_{\rm fluid}(n,\, S)=\partial \varepsilon_{\rm fluid}(n,\, S)/\partial n$ 
is the same function of $n$ and $S$ as in the system without the electromagnetic field
and $\delta \mu$ is 
\begin{equation}
\delta \mu= -\frac{1}{8\pi} 
\left( 
\frac{\partial \widehat{\varepsilon}(n,\, S)}{\partial n} \, E_{\alpha} E^{\alpha}+
\frac{\partial \widehat{\mu}(n,\, S)}{\partial n} \, H_{\alpha} H^{\alpha}
\right).
\label{dmu}
\end{equation} 
Similar formulas can also be written for the temperature, 
$T=T_{\rm fluid}(n, S)+\delta T$,
where
\begin{equation}
\delta T= -\frac{1}{8\pi} 
\left( 
\frac{\partial \widehat{\varepsilon}(n,\, S)}{\partial S} \, E_{\alpha} E^{\alpha}+
\frac{\partial \widehat{\mu}(n,\, S)}{\partial S} \, H_{\alpha} H^{\alpha}
\right).
\label{dT1}
\end{equation} 
Substituting Eq.\ (\ref{mu}) and similar equation for $T$
into Eq.\ (\ref{Pres4}),
we arrive at
\begin{equation}
P=P_{\rm fluid}+\delta \mu \,  n + \delta T \, S
-\frac{1}{8\pi}
\left(
E_\alpha D^{\alpha}+H_\alpha B^{\alpha}
\right),
\label{Pres5}
\end{equation}
where $P_{\rm fluid} = -\varepsilon_{\rm fluid}+\mu_{\rm fluid} n + T_{\rm fluid} S$.
Now, using equations derived above one 
can present Eq.\ (\ref{Tmunu5}) in the form
\begin{equation}
T^{\mu\nu}=
T^{\mu\nu}_{\rm (fluid)}+ \mathcal{T}^{\mu\nu}_{({\rm EM})},
\label{Tmunu4}
\end{equation}
where $T^{\mu\nu}_{\rm (fluid)}
= (P_{\rm fluid}+\varepsilon_{\rm fluid}) \, u^{\mu} u^{\nu} + P_{\rm fluid}\,  g^{\mu \nu}$ 
is the fluid energy-momentum tensor 
(the same as in the absence of electromagnetic field)
and $\mathcal{T}^{\mu\nu}_{({\rm EM})}$ is the 
electromagnetic tensor in the {\it medium},
\begin{eqnarray}
\mathcal{T}^{\mu}_{({\rm EM}) \, \nu} &=& 
\perp^\mu_{\,\,\,\nu} (\delta \mu \, n +\delta T \, S)
-\frac{1}{8\pi} g^{\mu}_{\,\,\,\nu} \, \left(E_\alpha D^{\alpha}+H_\alpha B^{\alpha} \right)
+\mathcal{T}^{\mu}_{({\rm E})\,\nu}	
+\mathcal{T}^{\mu}_{({\rm M})\, \nu}
\nonumber\\	
&=&	
\perp^\mu_{\,\,\,\nu} (\delta \mu \, n +\delta T \, S)+
\frac{1}{8\pi} g^{\mu}_{\,\,\,\nu} \, \left(E_\alpha D^{\alpha} - H_\alpha B^{\alpha} \right)
\nonumber\\
&+&\frac{1}{4\pi} \left[ 
G^{\mu\alpha} F_{\nu \alpha}
+ u^{\mu}u^{\gamma}
\perp_{\nu\beta} \left(F^{\alpha\beta} G_{\alpha\gamma} - G^{\alpha\beta} F_{\alpha\gamma} \right)
\right].
\label{TmunuMedium1}
\end{eqnarray}
It is easily checked 
that this tensor equals 
to the so called Abraham tensor, $T^{\mu\nu}_{\rm (Abraham)}$ \cite{ginzburg73, gu76, toptygin15},
\begin{equation}
\mathcal{T}^{\mu\nu}_{({\rm EM})} \equiv T^{\mu\nu}_{\rm (Abraham)} =
T^{\mu\nu}_{\rm (Minkowski)} + \left(g_{\rm (A)}^\mu - g_{\rm (M)}^\mu \right)  u^{\nu},
\label{TmunuMedium2}
\end{equation}
where $T^{\mu\nu}_{\rm (Minkowski)}$
is the Minkowski tensor \cite{toptygin15},
\begin{eqnarray}
T^{\mu\nu}_{({\rm Minkowski})} &\equiv&  
\perp^{\mu\nu} (\delta \mu \, n +\delta T \, S)
\nonumber\\
&+& \frac{1}{4\pi}\left(
F^\mu_{\,\,\,\,\gamma} G^{\nu\gamma} 
- \frac{1}{4} \, g^{\mu\nu} \, F_{\gamma\delta}G^{\gamma\delta}
\right).
\label{TmunuMinkowski}
\end{eqnarray}
and the four-vectors 
$g_{\rm (A)}^{\mu}$ and $g_{\rm (M)}^{\mu}$ are
\begin{eqnarray}
g^{\mu}_{\rm (A)} &=& \frac{1}{4\pi}\,  \epsilon^{\mu\nu\alpha\beta}\, u_{\nu} \, E_{\alpha} H_{\beta},
\label{ga}\\
g^{\mu}_{\rm (M)} &=& \frac{1}{4\pi}\,  \epsilon^{\mu\nu\alpha\beta}\, u_{\nu} \, D_{\alpha} B_{\beta}.
\label{gM}
\end{eqnarray}
The latter four-vectors reduce, in the comoving frame, to
\begin{eqnarray}
(0,\, {\pmb g}_{\rm A})=
\left(0, \, \frac{{\pmb E}\times {\pmb H}}{4\pi} \right),
\label{ga2}\\
(0,\, {\pmb g}_{\rm M})=\left(0, \, \frac{{\pmb D}\times {\pmb B}}{4\pi}\right),
\label{gM2}
\end{eqnarray}
where 
${\pmb g}_{\rm A}$ is the so called Abraham momentum density 
(it coincides with the energy flux density) 
and ${\pmb g}_{\rm M}$ is the Minkowski momentum density.
In the comoving frame the tensor 
$\mathcal{T}^{\mu\nu}_{({\rm EM})}$ [$= T^{\mu\nu}_{\rm (Abraham)}$]
can be schematically presented as
\begin{equation}
\mathcal{T}^{\mu\nu}_{\rm (EM)} 
= T^{\mu\nu}_{\rm (Abraham)}=
\left( 
\begin{array}{cc}
	\varepsilon_{\rm EM} & {\pmb g}_{\rm A} \\ 
	{\pmb g}_{\rm A} & -\sigma^{lm}
 \end{array} 
 \right),
\label{TAbraham}
\end{equation}
where $\varepsilon_{\rm EM}$ is the energy density
and $\sigma^{lm}$ is the stress tensor of the electromagnetic field ($l$, $m=1$, $2$, $3$),
\begin{eqnarray}
\varepsilon_{\rm EM} &=& \frac{1}{8\pi} \left( \widehat{\varepsilon} \, {\pmb E}^2 + \widehat{\mu} \, {\pmb H}^2 \right),
\label{W}\\
\sigma^{lm}&=& \frac{1}{4\pi}\left(E^l D^m + H^l B^m \right)
\nonumber\\
&-& 
\left[
\frac{\pmb{E}^2}{8 \pi} \left( \widehat{\varepsilon}-n \frac{\partial\widehat{\varepsilon}}{\partial n}
- S \frac{\partial\widehat{\varepsilon}}{\partial S}\right)
+ \frac{{\pmb H}^2}{8 \pi} \left( \widehat{\mu}-n \frac{\partial\widehat{\mu}}{\partial n}
- S \frac{\partial\widehat{\mu}}{\partial S}\right)
\right] \delta^{lm},
\label{sigmalm}
\end{eqnarray}
and $\delta^{lm}$ is the Kronecker symbol. 
To obtain Eq.\ (\ref{sigmalm}), we express $\delta \mu$ and $\delta T$ 
with the help of Eqs.\ (\ref{dmu}) and (\ref{dT1}).
Usually, one accounts only for the dependence of
$\widehat{\varepsilon}$ and $\widehat{\mu}$ on $n$ \cite{toptygin15}.
In the latter case Eq.\ (\ref{TAbraham})
reduces to the standard equation for Abraham tensor (see, e.g., Refs.\ \cite{gu76, toptygin15}).

\section{General formulas for isolated neutron and proton vortices}
\label{vortex1}

Here we briefly review the properties of isolated neutron and proton vortices 
taking into account the entrainment effect \cite{ab76} 
and closely following Refs.\ \cite{als84,mendell91a}, GAS11, and G16.
Note, however, that our consideration differs from that 
in Refs.\ \cite{als84,mendell91a} and GAS11
in three aspects: (i) we use a bit different (but equivalent) 
formulation of superfluid hydrodynamics;
(ii) we consider {\it relativistic} $npe$-mixture, 
and thus employ
relativistic entrainment matrix instead 
of its nonrelativistic counterpart \cite{ab76};
(iii) we do not assume the zero-temperature approximation.
Although below we make use of the London equations, 
one should bear in mind that it is not a very good approximation 
when the particle coherence length becomes comparable 
to their London penetration depth \cite{ll80, degennes99}.

\subsection{London equations and their solution}
\label{London}

Assume that a neutron $i=n$ or proton $i=p$ vortex is at rest
in the chosen coordinate frame
and there are no external (superfluid and normal) 
particle currents and magnetic field at the spatial infinity.
We also assume that all the velocities generated by the vortex are nonrelativistic
(but, at the same time, equation of state is relativistic),
so that one can use nonrelativistic expressions for, 
e.g., particle current densities.
All equations below are written in dimensional units.

Consider, for example, a proton vortex ($i=p$; the case $i=n$ can then be obtained 
by exchanging $p \rightleftharpoons n$ in all formulas).
In the presence of the vortex $p$ the gradient of the scalar $\phi_p$,
which is proportional to the wave-function phase $\Phi_p$ 
of the Cooper-pair condensate ($\phi_p=\Phi_p/2$),
is given by (e.g., G16)
\begin{equation}
\partial^a \phi_p=\frac{{\pmb e}_\varphi}{2r},
\label{phii}
\end{equation}
where ${\rm {\pmb e}}_{\varphi}$ is the unit vector in the azimuthal direction 
($\varphi$ is the polar angle); $r$ is the distance from the vortex;
and $a=1$, $2$, $3$ is the space index.
Using Eq.\ (\ref{def2}) one then has 
\begin{equation}
w^a_{(p)}=\hbar c \,\, \partial^a \phi_p-e_p A^a,
\label{wi}
\end{equation}
where we make use of the fact that $u^a=(0,\,0,\,0)$.
Similarly, for neutrons one has
\begin{eqnarray}
\partial^a \phi_n &=& 0,
\label{phik}\\
w^a_{(n)} &=& -e_n A^a,
\label{wk}
\end{eqnarray}
(we do not set $e_n=0$ in order to rewrite easily these formulas 
for neutron vortex if necessary),
so that the total electric current density is [see Eq.\ (\ref{jfree2})]
\begin{equation}
{\pmb J}_{\rm free}
=c \, e_i Y_{ik} w^a_{(k)}=a_1 A^a + a_2 \partial^a \phi_p,
\label{J}
\end{equation}
where the parameters $a_1$ and $a_2$
\begin{eqnarray}
a_1 &=& -c \, (e_n^2 Y_{nn}+2 e_n e_p Y_{np}+ e_p^2 Y_{pp}),
\label{a1}\\
a_2 &=& \hbar c^2 \left(e_n Y_{np}+e_p Y_{pp}\right)
\label{a2}
\end{eqnarray}
are constants since we neglect small dependence of $Y_{ik}$ on $r$ 
(see, e.g., Ref.\ \cite{khalatnikov00} and G16 where a similar approximation is discussed).
Now, using Maxwell's equations (\ref{divB1}) and (\ref{rotB1}) 
with ${\pmb H}={\pmb B}$, one arrives at the following equation 
for the vortex magnetic field ${\pmb B}$
\begin{equation}
-\Delta {\pmb B}=\frac{4\pi}{c}\left[a_1 {\pmb B}+\pi a_2 \, {\pmb e}_z \, \delta(r)\right],
\label{Beq}
\end{equation}
or
\begin{equation}
\Delta {\pmb B}- \frac{1}{\delta_p^2} \, {\pmb B} = 
- \frac{\hat{\phi}_{p0}}{\delta_p^2} \,\, {\pmb e}_z \, \delta(r),
\label{Beq2}
\end{equation}
where $\delta(r)$ is the two-dimensional delta-function in polar coordinate system $(r,\, \phi)$;
${\pmb e}_z$ is the unit vector along the vortex axis; and
\begin{eqnarray}
\frac{1}{\delta_p^2} &\equiv& - \frac{4\pi a_1}{c}=4 \pi \, \left(e_n^2 Y_{nn}+2 e_n e_p Y_{np}+ e_p^2 Y_{pp}\right) ,
\label{lambdap}\\
\hat{\phi}_{p0} &\equiv& -\frac{\pi a_2}{a_1}=\frac{\pi \, \hbar c \, 
	\left(e_n Y_{np}+e_p Y_{pp}\right)}{e_n^2 Y_{nn}+2 e_n e_p Y_{np}+ e_p^2 Y_{pp}}.
\label{phi0p}
\end{eqnarray}
Here $\delta_p$ is the London penetration depth and
$\hat{\phi}_{p0}$ is the magnetic flux associated with the vortex (see below).
The nonrelativistic limit of these equations can be reproduced if
one takes into account that then 
$Y_{ik} \rightarrow \rho_{ik}/(m_i m_k c^2)$,
where $\rho_{ik}$ is the entrainment (or mass-density) matrix \cite{ab76, bjk96, gh05, ch06, gusakov10}.
Equation (\ref{Beq2}) can easily be solved \cite{ll80}, 
the result is 
\begin{equation}
{\pmb B}({\pmb r}) = \frac{\hat{\phi}_{p0}}{2\pi \delta_p^2} \,\, K_0\left(\frac{r}{\delta_p}\right) \, {\pmb e}_z,
\label{Beqsolve}
\end{equation}
where $K_0(r)$ is the MacDonald function.
One can verify that, indeed, 
$\hat{\phi}_{p0}$ is the total vortex magnetic flux,
$\int_0^\infty B(r) \, 2\pi r dr = \hat{\phi}_{p0}$.
Using (\ref{Beqsolve}), one finds:
${\rm curl} \, {\pmb B}= \hat{\phi}_{p0}/(2 \pi \delta_p^3)\,  K_1(r/\delta_p) \, {\pmb e}_\varphi$,
and hence from Eqs.\ (\ref{rotB1}) and (\ref{J}) 
\begin{equation}
{\pmb A}({\pmb r}) = \frac{\hat{\phi}_{p0}}{2\pi } \left[ \frac{1}{r}-\frac{1}{\delta_p} 
\, K_1\left(\frac{r}{\delta_p}\right)  \right] \,  {\pmb e}_{\varphi},
\label{Apot}
\end{equation}
so that Eqs.\ (\ref{wi}) and (\ref{wk}) can be rewritten as
\begin{eqnarray}
w^a_{(p)} &=& \frac{\hbar c}{2r} \left(1-\frac{e_p\hat{\phi}_{p0}}{\pi \, \hbar c}\right) 
\,{\pmb e}_{\varphi} + \frac{e_p \, \hat{\phi}_{p0}}{2 \pi \, \delta_p} \,\, K_1\left(\frac{r}{\delta_p }\right) \,{\pmb e}_{\varphi},
\label{wi2}\\
w^a_{(n)} &=& -\frac{e_n \, \hat{\phi}_{p0}}{2\pi } \left[ \frac{1}{r}-\frac{1}{\delta_p} 
\, K_1\left(\frac{r}{\delta_p}\right)  \right] \,  {\pmb e}_{\varphi}.
\label{wk2}
\end{eqnarray}
For neutron vortex 
similar formulas 
can be obtained by exchanging 
$p \rightleftharpoons n$ in Eqs.\ (\ref{phii})--(\ref{wk2}).
Note that, in the case of protons,
the first term in the r.h.s.\ of Eq.\ (\ref{wi2}) equals zero.

\subsection{Vortex energy}
\label{energyapp}

Neglecting a small contribution from the vortex core, 
the general expression for the vortex energy per unit length is 
\begin{equation}
\hat{E}_{\rm V} = \int \frac{1}{2} \left[
Y_{nn} \, {\pmb w}_{n}^2+
2 Y_{np} \, {\pmb w}_{n}{\pmb w}_{p}
+Y_{pp} \, {\pmb w}_{p}^2
\right] r dr d\varphi
+\int \frac{B^2}{8\pi}\,  r dr d\varphi,
\label{Energy}
\end{equation}
where ${\pmb w}_{i}=[w^1_{(i)},\, w^2_{(i)},\, w^3_{(i)}]$.
The first integral in this equation is the kinetic energy of superfluid currents 
\cite{mendell91a,kg09}; 
the second
integral is the magnetic energy,
it is generally smaller (e.g., GAS11).
Equations (\ref{Beqsolve}), (\ref{wi2}), and (\ref{wk2})
allow one to calculate 
the integrals in Eq.\ (\ref{Energy})
and to obtain the following approximate expressions
for, respectively, proton $\hat{E}_{{\rm V}p}$ and neutron $\hat{E}_{{\rm V}n}$ vortex energies per unit length,
\begin{eqnarray}
\hat{E}_{{\rm V}p} &\approx& \frac{\pi}{4} \,\, \hbar^2 c^2\,\, Y_{pp} \,\, {\rm ln}\left(\frac{\delta_p}{\xi_p}\right),
\label{Evp}\\
\hat{E}_{{\rm V}n} &\approx& \frac{\pi}{4} \,\, \hbar^2 c^2\,\,
\frac{(Y_{nn} Y_{pp}-Y_{np}^2)}{Y_{pp}}\,\,
{\rm ln}\left(\frac{b_n}{\xi_n}\right).
\label{Evn}
\end{eqnarray}
In these formulas $\xi_p$ and $\xi_n$ are, respectively the proton and neutron 
coherence lengths \cite{mendell91a} (effective sizes of the vortex cores)
and $b_n$ is some ``external'' radius of the order 
of the typical intervortex spacing (see, e.g., Refs.\ \cite{khalatnikov00} and G16).
In the nonrelativistic limit these formulas reduce to the corresponding expressions
(A12) and (A18) of Mendell \cite{mendell91a}.

Equations (\ref{Evp}) and (\ref{Evn})
are derived under assumption that 
a neutron (proton)
vortex is at rest in the comoving frame 
[i.e., in the frame in which $u^{\mu}=(1,\,0,\,0,\,0)$].
As it is argued in G16 in application to uncharged superfluids, 
the same equations
also apply to
moving vortices, 
provided that 
the difference between the macroscopic (smooth-averaged) 
normal and superfluid velocities in the system is much smaller than the speed of light $c$.
The latter condition is always satisfied in neutron stars (see G16 for details).
Thus, 
it is justifiable to 
assume 
that
Eqs.\ (\ref{Evp}) and (\ref{Evn}) represent
correct vortex energies, 
independently of whether vortices move 
or not.

\section{Bound charges in the presence of vortices}
\label{why}

The aim of this appendix is to explain why 
the displacement field ${\pmb D}$ is not generally equal 
to the electric field ${\pmb E}$ in the system with vortices.
In what follows it is assumed that we sit in the comoving frame, 
i.e. the frame associated with the normal liquid component. 
Consider, 
for example,
a single proton vortex directed along the axis $z$ 
of the Cartesian coordinate system $xyz$ 
and moving with the velocity 
${\pmb v}_{\rm L}=v_{{\rm L} x} \,{\pmb e}_x + v_{{\rm L} z} \,{\pmb e}_z$,
where
${\pmb e}_x$ and ${\pmb e}_z$ are the unit vectors along the axes $x$ and $z$, respectively.
In the rest frame of the vortex its magnetic field ${\pmb B}(r)$ is given by Eq.\ (\ref{Beqsolve}).
Correspondingly, as follows from Eq.\ (\ref{rotE1}), in the comoving frame 
it generates the electric field (e.g., Ref.\ \cite{gk75})
\begin{equation}
{\pmb E}=-\frac{1}{c} \,\, {\pmb v}_{\rm L} \times {\pmb B}(r)
\label{Evort}
\end{equation}
(we assume that $|{\pmb v}_{\rm L}|\ll c$, 
which is always the case \cite{gk13}; 
the same formula can be obtained by making Lorentz transformation 
from the vortex rest frame to the comoving frame).
An associated charge density, $\rho_{\rm c}$, 
induced in that frame, 
is found from Maxwell's equation 
${\rm div} \, {\pmb E}=4\pi \rho_{\rm c}$,
\begin{equation}
\rho_{\rm c}= \frac{v_{{\rm L}x}}{4\pi c}\, 
\frac{d B(r)}{dr}\, {\rm sin} \varphi,
\label{rhoc}
\end{equation}
where $\varphi$ is the polar angle in the $xy$-plane.
Correspondingly, the dipole moment of the vortex segment of length $\Delta z$
is given by
\begin{equation}
{\pmb P}_{\rm V}=\int {\pmb r} \, \rho_{\rm c} \, dV = - \frac{v_{{\rm L}x}}{4\pi c} 
\, \hat{\phi}_{p0} \, \Delta z \, {\pmb e}_y,
\label{dipole}
\end{equation}
where $\hat{\phi}_{p0}$ is introduced in Eq.\ (\ref{phi0p2}) 
and ${\pmb e}_y$ is the unit vector along $y$.
Now, assuming that there are many vortices moving with one and the same velocity 
${\pmb v}_{\rm L}$, 
the dipole moment of the unit volume is
\begin{equation}
{\pmb P}= \frac{{\pmb P}_{\rm V} N_{{\rm V}p}}{\Delta z}=
- \frac{v_{{\rm L}x}}{4\pi^2\, c \,b_p^2} 
\, \hat{\phi}_{p0} \, {\pmb e}_y
\label{dipole2}
\end{equation}
[see Eq.\ (\ref{Nvi}) for a definition of $N_{{\rm V}p}$].
It is easily checked that ${\pmb P}$ 
and the average electric field ${\pmb E}$, generated by vortices, 
are related by the standard condition \cite{ll60}, 
${\pmb E} = -4 \pi \, {\pmb P}$, 
which should take place 
for any homogeneous system in which all currents are bound, so that
${\pmb D}=0$.
We come to conclusion that the electric field of moving vortices 
should be considered as produced by bound charges, 
similar to how their magnetic field is produced by (vortex) bound currents.
A further implication of this observation can be found in Appendix \ref{vortapp}.

\section{Determination of the phenomenological coefficients of Sec.\ \ref{symmetry} for two simple microscopic models}
\label{deterapp}

Our aim here will be to determine the exact form of Eq.\ (\ref{2ndlaw3}) 
(or, equivalently, to find an expression for $d\varepsilon_{\rm add}$)
in two situations considered above (intermediate state and ``vortex'' state of $npe$-mixture).
This aim can be achieved by specifying a microphysics model for the energy density
of the system.
Below, for illustration, we consider two very simple microphysics models 
(in particular, the model, considered in Sec.\ \ref{vortapp} was studied in GAS11),
but one should bear in mind that the very same approach can be used to formulate
dynamic equations for more elaborated models.

\subsection{Intermediate state of a nonrotating $npe$-mixture (type-I proton superconductivity)}
\label{interapp}

Assume we are sitting in the normal-liquid (comoving) frame
in which nonsuperconducting domains (flux tubes) are at rest.
Let us calculate the coefficient $\gamma$ in Eq.\ (\ref{Hmux2}),
which allow us to 
determine $d \varepsilon_{\rm add}$ from Eq.\ (\ref{deadd4}).
In what follows, instead of $\varepsilon$ it will be more convenient to
deal with
the (Helmholtz) free energy density, $F \equiv \varepsilon-T S$.

The magnitude of the field in a flux tube
coincides with the critical thermodynamic field $H_{\rm c}$ \cite{ll60},
it is directed along the average magnetic induction ${\pmb B}$, 
and can be found from
the following approximate formula \cite{degennes99},
\begin{equation}
F_{\rm nonsp} - F_{\rm sp} \approx \frac{H_{\rm c}^2}{8\pi},
\label{equation}
\end{equation}
where $F_{\rm nonsp}$ 
is the free energy density of nonsuperconducting matter
in the flux tube %
%
\footnote{It does not include the energy of the magnetic field \cite{degennes99}.}
%
and $F_{\rm sp}$  is the free energy density 
of the surrounding (superconducting) matter,
it is the same function of thermodynamic quantities as in the absence of the magnetic field.

Now, introducing the volume fraction occupied by nonsuperconducting domains, $x_{\rm nonsp}$,
and following the consideration of Refs.\ \cite{degennes99, ll60}
(in particular, neglecting all striction effects),
it is easy to obtain 
an expression for
the macroscopically averaged free energy density $F$ of npe-mixture in the intermediate state, 
\begin{equation}
F \approx F_{\rm sp}
+ \frac{H_{\rm c}^2}{4\pi} \, x_{\rm nonsp}.
\label{F}
\end{equation}
On the other hand, magnetic flux conservation requires 
that the average magnetic induction ${\pmb B}$
to be given by $|{\pmb B}|=H_{\rm c} x_{\rm nonsp}$.
Hence, Eq.\ (\ref{F}) can be represented as \cite{ll60}
\begin{equation}
F \approx F_{\rm sp}
+ \frac{H_{\rm c} |{\pmb B}|}{4\pi}=
F_{\rm sp}
+
\frac{H_{\rm c}}{4\pi} \, (B_{\mu}B^{\mu})^{1/2}.
\label{F2}	
\end{equation}
The latter equality is written in an explicitly Lorentz-invariant form; 
$B^{\mu}$ is given by Eq.\ (\ref{Bmux}).
Now, using Eqs.\ (\ref{2ndlaw3}), (\ref{deadd4}), (\ref{F2}), 
and the definition $F\equiv \varepsilon-TS$,
one can find that the macroscopic parameter 
$H^{\mu}$
of the phenomenological theory of Sec.\ \ref{TypeIsymmetry} is
\begin{eqnarray}
H^{\mu} &=&\gamma B^{\mu},
\label{Hmux3}
\end{eqnarray}
where 
\begin{equation}
\gamma = \frac{H_{\rm c}}{(B_{\mu}B^{\mu})^{1/2}}.
\label{gamma2}
\end{equation}
%

\vspace{0.2 cm}
\noindent
%
{\bf Remark 1.} ---
The model discussed here and in Sec.\ \ref{TypeIsymmetry}
is designed at describing nonrotating $npe$-mixture in the intermediate state.
Generalization of the model to 
allow for
rotation and neutron vortices
is rather straightforward and can be done along the lines discussed 
in Appendix \ref{vortapp}.

\subsection{The $npe$-mixture with neutron and proton vortices (type-II proton superconductivity)}
\label{vortapp}

We follow here the approach similar to that described
in section 4.2 of GAS11 and 
in Appendix D of G16.
We work in the comoving frame and  
neglect vortex-vortex interactions in all calculations.
Assume we have a bunch of parallel neutron or proton vortices 
with the intervortex spacing 
$b_i$ ($i=n$ or $p$).
The parameter $b_i$ is related to the average number of vortices
$N_{{\rm V}i}$ per unit area by the formula 
(see, e.g., Ref.\ \cite{khalatnikov00} and G16),
\begin{equation}
\pi b_i^2 =\frac{1}{N_{{\rm V}i}}.
\label{Nvi}
\end{equation}
On the other hand, 
as follows from Eqs.\ (\ref{int33}) and (\ref{Vmagn}) 
[cf.\ Eq.\ (D9) of G16],
\begin{equation}
N_{{\rm V}i} = \frac{|\epsilon^{abc}\,\mathcal{V}_{(i)bc}|}{2\pi\hbar}
= \frac{1}{\pi\hbar}\, |\mathcal{V}^a_{({\rm M}i)}|, 
\label{Nvi1}
\end{equation}
where $a$, $b$, and $c$ are the space indices and we use dimensional units.
To obtain this formula we perform integration in Eq.\ (\ref{int33}) 
over the unit area in the plane 
perpendicular to vortex lines.
The areal density $N_{{\rm V}i}$ is {\it defined} 
in the comoving frame. 
It is thus a Lorentz invariant 
and it can be rewritten in an explicitly Lorentz-invariant form as
\begin{equation}
N_{{\rm V}i} = \frac{1}{\pi\hbar}\, \sqrt{\mathcal{V}_{({\rm M}i)\mu}\mathcal{V}^\mu_{({\rm M}i)}}
= \frac{1}{\pi\hbar}\, \mathcal{V}_{({\rm M}i)}
\label{Nvi2}
\end{equation}
[see Eq.\ (\ref{VM}) for the definition of $\mathcal{V}_{({\rm M}i)}$].
For an uncharged fluid 
$\mathcal{V}_{({\rm M}i)}$
reduce, in the non-relativistic limit,
to $m_i \,|{\rm curl}\, {\pmb V}_{{\rm s}i}|$, 
where $m_i$ is the mass of particle species $i$
and ${\pmb V}_{{\rm s}i}$ is the superfluid velocity.

Using Eqs.\ (\ref{Nvi}), (\ref{Nvi2}) and (\ref{Evp}), (\ref{Evn}),
the vortex energy density $\varepsilon_{{\rm vortex} \, i}$
can be presented as~%
%
\footnote{
\label{20app}
Strictly speaking, this is the vortex energy obtained under assumption that 
the vortex is at rest in the comoving frame.
Thus, it neglects, for example, the contribution to the energy density 
from the electric field generated by a moving vortex
(see Appendix \ref{why}). 
All such contributions are small and can be ignored, 
as it is emphasized
in the end of Appendix \ref{energyapp}.
}
%
\begin{equation}
\varepsilon_{{\rm vortex} \, i} = \frac{\hat{E}_{{\rm V}i}}{\pi b_i^2}=
\frac{\hat{E}_{{\rm V}i}}{\pi\hbar} \, \mathcal{V}_{({\rm M}i)} 
\equiv \frac{ \lambda_i}{m_i} \, \mathcal{V}_{({\rm M}i)},
\label{ev}
\end{equation}
where 
\begin{eqnarray}
\lambda_p &=& \frac{1}{4} \,\, \hbar c^2\, m_p \, Y_{pp} \,\, {\rm ln}\left(\frac{\delta_p}{\xi_p}\right),
\label{lambdap1}\\
\lambda_n &=& \frac{1}{4} \,\, \hbar c^2\,m_n\,
\frac{(Y_{nn} Y_{pp}-Y_{np}^2)}{Y_{pp}}\,\,
{\rm ln}\left(\frac{b_n}{\xi_n}\right).
\label{lambdan1}
\end{eqnarray}
In the absence of entrainment ($Y_{np}=0$)
or for a one-component liquid 
Eq.\ (\ref{lambdan1})
reduces to the parameter $\lambda$ 
defined in Eq.\ (D10) of G16. 
This parameter is, in turn, the relativistic generalization
of the parameter $\lambda$ introduced in Refs.\ \cite{bk61,khalatnikov00}.

The contribution of vortex magnetic field ${\pmb B}_{{\rm V}i}$ 
to the total magnetic induction
can be found the same way as $\varepsilon_{{\rm vortex} \, i}$ [cf. Eq.\ (50) of GAS11],
\begin{equation}
{\pmb B}_{{\rm V}i}=\hat{\phi}_{i0} N_{{\rm V}i} \,\, 
\frac{\mathcal{V}^a_{({\rm M}i)}}{\mathcal{V}_{({\rm M}i)}}
=\frac{\hat{\phi}_{i0}}{\pi \hbar}  \,\,\mathcal{V}^a_{({\rm M}i)},
\label{BVi}
\end{equation}
where $\mathcal{V}^a_{({\rm M}i)}/\mathcal{V}_{({\rm M}i)}$
is the unit vector along the local direction of vortex lines,
while $\hat{\phi}_{p0}$ and $\hat{\phi}_{n0}$ are [see Eq.\ (\ref{phi0p})]
\begin{eqnarray}
\hat{\phi}_{p0} &=& \frac{\pi \hbar c}{e_p},
\label{phi0p2}\\
\hat{\phi}_{n0} &=&\frac{\pi \hbar c}{e_p} \, \frac{Y_{np}}{Y_{pp}}.
\label{phi0n2}
\end{eqnarray}
Similarly, the contribution of the vortex electric field ${\pmb E}_{{\rm V}i}$, 
to the (averaged) electric field ${\pmb E}$
is (see Appendix \ref{why})
\begin{equation}
{\pmb E}_{{\rm V}i}=-\frac{1}{c} \,\, {\pmb v}_{{\rm L}i} \times {\pmb B}_{{\rm V}i}
= \frac{\hat{\phi}_{i0}}{\pi \hbar c} \,   \mathbfcal{V}_{{\rm M}i}\times {\pmb v}_{{\rm L}i}
=\frac{\hat{\phi}_{i0}}{\pi \hbar } \,  \mathbfcal{V}_{{\rm E}i},
\label{Evi}
\end{equation}
where ${\pmb v}_{{\rm L}i}$ is the velocity of vortex species $i$. 
To obtain the last two equalities in the r.h.s.\ of Eq.\ (\ref{Evi}) 
we made use of Eqs.\ (\ref{important}) and (\ref{BVi}).

Having determined $\varepsilon_{{\rm vortex}\,i}$, 
our next step
will be to write down the total energy density $\varepsilon$
of the system in the comoving frame.
As it is discussed in detail in GAS11, it is the sum of five
%
\footnote{GAS11 considered only the first four of these terms and ignored the last one
since that reference assumed (incorrectly) that there are no bound charges in the system.}
%
``noninterfering'' terms 
(see also G16 for a similar discussion of $\varepsilon$ in an uncharged fluid),
\begin{equation}
\varepsilon=
\varepsilon_{\rm fluid}(n_n,\, n_p,\, n_e,\, S,\, w_{(i)\mu} w^\mu_{(k)})
+\varepsilon_{{\rm vortex}\,n}+\varepsilon_{{\rm vortex}\,p}
+\frac{{\pmb B}_{\rm L}^2}{8\pi} 
+\frac{{\pmb E}_{\rm L}^2}{8\pi}.
\label{energydensity}
\end{equation}
The first term here is the same as in the absence of vortices and magnetic field
in the system; it consists of the internal energy of the fluid at rest plus 
kinetic energy of superfluid currents (i.e., terms depending on $w_{(i)\mu} w^\mu_{(k)}$).
The differential of $\varepsilon_{\rm fluid}$ contribute only to the first four terms in Eq.\ (\ref{2ndlaw3})
and do not affect $d\varepsilon_{\rm add}$.
Thus, this term is not interesting for us here.
The second and third terms account for the vortex energies, 
including the magnetic energy of vortices.
Further, the fourth term represents the magnetic energy density of the so called ``London field'',
which is not associated with vortices. 
The London field can be non-zero even far from vortices and 
for our model it equals
\begin{equation}
{\pmb B}_{\rm L} = {\pmb B}-{\pmb B}_{{\rm V}n}-{\pmb B}_{{\rm V}p}
\label{BLondon}
\end{equation}
[see Eq.\ (\ref{BVi}) for the definition of vortex contribution to magnetic induction].
Generally, this field is very small. 
For example, for a uniformly rotating one-component 
vortex-free superconductor ${\pmb B}_{\rm L} \approx -2 m c \, {\pmb \Omega}/e=
-2 \times 10^{-2} \, [{\pmb \Omega}/(100\, {\rm s}^{-1})] \,  {\rm G}$,
where ${\pmb \Omega}$ is the spin frequency, and to make the estimate we take $m=m_p$ and $e=e_p$ 
(see, e.g., Ref.\ \cite{mendell91a} and GAS11 for more details).
Finally, the last term in Eq.\ (\ref{energydensity})
is similar to the fourth term, but describes the {\it electric} energy density
of matter, not associated with vortices.
Similarly to Eq.\ (\ref{BLondon}), it can be presented as
\begin{equation}
{\pmb E}_{\rm L} = {\pmb E}-{\pmb E}_{{\rm V}n}-{\pmb E}_{{\rm V}p}.
\label{ELondon}
\end{equation}

The two last terms in the r.h.s.\ of Eq.\ (\ref{energydensity}) can be rewritten
in the manifestly Lorentz-invariant form, 
${\pmb B}_{\rm L}^2/(8\pi)=B_{({\rm L})\mu}B^{\mu}_{({\rm L})}/(8\pi)$
and 
${\pmb E}_{\rm L}^2/(8\pi)=E_{({\rm L})\mu}E^{\mu}_{({\rm L})}/(8\pi)$,
if we introduce the London field four-vectors $B^{\mu}_{({\rm L})}$ 
and $E^{\mu}_{({\rm L})}$,
\begin{eqnarray}
B^{\mu}_{(\rm L)} &\equiv& B^{\mu}-B^{\mu}_{({\rm V}n)}-B^{\mu}_{({\rm V}p)},
\label{BLondon2}\\
E^{\mu}_{(\rm L)} &\equiv& E^{\mu}-E^{\mu}_{({\rm V}n)}-E^{\mu}_{({\rm V}p)},
\label{ELondon2}
\end{eqnarray}
where the corresponding vortex-related four-vectors are defined as
\begin{eqnarray}
B^{\mu}_{({\rm V}i)} &\equiv&  \frac{\hat{\phi}_{i0}}{\pi \hbar}  \,\,\mathcal{V}^\mu_{({\rm M}i)},
\label{BVP2}\\
E^{\mu}_{({\rm V}i)} &\equiv&  \frac{\hat{\phi}_{i0}}{\pi \hbar}  \,\,\mathcal{V}^\mu_{({\rm E}i)}.
\label{EVP2}
\end{eqnarray}
It is easily verified that in the comoving frame the time components of 
these four-vectors $B^{\mu}_{(\rm L)}$, 
$B^{\mu}_{({\rm V}i)}$, $E^{\mu}_{(\rm L)}$, and $E^{\mu}_{({\rm V}i)}$
are all zero, 
while their spatial components coincide with those of 
the 3D-vectors 
${\pmb B}_{\rm L}$, 
${\pmb B}_{{\rm V}i}$, ${\pmb E}_{\rm L}$, and ${\pmb E}_{{\rm V}i}$, 
respectively [see Eqs.\ (\ref{BVi}), (\ref{Evi}), (\ref{BLondon}), and (\ref{ELondon})]
%
\footnote{Note also that when protons are normal one has 
$Y_{np}=0$ \cite{gkh09b}, hence $\hat{\phi}_{n0}=0$ 
and, consequently, $B^{\mu}_{({\rm V}n)}=E^{\mu}_{({\rm V}n)}=0$.}.
%

Using these definitions as well as 
Eqs.\ (\ref{ev}) and (\ref{energydensity}), 
the second law of thermodynamics (\ref{2ndlaw3}) takes the form
\begin{equation}
d \varepsilon = T \, d S + \mu_i \, d n_i + \mu_{e} \, d n_{e} 
+ { Y_{ik} \over 2} \, d \left( w^{\alpha}_{(i)} w_{(k) \alpha} \right)
+d \varepsilon_{\rm add},
\label{2ndlaw5}
\end{equation}
where 
\begin{eqnarray}
T &=&\frac{\partial \varepsilon_{\rm fluid}}{\partial S}
+\sum_{k=n,\, p} \left\{\frac{1}{m_k}\, 
\frac{\partial \lambda_k}{\partial S} \, \mathcal{V}_{({\rm M}k)}
-\underline{\frac{1}{4 \pi^2 \hbar} \,
\frac{\partial \hat{\phi}_{k0}}{\partial S} \,
B_{({\rm L})\mu} \,
 \mathcal{V}^{\mu}_{({\rm M}k)}
}
 \right\},
\label{T}\\
\mu_i &=&\frac{\partial \varepsilon_{\rm fluid}}{\partial n_i}
+\sum_{k=n,\, p} \left\{\frac{1}{m_k}\, 
\frac{\partial \lambda_k}{\partial n_i} \, \mathcal{V}_{({\rm M}k)}
-\underline{\frac{1}{4 \pi^2 \hbar} \,
\frac{\partial \hat{\phi}_{k0}}{\partial n_i} \,
B_{({\rm L})\mu} \,
\mathcal{V}^{\mu}_{({\rm M}k)}
}
\right\},
\label{mui}\\
\mu_e &=&\frac{\partial \varepsilon_{\rm fluid}}{\partial n_e},
\label{mue}\\
Y_{ik} &=& 2 \, \frac{\partial \varepsilon_{\rm fluid}}{\partial (w^{\alpha}_{(i)} w_{(k) \alpha})}
\nonumber\\
&+&2 \sum_{l=n,\, p} 
\left\{
\underline{\underline{
\frac{1}{m_l}\, 
\frac{\partial \lambda_l}{\partial [w^{\alpha}_{(i)} w_{(k) \alpha}]} \, \mathcal{V}_{({\rm M}l)}
}}
-
\underline{
\frac{1}{4 \pi^2 \hbar} \,
\frac{\partial \hat{\phi}_{l0}}{\partial [w^{\alpha}_{(i)} w_{(k) \alpha}]} \,
B_{({\rm L})\mu} \,
\mathcal{V}^{\mu}_{({\rm M}l)}
}
\right\},
\label{Yik}\\
d\varepsilon_{\rm add} &=& \sum_{k=n,\, p} \frac{\lambda_k}{m_k \, \mathcal{V}_{({\rm M}k)}} 
\mathcal{V}_{({\rm M}k)\mu} d \mathcal{V}^{\mu}_{({\rm M}k)}
\nonumber\\
&+& \frac{1}{4\pi} \, B_{({\rm L})\mu} \,
\left[d B^{\mu}-\frac{\hat{\phi}_{n0}}{\pi\hbar} \, d \mathcal{V}^{\mu}_{({\rm M}n)}
-\frac{\hat{\phi}_{p0}}{\pi\hbar} \, d \mathcal{V}^{\mu}_{({\rm M}p)}\right]
\nonumber\\
&+& \frac{1}{4\pi} 
\left[E^{\mu}-\frac{\hat{\phi}_{n0}}{\pi\hbar} \, \mathcal{V}^{\mu}_{({\rm E}n)}
-\frac{\hat{\phi}_{p0}}{\pi\hbar} \,  \mathcal{V}^{\mu}_{({\rm E}p)}\right] d E_{({\rm L})\mu}.
\label{deadd6}
\end{eqnarray}
In Eqs.\ (\ref{T})--(\ref{deadd6})
the parameters $\varepsilon_{\rm fluid}$, $\lambda_i$, and $\hat{\phi}_{i0}$
should be treated as the same functions of 
$S$, $n_i$, $n_e$, and $w^{\alpha}_{(i)} w_{(k) \alpha}$
as in the absence of vortices and the magnetic field.
The underlined terms there are generally small and can be neglected.
The terms underlined once are small because they depend 
on the tiny London field $B^{\mu}_{({\rm L})}$ 
[see Eq.\ (\ref{BLondon2})];
the term underlined twice is small because $\lambda_i$ is a very weak function of 
$w^{\alpha}_{(i)} w_{(k) \alpha}$ 
in the regime when the dependence of $Y_{ik}$ on the 
difference between the velocities of superfluid and normal liquid components can be neglected
(e.g., G16).
The second term in Eq.\ (\ref{deadd6}) also depends on $B^{\mu}_{({\rm L})}$ 
and can, in principle, be omitted.
However, we keep it in what follows because it is 
this term which makes $H^{\mu}$ non-zero.
Comparing (\ref{deadd6}) with the 
general expression (\ref{deadd5}) for $d \varepsilon_{\rm add}$
and using Eqs.\ (\ref{Hmu2})--(\ref{WmuE3}), 
one finds 
\begin{eqnarray}
\gammaM &=& \gammaEtilde =1,
\label{gamma}\\
\GammaMi &=&  \GammaEitilde = - \frac{\hat{\phi}_{i0}}{4\pi^2\hbar},
\label{Gammai2}\\
\GammaMik &=& \frac{\lambda_i}{m_i \mathcal{V}_{({\rm M}i)}}\delta_{ik} 
	+ \frac{\hat{\phi}_{i0}\, \hat{\phi}_{k0}}{4\pi^3 \hbar^2},
\label{Gammaik2}\\
\GammaEiktilde &=& 
\frac{\hat{\phi}_{i0}\, \hat{\phi}_{k0}}{4\pi^3 \hbar^2}.
\label{Gammaik3}
\end{eqnarray}
The latter equation differs from its magnetic counterpart, Eq.\ (\ref{Gammaik2}),
because we neglected the electric field contribution 
to 
the vortex energy,
$\varepsilon_{{\rm vortex} \, i}$.
From Eqs.\ (\ref{Hmu2}), (\ref{Emu3}), (\ref{BLondon2}), and (\ref{ELondon2}) 
it then follows that $H^{\mu}=B^{\mu}_{({\rm L})}$ and 
$D^{\mu}=E^{\mu}_{({\rm L})}$.
The first of these equalities was earlier discussed in GAS11.

\vspace{0.2 cm}
\noindent
%
{\bf Remark 1.} ---
The results obtained above allow us to 
make a few useful estimates. 
First of all, 
since the total number of neutron vortices in a star
is by more than ten orders of magnitude smaller than the total number 
of proton vortices 
(for a typical neutron star with $B \sim 10^{12}$~G 
and a period $P \sim 0.1$~s, see, e.g., GAS11),
one can neglect ${\pmb B}_{{\rm V}n}$ 
and ${\pmb E}_{{\rm V}n}$ in comparison to, respectively, ${\pmb B}_{{\rm V}p}$
and ${\pmb E}_{{\rm V}p}$ in Eqs.\ (\ref{BLondon}) and (\ref{ELondon}), 
and write
\begin{eqnarray}
{\pmb B} &=& {\pmb H} + {\pmb B}_{{\rm V}n} + {\pmb B}_{{\rm V}p} \approx {\pmb B}_{{\rm V}p},
\label{Bapp}\\
{\pmb E} &=& {\pmb D} + {\pmb E}_{{\rm V}n} + {\pmb E}_{{\rm V}p} \approx {\pmb D} + {\pmb E}_{{\rm V}p}.
\label{Eapp}
\end{eqnarray}
Here we also neglect ${\pmb H}$ 
in Eq.\ (\ref{Bapp})
since typically 
$|{\pmb H}|\sim 2 \times 10^{-2} 
\, [\Omega/(100\, {\rm s}^{-1})] \, {\rm G} \ll |{\pmb B}|$,
as discussed in the text above.
Second, note that for a static or very weakly perturbed neutron star 
[i.e., a star for which ${\pmb v}_{{\rm L}p}$ is so small, 
that ${\pmb E}_{{\rm V}p}$ in Eq.\ (\ref{Eapp}) 
can be neglected, see Eq.\ (\ref{Evi})],
one can estimate $|{\pmb E}|$ (and $|{\pmb D}|$) as
$|{\pmb D}| \approx |{\pmb E}| \sim |{\pmb \nabla} \mu_e|/e_p \sim 1$~g$^{1/2}$~cm$^{-1/2}$~s$^{-1}$.
The latter estimate allows one to find an approximate proton vortex velocity $v_{{\rm L}p0}$
at which  $|{\pmb E}_{{\rm V}p}|$ becomes comparable
to $|{\pmb D}|$. 
Using Eq.\ (\ref{Evi}), one finds 
$v_{{\rm L}p0} \sim c \, 
|{\pmb \nabla} \mu_e|/(e_p |{\pmb B}_{{\rm V}p}|) 
\sim 3 \times 10^{-2}$~cm~s$^{-1}$ (we take $|{\pmb B}_{{\rm V}p}| \approx |{\pmb B}| = 10^{12}$~G).
Thus, for example, at $|{\pmb v}_{{\rm L}p}|\gg v_{{\rm L}p0}$
one has:
${\pmb E}\approx -(1/c) \, {\pmb v}_{{\rm L}p}\times {\pmb B}$, 
so that
$|{\pmb H}| \lesssim |{\pmb D}| \ll |{\pmb E}| \ll |{\pmb B}|$.
Correspondingly, in the opposite limit 
$|{\pmb H}| \lesssim |{\pmb D}| \approx |{\pmb E}| \ll |{\pmb B}|$.

\section{Summary of results: full system of relativistic equations describing dynamics of superfluid-superconducting neutron stars}
\label{sumapp}

Here we present the full system of dynamic equations discussed in the main text.
For the reader's convenience this appendix is made self-contained.
In the present paper we are mainly interested in nondissipative equations 
(the only dissipative mechanism,
which is 
accounted for,
is the mutual friction, see below).
Thus, we assume
that neutron and proton thermal excitations 
(Bogoliubov quasiparticles), 
as well as electrons
move with one and the same four-velocity $u^{\mu}$, 
normalized by the condition
$u_{\mu}u^{\mu}=-1$. 

Superfluid degrees of freedom are characterized by the 
four-vectors
$w^{\mu}_{(i)}$ ($i=n$, $p$), 
which are closely related to the superfluid velocities 
of the corresponding nonrelativistic theory (see Appendix \ref{nonrel}),
and are orthogonal to $u^{\mu}$,
\begin{equation}
u_{\mu} w^{\mu}_{(i)} =0.
\label{uwapp}
\end{equation}
Other important  parameters of the theory include
the vorticity tensors $\mathcal{V}^{\mu\nu}_{(i)}$,
\begin{eqnarray}
\mathcal{V}^{\mu\nu}_{(i)} &\equiv & 
\partial^\mu\left[ w^\nu_{(i)}+\mu_i u^{\nu}\right]
-\partial^\nu\left[ w^\mu_{(i)}+\mu_i u^{\mu}\right]+e_i F^{\mu\nu},
\label{Vmunuapp}
\end{eqnarray}
and the 
electromagnetic tensors $F^{\alpha\beta}$ and $G^{\alpha\beta}$
[see Eqs.\ (\ref{Fik}) and (\ref{Gik})], 
satisfying Maxwell's equations (\ref{11}) and (\ref{22}),
\begin{eqnarray}
\partial_{\alpha} 
\Fdual^{\alpha\beta}&=&0,
\label{11app}\\
\partial_\alpha G^{\alpha\beta}&=&-4\pi \, J_{({\rm free})}^\beta.
\label{22app}
\end{eqnarray}
In these formulas $e_i$ is the charge of nucleon species $i$;
\begin{equation}
J^\mu_{({\rm free})}= e_p (n_p-n_e) u^\mu + e_i \, Y_{ik}w^{\mu}_{(k)}
\label{jfreeapp}
\end{equation}
is the four-current density of free charges 
[see Eqs.\ (\ref{jfree2}) and (\ref{Jnorm})]; 
$\Fdual^{\mu\nu}$
is the tensor dual to $F^{\mu\nu}$ (see Appendix \ref{notation});
the thermodynamic parameters $n_e$, $n_p$, and $Y_{ik}$ are defined in what follows. 
In addition to the tensors 
$\mathcal{V}^{\mu\nu}_{(i)}$, $F^{\alpha\beta}$, and $G^{\alpha\beta}$
it is convenient to introduce the four-vectors 
[see Eqs.\ (\ref{Emux})--(\ref{Hmux}),  (\ref{Velectr}), and (\ref{Vmagn})]
\begin{eqnarray}
\mathcal{V}^{\mu}_{({\rm E}i)} &\equiv& 
u_\nu \mathcal{V}^{\mu\nu}_{(i)},
\label{Velectrapp}\\
\mathcal{V}^{\mu}_{({\rm M}i)} &\equiv&  \frac{1}{2} \, \epsilon^{\mu \nu \alpha \beta} \, u_{\nu} \, \mathcal{V}_{(i) \alpha \beta},
\label{Vmagnapp}\\
E^\mu &\equiv&  u_{\nu} F^{\mu\nu},
\label{Emuxapp}\\
D^{\mu}&\equiv& u_{\nu} G^{\mu\nu},
\label{Dmuxapp}\\
B^{\mu} &\equiv&  
\frac{1}{2} \, \epsilon^{\mu \nu \lambda \eta} \, u_{\nu} \, F_{\lambda \eta},
\label{Bmuxapp}\\
H^{\mu} &\equiv&  
\frac{1}{2} \, \epsilon^{\mu \nu \lambda \eta} \, u_{\nu} \, G_{\lambda \eta}.
\label{Hmuxapp}
\end{eqnarray}
In the {\it comoving frame} in which the normal liquid component is at rest 
[i.e., $u^{\mu}=(1,\, 0,\, 0,\, 0)$]
the space components of the four-vectors  
$E^{\mu}$, $D^{\mu}$, $B^{\mu}$, and $H^{\mu}$ reduce to
the electric field, displacement field, magnetic induction, and magnetic field, respectively.

The equations describing dynamics of superfluid-superconducting $npe$-mixture
consist of: (i) Maxwell's equations (\ref{11app}) and (\ref{22app});
(ii) the particle and energy-momentum conservations,
\begin{eqnarray}
\partial_{\mu}j^{\mu}_{(j)} &=&0,
\label{jmueqapp}\\
\partial_{\mu}T^{\mu\nu}&=&0
\label{Tmunueqapp}
\end{eqnarray}
with
\begin{eqnarray}
j^{\mu}_{(i)} &=& n_i u^{\mu} + Y_{ik} w^{\mu}_{(k)},
\label{jiapp}\\
j^{\mu}_{({e})} &=& n_{e} u^{\mu},
\label{jeapp}
\end{eqnarray}
and
\begin{equation}
T^{\mu \nu} = (P+\varepsilon) \, u^{\mu} u^{\nu} + P g^{\mu \nu} 
+ Y_{ik} \left( w^{\mu}_{(i)} w^{\nu}_{(k)} + \mu_i \, w^{\mu}_{(k)} u^{\nu} 
+ \mu_k \, w^{\nu}_{(i)} u^{\mu} \right) + \Delta T^{\mu\nu};
\label{Tmunu6app}
\end{equation}
(iii) the second law of thermodynamics 
[note that all the thermodynamic quantities
are measured in the comoving frame, where $u^{\mu}=(1,\,0,\,0,\,0)$], 
\begin{equation}
d \varepsilon = T \, d S + \mu_i \, d n_i + \mu_{e} \, d n_{e} 
+ { Y_{ik} \over 2} \, d \left( w^{\alpha}_{(i)} w_{(k) \alpha} \right)
+d \varepsilon_{\rm add};
\label{2ndlaw6app}
\end{equation}
and (iv) the superfluid equations, 
which 
will be discussed a bit later. 
In Eqs.\ (\ref{jmueqapp})--(\ref{2ndlaw6app})
$n_j$ and  $\mu_j$ are, respectively, 
the number density and relativistic chemical potential of 
particle species $j=n$, $p$, $e$; $T$, $S$, $\varepsilon$, and 
$P=-\varepsilon+\mu_e n_e + \mu_i n_i + TS$ 
are the temperature, entropy density, energy density, and pressure,
respectively.
Note that all the thermodynamic quantities are defined (measured) in the comoving frame.
Finally, $Y_{ik}$ is the relativistic entrainment matrix \cite{ga06,gusakov07, gkh09a, gkh09b, ghk14}
and $g^{\mu\nu} = {\rm diag}(-1,\, 1,\, 1,\, 1)$ is the metric tensor.

The corrections $\Delta T^{\mu\nu}$ and
$d \varepsilon_{\rm add}$ 
in Eqs.\ (\ref{Tmunu6app}) and (\ref{2ndlaw6app})
appear due to the electromagnetic and vortex contributions to the energy-momentum tensor 
and energy density, 
and differ
depending on the assumed type (I or II) of the proton superconductivity.
The same is also true for superfluid equations, 
thus they should be discussed separately for each case.

\subsection{Vortex-free $npe$-mixture in the intermediate state (type-I proton superconductivity)}
\label{TypeIapp}

Assuming that protons in the $npe$-mixture  
form a type-I superconductor in the intermediate state 
and that neutrons are superfluid, 
one has the following formulas for 
$\Delta T^{\mu\nu}$ and $d \varepsilon_{\rm add}$ (see Sec.\ \ref{TypeI})
\begin{eqnarray}
\Delta T^{\mu\nu} &=& \mathcal{T}^{\mu\nu}_{({\rm E})}+
\mathcal{T}^{\mu\nu}_{({\rm M})},
\label{DeltaTmunuapp}\\
d \varepsilon_{\rm add} &=& \frac{1}{4\pi} \, E_{\mu} d D^{\mu} + \frac{1}{4\pi} \, H_{\mu} d B^{\mu},
\label{deadd1app}
\end{eqnarray}
where
\begin{eqnarray}
\mathcal{T}^{\mu\nu}_{({\rm E})} &=& \frac{1}{4\pi} \, 
\left(
\perp^{\mu\nu} D^{\alpha}E_{\alpha}-D^{\mu}E^{\nu}
\right),
\label{Te1app}\\
\mathcal{T}^{\mu\nu}_{({\rm M})} &=& 
\frac{1}{4\pi} 
\left(
\Gperp^{\mu\alpha}\Fperp^{\nu}_{\,\,\,\alpha} + u^{\nu} \Gperp^{\mu\alpha}E_{\alpha}
+u^{\mu}\Gperp^{\nu\alpha}E_{\alpha}
\right),
\label{Tm1app}
\end{eqnarray}
and 
\begin{eqnarray}
\Gperp^{\mu\nu} &=& \epsilon^{\alpha\beta\mu\nu} \, 
u_{\beta}\, H_{\alpha}
\label{Gperp}
\end{eqnarray}
(see Appendix \ref{notation}). 
In turn, superfluid equations for protons and neutrons take the form 
[see Eqs.\ (\ref{sfleq3}) and (\ref{sfleq3drag}); 
we assume that there are no neutron
vortices in the system]
\begin{eqnarray}
\mathcal{V}^{\mu\nu}_{(n)} &=&0,
\label{Vappn}\\
u_\mu \mathcal{V}^{\mu\nu}_{(p)}&=&0.
\label{Vappp}
\end{eqnarray}
These equations should be supplemented by the two 
conditions relating 
the four-vectors $D^{\mu}$ with $E^{\mu}$
and $H^{\mu}$ with $B^{\mu}$. 
These 
conditions 
are obtained in Sec.\ \ref{TypeIsymmetry} and in Appendix \ref{interapp}.

\subsection{$npe$-mixture in the presence of neutron and proton vortices (type-II proton superconductivity)}
\label{TypeIIapp}

Assume now that protons form a type-II superconductor and consider
$npe$-mixture in the mixed state, allowing for the presence 
of both neutron and proton vortices.
The corrections $\Delta T^{\mu\nu}$ 
and $d \varepsilon_{\rm add}$ 
are then given by (see Sec.\ \ref{TypeII})
\begin{eqnarray}
d \varepsilon_{\rm add} &=& 
\frac{1}{4\pi} \, E_{\mu} d D^{\mu} + \frac{1}{4\pi} \, H_{\mu} d B^{\mu}
+ \mathcal{V}^{\mu}_{({\rm E}i)} d \mathcal{W}_{({\rm E}i)\mu}
+ \mathcal{W}_{({\rm M}i)\mu} d \mathcal{V}^{\mu}_{({\rm M}i)},
\label{deadd2app}\\
\Delta T^{\mu\nu}&=&
\mathcal{T}^{\mu\nu}_{({\rm E})}+
\mathcal{T}^{\mu\nu}_{({\rm M})} 
+\mathcal{T}^{\mu\nu}_{({\rm VE})}
+\mathcal{T}^{\mu\nu}_{({\rm VM})},
\label{DTmunuapp}
\end{eqnarray}
where
$\mathcal{W}^\mu_{({\rm E}i)}$ 
and 
$\mathcal{W}^\mu_{({\rm M}i)}$ 
are
the four-vectors analogous to $D^{\mu}$ and $H^{\mu}$, respectively;
their relation to the four-vectors 
$\mathcal{V}^{\mu}_{({\rm E}i)}$, $E^{\mu}$,
$\mathcal{V}^{\mu}_{({\rm M}i)}$, and $B^{\mu}$
is explored
in Sec.\ \ref{TypeIIsymmetry} 
and (for a particular model) in Appendix \ref{vortapp}.
In Eq.\ (\ref{DTmunuapp}) 
$\mathcal{T}^{\mu\nu}_{({\rm E})}$ and
$\mathcal{T}^{\mu\nu}_{({\rm M})}$ are given by Eqs.\ (\ref{Te1app}) and (\ref{Tm1app}), 
respectively, while
$\mathcal{T}^{\mu\nu}_{({\rm VE})}$ and $\mathcal{T}^{\mu\nu}_{({\rm VM})}$ 
are
\begin{eqnarray}
\mathcal{T}^{\mu\nu}_{({\rm VE})} &=&  
\perp^{\mu\nu} \mathcal{W}_{({\rm E}i)}^{\alpha}\mathcal{V}_{({\rm E}i)\alpha }
-\mathcal{W}_{({\rm E}i)}^{\mu} \mathcal{V}^{\nu}_{({\rm E}i)},
\label{vortexEapp}\\
\mathcal{T}^{\mu\nu}_{({\rm VM})} &=& 
\Wperp^{\mu\alpha}_{(i)}\Vperp^{\nu}_{(i)\,\alpha} + u^{\nu} \Wperp^{\mu\alpha}_{(i)} \mathcal{V}_{({\rm E}i)\alpha}
+u^{\mu}\Wperp^{\nu\alpha}_{(i)}\mathcal{V}_{({\rm E}i)\alpha},
\label{vortexapp}
\end{eqnarray}
where
\begin{eqnarray}
\Wperp^{\mu\nu}_{(i)} &=& \epsilon^{\alpha\beta\mu\nu} u_{\beta} \, 
\mathcal{\mathcal{W}}_{({\rm M}i)\, \alpha}.
\label{Wperpapp}
\end{eqnarray}
The superfluid equations for neutrons ($i=n$) and protons ($i=p$) 
take the form
\begin{equation}
u^\nu \mathcal{V}_{(i)\mu\nu}= \mu_i n_i \, f_{(i)\mu},
\label{sflrot1app}
\end{equation}
where
\begin{equation}
f^{\mu}_{(i)} = \alpha_i \perp^{\mu\nu} \mathcal{V}_{(i)\nu\lambda} \, W_{(i)\delta} \perp^{\lambda \delta}
+ \frac{\beta_i-\gamma_i}{\mathcal{V}_{({\rm M}i)}}  \perp^{\mu\eta} \perp^{\nu\sigma}  \mathcal{V}_{(i)\eta\sigma} \mathcal{V}_{(i)\lambda\nu}  \,
W_{(i)\delta} \perp^{\lambda \delta}
+\gamma_i  \mathcal{V}_{({\rm M}i)} \, W_{(i)\delta} \perp^{\mu \delta},
\label{f3app}
\end{equation}
(see a Remark 1 in Sec.\ \ref{TypeII}). 
In Eq.\ (\ref{f3app}) 
$\perp^{\mu\nu}=g^{\mu\nu}+u^{\mu}u^{\nu}$;
$\alpha_i$ is a non-dissipative mutual friction coefficient; 
$\beta_i\geq0$ and $\gamma_i\geq 0$ are the positive 
dissipative mutual friction coefficients, and %
%
\footnote{
\label{Wparapp}	
The tensor $\mathcal{W}^{\mu \nu}_{(i)}$ in Eq.\ (\ref{Wmuapp}) equals
$\mathcal{W}^{\mu \nu}_{(i)}=\Wpar^{\mu\nu}_{(i)} + \Wperp^{\mu\nu}_{(i)}$, 
where $\Wpar^{\mu\nu}_{(i)} = 
-u^{\nu} \mathcal{W}^{\mu}_{({\rm E}i)}+u^{\mu} \mathcal{W}^{\nu}_{({\rm E}i)}$ 
(see Appendix \ref{notation}).	
}
%
\begin{eqnarray}
W^{\mu}_{(i)} &\equiv& \frac{1}{n_i}\left[
Y_{ik} w^\mu_{(k)} 
+ \partial_{\alpha} \mathcal{W}^{\mu \alpha}_{(i)}
\right],
\label{Wmuapp}\\
\mathcal{V}_{({\rm M}i)} &\equiv& \sqrt{\mathcal{V}^{\mu}_{({\rm M}i)}\mathcal{V}_{({\rm M}i)\mu}}.
\label{VMapp}
\end{eqnarray}
Recalling the definition (\ref{Velectrapp}), 
one sees that 
Eq.\ (\ref{sflrot1app}) 
is simply the statement that
\begin{equation}
\mathcal{V}^\mu_{({\rm E}i)} = \mu_i n_i \, f_{(i)}^{\mu}.
\label{sflrot1app2}
\end{equation}

As in Appendix \ref{TypeIapp}, the dynamic equations formulated here
should be supplemented with the expressions relating the vectors 
$D^{\mu}$, $H^{\mu}$, 
$\mathcal{W}^{\mu}_{({\rm E}i)}$,
$\mathcal{W}^{\mu}_{({\rm M}i)}$ with 
$E^{\mu}$, $B^{\mu}$, $\mathcal{V}^{\mu}_{({\rm E}i)}$, $\mathcal{V}^{\mu}_{({\rm M}i)}$.
These expressions are discussed in Sec.\ \ref{TypeIIsymmetry} and in Appendix \ref{vortapp}.

\section{Nonrelativistic limit of ``magnetohydrodynamic'' equations of Sec.\ \ref{MHDapprox}}
\label{nonrel}

Here we present the nonrelativistic limit of the simplified dynamic
equations discussed in Sec.\ \ref{MHDapprox}.
In what follows, unless otherwise stated,
all the 3D-vectors appearing in the text 
(shown in boldface)
are defined in the {\it laboratory} frame.
As in other parts of the paper, 
indices $i$ and $k$ 
refer to nucleons: $i$, $k=n$, $p$;
other Latin letters are the space indices;
we use dimensional units in this Appendix.

The four-vector $u^{\mu}$ is related to the normal
velocity ${\pmb V}_{{\rm norm} }$ 
of the nonrelativistic superfluid hydrodynamics
by the standard formula,
\begin{equation}
u^{\mu} \equiv (u^0,\,{\pmb u}) = \left(\frac{1}{\sqrt{1-{\pmb V}_{{\rm norm}}^2/c^2}},\, 
\frac{{\pmb V}_{{\rm norm}}}{c\sqrt{1-{\pmb V}_{{\rm norm}}^2/c^2}} \right).
\label{umuNR}
\end{equation}
Instead of the four-vector $w^{\mu}_{(i)}\equiv (w^0_{(i)},\, {\pmb w}_{i})$
it is convenient to introduce the superfluid four-velocity 
$V^{\mu}_{({\rm s}i)}\equiv(V^{0}_{({\rm s}i)},\,{\pmb V}_{{\rm s}i} )$, 
such that 
\begin{equation}
w^{\mu}_{(i)}=m_i c \,  V^{\mu}_{({\rm s}i)} -\mu_i u^{\mu}.
\label{wiapp}
\end{equation}
As shown in G16 (see also Ref.\ \cite{ga06}),
the spatial component ${\pmb V}_{{\rm s}i}$ of this four-vector
is the superfluid velocity of the nonrelativistic theory %
%
\footnote{Note that $V^{\mu}_{({\rm s}i)}$ is measured in cm/s while $u^{\mu}$
is dimensionless [see Eq.\ (\ref{umuNR})].}.
%
Using Eq.\ (\ref{uw}) and the definition (\ref{wiapp}) 
one finds the following equation for $V^{\mu}_{({\rm s}i)}$,
$u_{\mu}  V^{\mu}_{({\rm s}i)}=-\mu_i/(m_i c)$,
from which the time component $V^0_{({\rm s}i)}$ 
is
\begin{equation}
V^0_{({\rm s}i)}=
\frac{\mu_i}{m_i c \, u^0}+\frac{{\pmb u} \, {\pmb V}_{{\rm s}i}}{u^0}.
\label{w0app}
\end{equation}
In terms of $V^{\mu}_{({\rm s}i)}$ the vorticity tensor (\ref{Vmunu})
can be rewritten as
\begin{eqnarray}
\mathcal{V}^{\mu\nu}_{(i)} &\equiv &\frac{1}{c}\left\{
\partial^\mu\left[ w^\nu_{(i)}+\mu_i u^{\nu}\right]
-\partial^\nu\left[ w^\mu_{(i)}+\mu_i u^{\mu}\right]+e_i F^{\mu\nu} \right\}
\nonumber\\
&=& 
m_i \left[\partial^\mu V^{\nu}_{({\rm s}i)} - \partial^\nu V^{\mu}_{({\rm s}i)}\right]
+\frac{e_i}{c}\, F^{\mu\nu},
\label{Vmunuapp2}
\end{eqnarray}
while the electric vector $\mathcal{V}^{\mu}_{({\rm E}i)}$  
is given by Eq.\ (\ref{nuE}),
and the 
magnetic vector 
$\mathcal{V}^{\mu}_{({\rm M}i)}$ 
is [see Eq.\ (\ref{Vmagnapp})]
\begin{equation}
\mathcal{V}^{\mu}_{({\rm M}i)} = 
 \frac{1}{2} \, \epsilon^{\mu \nu \alpha \beta} \, u_{\nu} \, 
m_i \, 
  \left[\partial_\alpha V_{({\rm s}i)\beta} - \partial_\beta V_{({\rm s}i)\alpha}\right]
 +\frac{e_i}{c} \, B^{\mu}.
\label{VMapp2}
\end{equation}
and reduces to 
$\mathcal{V}^{\mu}_{({\rm M}i)}=(0,\, m_i\, {\pmb \omega_i})$ 
in the comoving frame, where we defined 
\begin{equation}
{\pmb \omega_i} \equiv   {\rm curl} \, {\pmb V}_{{\rm s}i}+\frac{e_i}{m_i c}\, {\pmb B}.
\label{omega}
\end{equation}
To leading order in ${\pmb V}_{\rm norm}/c$ 
the same expression 
$\mathcal{V}^{\mu}_{({\rm M}i)}=(0,\, m_i\, {\pmb \omega_i})$ 
is also valid in the laboratory frame 
(and this is also true 
for other ``magnetic'' vectors).
It remains to express the relativistic entrainment matrix, $Y_{ik}$, 
through its nonrelativistic counterpart, $\rho_{ik}$.
As shown, e.g., in Ref.\ \cite{ga06}, 
in the nonrelativistic limit they are related by the formula:
$\rho_{ik}=m_i m_k c^2 \, Y_{ik}$.

Using these definitions and relations the nonrelativistic 
version of the 
superfluid equation (\ref{sflrot1})
takes the form 
\begin{eqnarray}
\partial_t {\pmb V}_{{\rm s}i} + ({\pmb V}_{{\rm s}i}{\pmb \nabla}){\pmb V}_{{\rm s}i} 
+ {\pmb \nabla}\left[ \breve{\mu}_i 
- \frac{1}{2} \left| {\pmb V}_{{\rm s}i}-{\pmb V}_{\rm norm}\right|^2\right]
&=&-{\rm curl}{\pmb V}_{{\rm s}i}\times \left({\pmb V}_{\rm norm}-{\pmb V}_{{\rm s}i}\right) 
\nonumber\\
&-& n_i \, {\pmb f}_i
+ \frac{e_i}{m_i} \, \left({\pmb E+\frac{{\pmb V}_{\rm norm}}{c}\times {\pmb B}}\right),
\label{spatialNR7}
\end{eqnarray}
where 
$\breve{\mu}_i \equiv (\mu_i-m_i c^2)/m_i$
and
\begin{eqnarray}
{\pmb f}_i=-\alpha_i\, m_i \, [{\pmb \omega}_i\times {\pmb W}_i]
-\beta_i\, m_i\,  {\pmb {\rm e}_i} \times[{\pmb \omega}_i \times {\pmb W}_i]+
\gamma_i\, m_i \,  {\pmb {\rm e}_i} ({\pmb W}_i \, {\pmb \omega}_i).
\label{fifiNR}
\end{eqnarray}
In the latter formula 
${\pmb e}_i={\pmb \omega}_i/|{\pmb \omega}_i|$; 
${\pmb W}_i$ is the spatial part of the four-vector $W^{\mu}_{(i)}$,
which is, in the dimensional form [see footnote \ref{Wparapp} and Eqs.\ (\ref{Wperpapp}), (\ref{Wmu})], 
\begin{equation}
W^\mu_{(i)} = \frac{1}{n_i} \left[c \, Y_{ik} w^{\mu}_{(k)} 
+\partial_{\alpha}\left(\epsilon^{\delta\beta\mu\alpha} u_{\beta} \, 
\mathcal{\mathcal{W}}_{({\rm M}i)\, \delta}+\Wpar^{\mu\alpha}_{(i)} \right)
\right].
\label{WNRapp}
\end{equation}

We have not made yet
any simplifying assumption 
about the value of the magnetic induction ${\pmb B}$,
so up until now our nonrelativistic 
equations are quite general.
Now let us make full use of
simplifications
of Sec.\ \ref{MHDapprox} 
%
\footnote{We remind the reader that Sec.\ \ref{MHDapprox} utilizes
the model of noninteracting vortices discussed in Appendix \ref{vortapp}. }.
%
Employing Eqs.\ (\ref{BVprelation}) and (\ref{Vmunu2}),
Eq.\ (\ref{omega}) can be presented as
\begin{eqnarray}
{\pmb \omega}_n={\rm curl} \, {\pmb V}_{{\rm s}n},
\label{omegan}\\
{\pmb \omega_p} \approx \frac{e_p}{m_p c}\, {\pmb B}.
\label{omegap}
\end{eqnarray}
In turn, Eq.\ (\ref{WM4}) becomes
\begin{eqnarray}
\mathcal{W}^{\mu}_{({\rm M}n)} &\equiv& 
(0,\,  \mathbfcal{W}_{{\rm M}n}) 
=\frac{\lambda_n}{m_n \mathcal{V}_{({\rm M}n)}} \,
\mathcal{V}^{\mu}_{({\rm M}n)}
= \frac{\lambda_n}{m_n \omega_n} \,
(0, \, \pmb{\omega}_n),
\label{WM5appn}\\
\mathcal{W}^{\mu}_{({\rm M}p)} &\equiv& 
(0,\,  \mathbfcal{W}_{{\rm M}p}) 
\approx \frac{\lambda_p}{m_p \mathcal{V}_{({\rm M}p)}} \,
\mathcal{V}^{\mu}_{({\rm M}p)}
\approx \frac{\lambda_p}{m_p B} \,
(0, \, \pmb{B}).
\label{WM5appp}
\end{eqnarray}
As it was argued in Sec.\ \ref{MHDapprox}, the term depending on $\Wpar^{\mu\alpha}_{(i)}$
in Eq.\ (\ref{WNRapp}) is small and can be omitted. 
Thus, the resulting nonrelativistic expression for $W^{\mu}_{(i)}$
is given by (see Appendix C of G16 for a similar equation)
\begin{equation}
{\pmb W}_i= \frac{1}{n_i} 
\left[ \sum_{k=n, \, p} \frac{\rho_{ik}}{m_i}\left({\pmb V}_{{\rm s}k} 
-{\pmb V}_{\rm norm}\right)  
+ {\rm curl} \,  \mathbfcal{W}_{{\rm M}i}
\right],
\label{Wvectapp}
\end{equation}
where the vectors $\mathbfcal{W}_{{\rm M}i}$ 
are defined in Eqs.\ (\ref{WM5appn})--(\ref{WM5appp}).
Equations (\ref{omegan})--(\ref{omegap}) and (\ref{Wvectapp})
should be used to calculate ${\pmb f}_i$ [see Eq.\ (\ref{fifiNR})].
Equation (\ref{Wvectapp}) can be further simplified in the case of protons ($i=p$)
if we note that the conditions (\ref{quasi}) and (\ref{jfree4})
can be rewritten as
\begin{eqnarray}
&& n_e = n_p,
\label{quasiapp}\\
&&\sum_{k=n, \, p} \rho_{pk} \left({\pmb V}_{{\rm s}k} 
-{\pmb V}_{\rm norm}\right)    = 0.
\label{jfree4app}
\end{eqnarray}
Using Eqs.\ (\ref{Wvectapp}) and (\ref{jfree4app}), one obtains
\begin{equation}
{\pmb W}_p= \frac{1}{n_p}\,  {\rm curl} \,  \mathbfcal{W}_{{\rm M}p}.
\label{Wvectapp1}
\end{equation}

Next, within the 
magnetohydrodynamic
approximation
adopted here, 
the vortex-related corrections (\ref{deadd7}) and (\ref{dTmunu2})
to, respectively,
the second law of thermodynamics (\ref{2ndlaw3})
and 
the energy-momentum tensor (\ref{Tmunu3})
are given, in the nonrelativistic limit, by 
\begin{eqnarray}
d \varepsilon_{\rm add} &\approx& 
\sum_{i=n,\, p} \frac{\lambda_i}{m_i \mathcal{V}_{({\rm M}i)}} \, \mathcal{V}_{({\rm M}i)\mu}
d \mathcal{V}^{\mu}_{({\rm M}i)}
= \frac{\lambda_n}{|{\pmb \omega}_n|} \, {\pmb \omega}_n \, d{\pmb \omega}_n
+\frac{\lambda_p}{|{\pmb \omega}_p|} \, {\pmb \omega}_p \, d{\pmb \omega}_p,
\label{deadd7app}\\
\Delta T^{\mu\nu} &=& \mathcal{T}^{\mu\nu}_{\rm (VM)} =
\left( 
\begin{array}{cc}
0 & {\pmb g}_n \\ 
{\pmb g}_n & \Pi^{lm}_{({\rm V}n)}
\end{array} 
\right)
+\left( 
\begin{array}{cc}
0 & {\pmb g}_p \\ 
{\pmb g}_p & \Pi^{lm}_{({\rm V}p)}
\end{array} 
\right),
\label{Tvapp1}
\end{eqnarray}
where
\begin{eqnarray}
{\pmb g}_i &=& \frac{1}{c} \, 
\left[ 
m_i n_i \, {\pmb f}_i 
+ \left( \mathbfcal{V}_{{\rm M}i} \times {\pmb V}_{\rm norm}\right)
\right]\times \mathbfcal{W}_{{\rm M}i},
\label{gapp}\\
\Pi^{lm}_{({\rm V}i)} &=&
\mathbfcal{V}_{{\rm M}i} \mathbfcal{W}_{{\rm M}i} \,  \delta^{lm}
-\mathcal{V}^l_{({\rm M}i)} \mathcal{W}^m_{({\rm M}i)}
\label{pilm3}
\end{eqnarray}
and $\mathbfcal{V}_{{\rm M}i}=m_i \,{\pmb \omega}_i$.
Using 
the definition for the critical magnetic field $H_{c1}$, 
$H_{c1} \equiv 4 \pi \hat{E}_{{\rm V}p}/\hat{\phi}_{p0}$ (see, e.g., Ref.\ \cite{ll80}),
as well as
Eqs.\ (\ref{ev}), (\ref{phi0n2}), (\ref{omegap}), (\ref{WM5appp}), and (\ref{pilm3}) 
it is easily demonstrated that the proton tensor $\Pi^{lm}_{({\rm V}p)}$ 
can be represented as
\begin{equation}
\Pi^{lm}_{({\rm V}p)} = \frac{H_{c1}}{4\pi} \left( B \, \delta^{lm} - \frac{B^l B^m}{B} \right). 
\label{Plmp}
\end{equation}

Note that $d \varepsilon_{\rm add}$ in Eq.\ (\ref{deadd7app})
can be considered as defined in the laboratory frame
up to corrections $\sim {\pmb V}_{\rm norm}/c$. 
All other parameters and equations of the theory
[e.g., continuity equations, 
the remaining parts of the second law of thermodynamics (\ref{2ndlaw3})
and the energy-momentum tensor (\ref{Tmunu3})] 
have the same form as in the standard (vortex-free) superfluid hydrodynamics 
(see, e.g., Refs.\ \cite{khalatnikov00, holm01, putterman74} and G16).
However, it is very important to point out 
that the temperature $T$ and chemical potential $\mu_i$ 
will be \underline{renormalized} in the presence of vortices 
according to Eqs.\ (\ref{T}) and (\ref{mui}).

\vspace{0.2 cm}
\noindent
%
{\bf Remark 1.} ---
Using the equations obtained above it is straightforward to derive 
the ``magnetic evolution'' equation.
To this aim let us take a curl of Eq.\ (\ref{spatialNR7}) written for protons ($i=p$).
Then, using Maxwell's equation (\ref{rotE1}) and 
neglecting the terms depending on 
${\rm curl} \,{\pmb V}_{{\rm s}p}$ in comparison 
to the similar terms depending on $e_p/(m_p c) \, {\pmb B}$
[our magnetohydrodynamic approximation; 
see a note after Eq.\ (\ref{Vmunu2})],
one gets
\begin{equation}
\frac{\partial {\pmb B}}{\partial t} + {\rm curl} 
\left[
\frac{m_p c}{e_p} \, n_p {\pmb f}_p + {\pmb B}\times{\pmb V}_{\rm norm}
\right]=0.
\label{MagnEvol}
\end{equation}
This equation can be further simplified if one neglects the small 
kinetic coefficient $\gamma_p$ in Eq.\ (\ref{fifiNR}).
Eq.\ (\ref{MagnEvol}) can then be rewritten as
(see also Ref.\ \cite{kg01} for a similar equation)
%
\begin{equation}
\frac{\partial {\pmb B}}{\partial t} + {\rm curl} \,
({\pmb B}\times {\pmb v}_{{\rm L}p})=0,
\label{MagnEvol2}
\end{equation}
where ${\pmb v}_{{\rm L}p}$ is the nonrelativistic velocity of proton vortices
[spatial part of the four-vector $v^{\mu}_{({\rm L}p)}$, see Eq.\ (\ref{VlX0})],
given by
\begin{equation}
{\pmb v}_{{\rm L}p} = 
{\pmb V}_{\rm norm} - \alpha_p \, m_p n_p \,  {\pmb W}_p - \frac{\beta_p}{B} \,
m_p n_p \,\, {\pmb B}\times {\pmb W}_p 
\label{Vlapp}
\end{equation}
with
\begin{equation}
{\pmb W}_p = \frac{1}{m_p n_p} \, 
{\rm curl} \left( \frac{\lambda_p}{B}\, {\pmb B}\right)
\label{Wp}
\end{equation}
[see Eqs.\ (\ref{WM5appp}) and (\ref{Wvectapp1})].
The physical meaning of Eq.\ (\ref{MagnEvol2}) is obvious:
It describes transport of the magnetic field 
(produced by the proton vortices) with the vortices.
A bit different equation has been recently obtained, 
in the approximation of vanishing temperature, in Ref.\ \cite{gagl15}
[see Eq.\ (67) there]
%
\footnote{The same equation follows from the Maxwell's equation (\ref{rotE1}) 
and Eqs.\ (161) and (162) of GAS11.}.
%
The magnetic field in that reference is transported with the velocity 
which {\it differs} from the vortex velocity ${\pmb v}_{{\rm L}p}$.
This is a puzzling result, since Ref.\ \cite{gagl15} explicitly assumes that 
the magnetic field is confined to proton vortices 
[see Eq.\ (65) in that reference]
and hence should be carried along with them.

Note, in passing, that the energy consideration 
of Ref.\ \cite{gagl15} 
does not look convincing.
In particular, Eq.\ (76) in that reference disagrees
with the result of Ref.\ \cite{ep77} for the free magnetic energy density $F_{\rm mag}$
(which must coincide with the magnetic energy density in the limit of $T=0$), 
see the formula after Eq.\ (16) in Ref.\ \cite{ep77}.


%

\end{document}